\documentclass[letterpaper,11pt]{article}
\pdfoutput=1

\usepackage{jheppub}

\usepackage{braket}
\usepackage{mathrsfs}

\usepackage{subfig}
\usepackage{xspace}
\usepackage{xcolor}
\usepackage[countmax]{subfloat}

\definecolor{darkblue}{rgb}{1,0,0}

\definecolor{darkgreen}{rgb}{0,0.5,0}

\newcommand{\ze}{ze}

\newcommand{\zcut}{z_\text{cut}}

\def\nn{\nonumber}

\newcommand{\glob}{\text{G}}

\newcommand{\ecf}[2]{e_{#1}^{(#2)}}

\DeclareRobustCommand{\Sec}[1]{Sec.~\ref{#1}}
\DeclareRobustCommand{\Secs}[2]{Secs.~\ref{#1} and \ref{#2}}
\DeclareRobustCommand{\App}[1]{App.~\ref{#1}}
\DeclareRobustCommand{\Tab}[1]{Table~\ref{#1}}

\DeclareRobustCommand{\Fig}[1]{Fig.~\ref{#1}}
\DeclareRobustCommand{\Figs}[2]{Figs.~\ref{#1} and \ref{#2}}
\DeclareRobustCommand{\Eq}[1]{Eq.~(\ref{#1})}
\DeclareRobustCommand{\Eqs}[2]{Eqs.~(\ref{#1}) and (\ref{#2})}
\DeclareRobustCommand{\Ref}[1]{Ref.~\cite{#1}}
\DeclareRobustCommand{\Refs}[1]{Refs.~\cite{#1}}

\newcommand{\pythia}[1]{\textsc{Pythia\xspace #1}}

\newcommand{\fastjet}[1]{\textsc{FastJet\xspace #1}}

\newcommand{\herwigpp}[1]{\textsc{Herwig++\xspace #1}}
\newcommand{\vincia}[1]{\textsc{Vincia\xspace #1}}

\newcommand{\eventtwo}{\texttt{EVENT2}}
\newcommand{\eerad}{\texttt{EERAD3}}

\def\cO{\mathcal{O}}

\def\slash{\!\!\!/}

\def\be{\begin{equation}}
\def\ee{\end{equation}}

\allowdisplaybreaks

\newcommand{\ppmatch}{D}

\title{
Factorization for groomed jet substructure beyond the next-to-leading logarithm
}

\author{Christopher Frye,}
\author{Andrew J.~Larkoski,}
\author{Matthew D.~Schwartz,}
\author{Kai Yan}

\affiliation{Center for the Fundamental Laws of Nature, Harvard University, Cambridge, MA 02138, USA}

\emailAdd{frye@physics.harvard.edu}
\emailAdd{larkoski@physics.harvard.edu}
\emailAdd{schwartz@physics.harvard.edu}
\emailAdd{kyan@physics.harvard.edu}

\abstract{
Jet grooming algorithms are widely used in experimental analyses at hadron colliders to 
remove contaminating radiation from within jets. 
While the algorithms perform a great service to the experiments, their intricate algorithmic 
structure and multiple parameters has frustrated precision theoretic understanding. 
In this paper, we demonstrate that one particular groomer  called {\it soft drop}
actually makes precision jet substructure easier. 
In particular, we derive a factorization formula for a large class of soft drop jet substructure
observables, including jet mass. 
The essential observation that allows for this factorization is that,
without the soft wide-angle radiation groomed by soft drop, all singular contributions are collinear.
The simplicity and universality of the collinear limit  in QCD allows us to show that
 to all orders, the normalized differential cross section has no contributions from 
non-global logarithms. 
It is also independent of process, up to the relative fraction of quark 
and gluon jets. In fact, soft drop allows us to define this fraction precisely. 
The factorization theorem also explains why soft drop observables are less sensitive to hadronization 
than their ungroomed counterparts.
Using the factorization theorem, we resum the soft drop jet mass to next-to-next-to-leading 
logarithmic accuracy. 
This requires calculating some clustering effects
that are closely related to corresponding effects found in jet veto calculations.  
We match 
our resummed calculation to 
fixed order results for both $e^+e^-\to$ dijets
and $pp\to Z+j$ events, producing the first jet substructure predictions (groomed or ungroomed) 
to this accuracy for the LHC.
}

\begin{document}
\maketitle

\section{Introduction}

The high luminosity proton collisions at the Large Hadron Collider (LHC) enable an unprecedented sensitivity to rare and high scale physics.  The cost of such high luminosities is the presence of significant amounts of pile-up radiation present in every event, arising from numerous secondary proton collisions per bunch crossing.  Pile-up is truly uncorrelated with the hard scattering and can contaminate any potential measurement.  This is particularly important for measurements made on jets, for which pile-up can effect a large systematic bias in observables like the jet mass.
 In searches for resonances that decay to boosted electroweak objects which have definite masses, pile-up can significantly degrade the ability to separate signal from background.  Over the past several years, numerous methods \cite{Cacciari:2008gn,Butterworth:2008iy,Ellis:2009me,Krohn:2009th,Soyez:2012hv,Dasgupta:2013ihk,Krohn:2013lba,Larkoski:2014wba,Berta:2014eza,Cacciari:2014gra,Bertolini:2014bba} have been developed for grooming jets and events for pile-up mitigation and removal, and are now standard experimental tools at both ATLAS and CMS experiments.  Especially in analyses of jets, measurements made at the LHC often involve some form of grooming.

With this motivation, it is imperative to understand these jet grooming techniques from first principles QCD.  
There have been a few studies of the theoretical aspects of jet groomers 
\cite{Dasgupta:2013ihk,Dasgupta:2013via,Larkoski:2014wba,Dasgupta:2015yua}, 
with predictions for jet-observable distributions calculated to next-to-leading logarithmic (NLL) accuracy 
for two widely used jet groomers: the modified mass drop tagger (mMDT) and soft drop.  
These explicit analytic studies showed that these jet groomers not only have desired experimental properties, 
but can also dramatically simplify theoretical calculations as compared to their ungroomed counterparts. 
Non-global logarithms (NGLs) that arise from correlations between in- and out-of-jet scales 
 have proven to be a significant obstruction to resummation of ungroomed jet observables to NLL accuracy and beyond.  
In particular, it was demonstrated by explicit calculation in \Refs{Dasgupta:2013ihk,Dasgupta:2013via,Larkoski:2014wba} 
that mMDT and soft drop groomers eliminate the leading non-global logarithms in jet mass 
distributions~\cite{Dasgupta:2001sh}.
mMDT and soft drop pave the way for systematically improvable resummed predictions of jet observables.

In this paper, we open the door to systematically improvable jet substructure calculations by presenting an all-orders factorization theorem for the soft-drop \cite{Larkoski:2014wba} groomed observables using soft-collinear effective theory (SCET) \cite{Bauer:2000yr,Bauer:2001ct,Bauer:2001yt,Bauer:2002nz}.  
An overview of the method we discuss here and some of our results were presented recently in \Ref{Frye:2016okc}.  This paper
provides a more detailed presentation of those results as well as a derivation of the factorization formula
and its remarkable properties.
  
The soft drop groomer walks through the branching history of a jet, discarding soft branches until a sufficiently hard branching is found.
This is enforced by effectively requiring
\begin{equation}\label{eq:sddef}
\frac{\min[E_i,E_j]}{E_i+E_j}> \zcut \left(\frac{\theta_{ij}}{R}\right)^\beta\,,
\end{equation}
where $E_i$ and $E_j$ are the energies of the particles in that step of the branching, $\theta_{ij}$ is their relative angle, and $R$ is the radius of the jet.  
$\zcut$ is a parameter that sets the scale of soft, wide angle emissions in the jet; the typical value is $\zcut = 0.1$.  $\beta$ is a parameter that controls the aggressiveness of the groomer: $\beta = \infty$ removes the groomer, $\beta = 0$ coincides with mMDT and is simply an energy cut, and $\beta < 0$ removes all soft and collinear singularities.  We will consider $\beta \geq 0$.  If \Eq{eq:sddef} is not satisfied, the softer of the two branches is removed from the jet, and the grooming procedure continues on the harder branch.  When \Eq{eq:sddef} is satisfied, the procedure terminates and the groomed jet is returned.  For concreteness, on this groomed jet, we measure the two-point energy correlation functions $\ecf{2}{\alpha}$ with angular exponent $\alpha>0$ \cite{Banfi:2004yd,Jankowiak:2011qa,Larkoski:2013eya}.

\begin{figure}
\begin{center}
\includegraphics[width=10cm]{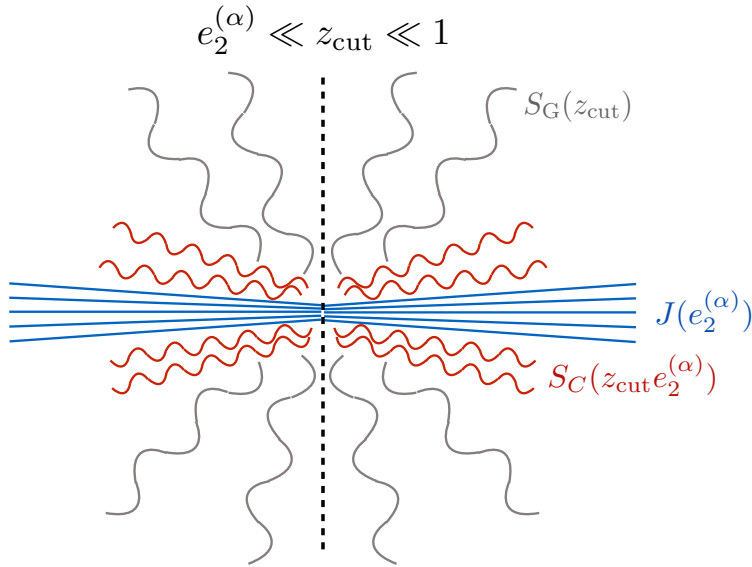}
\end{center}
\caption{Schematic of the modes in the factorization theorem for soft-drop groomed hemispheres in $e^+e^-\to $ dijets events.  $S_\glob(\zcut)$ denotes the soft wide-angle modes, $S_C(\zcut\ecf{2}{\alpha})$ denotes the collinear-soft modes, and $J(\ecf{2}{\alpha})$ denotes the jet modes.
}
\label{fig:factmodes}
\end{figure}

In $e^+e^-\to $ dijets events, the factorization formula we derive in this paper for soft-drop groomed left and right hemisphere jets is: 
\begin{equation}\label{eq:factee}
\hspace{-0.25cm}\frac{d^2\sigma}{d\ecf{2,L}{\alpha}\, d\ecf{2,R}{\alpha}} = 
H(Q^2) S_\glob(\zcut) \left[S_{C}(\zcut\ecf{2,L}{\alpha})\otimes J(\ecf{2,L}{\alpha})\right]\left[S_{C}(\zcut\ecf{2,R}{\alpha})\otimes J(\ecf{2,R}{\alpha})\right]\,.
\end{equation}
This factorization theorem applies when $\zcut \ll 1$ and the left- and right-hemisphere energy correlation functions 
are asymptotically small:
$\ecf{2,L}{\alpha},\ecf{2,R}{\alpha}\ll \zcut \ll1$.  We illustrate the physical configuration 
corresponding to this factorization theorem in \Fig{fig:factmodes}.  In \Eq{eq:factee}, $H(Q^2)$ 
is the hard function for $e^+e^-\to q\bar q$.  $S_\glob(\zcut)$ is the global soft function, which is only
sensitive to the scale set by $\zcut$ since all of its emissions fail soft drop.
$S_{C}(\zcut\ecf{2,L}{\alpha})$ is a soft function that is boosted along the direction of the 
jet in the left hemisphere; its corresponding modes are referred to as collinear-soft 
\cite{Bauer:2011uc,Procura:2014cba,Larkoski:2015zka,Larkoski:2015kga,Becher:2015hka,Chien:2015cka}.  
Emissions in $S_{C}(\zcut\ecf{2,L}{\alpha})$ may or may not pass the soft drop requirement 
and are therefore constrained by both $\zcut$ and $\ecf{2,L}{\alpha}$.  Importantly, this collinear-soft mode depends on only a single scale which we generically denote by $\zcut\ecf{2,L}{\alpha}$.
(For $\alpha\ne 2$ or $\beta >0$, the single scale is a different combination of $\zcut$ and $\ecf{2,L}{\alpha}$;
we simply call it  $\zcut\ecf{2,L}{\alpha}$ for notational brevity.)
 $J(\ecf{2,L}{\alpha})$ is 
the jet function for the left hemisphere jet, and all emissions in the jet function parametrically 
pass the soft drop requirement.  Thus, the jet function is independent of the scale set by $\zcut$, 
and only depends on $\ecf{2,L}{\alpha}$.  $\otimes$ denotes convolution in $\ecf{2,L}{\alpha}$, 
and a similar collinear-soft and jet factorization exists for the right hemisphere.

As we will explain in detail, there are several important consequences of this factorization formula.
Because the formula depends on the 
observables $\ecf{2,L}{\alpha},\ecf{2,R}{\alpha}$ only through collinear objects each of
which has a single scale, there are no non-global logarithms.
The elimination of the purely soft contribution also makes the  shape of soft-drop groomed jet shapes
largely independent of what else is going on in the event.
For example, 
the shape of the left hemisphere jet mass is independent of what is present in the right hemisphere.
Additionally, 
the scale associated with 
 the collinear-soft mode is parametrically larger than the soft scale associated with ungroomed masses, 
so non-perturbative corrections such as hadronization are correspondingly smaller.

This factorization theorem allows us to go beyond NLL accuracy to arbitrary accuracy.  In this paper, we show that next-to-next-to-leading logarithmic (NNLL) accuracy is readily achievable.
We focus on $\alpha=2$ where the two-point energy correlation 
function is equal to the squared jet mass (up to a trivial normalization). This lets us extract most of 
the necessary two-loop anomalous dimensions from the existing literature. 
For $\beta = 0$, the global soft function $S_\glob(\zcut)$  is closely related to the soft function 
with an energy veto \cite{vonManteuffel:2013vja,Chien:2015cka} which is known to two-loop order.  
There are additional clustering effects from the soft drop algorithm, but these are straightforward to calculate.
Interestingly, we find that the clustering effects in the soft drop groomer are intimately related 
to similar effects observed in jet veto 
calculations \cite{Banfi:2012yh,Becher:2012qa,Banfi:2012jm,Becher:2013xia,Stewart:2013faa}.
For $\beta = 1$, we compute the two-loop anomalous dimension of $S_\glob(\zcut)$ numerically
using the fixed-order code \eventtwo~\cite{Catani:1996vz}.\footnote{While we will not do it in this paper, one could use the 
results of \Ref{Bell:2015lsf} which calculates the anomalous dimension of the soft function for event-wide 
(recoil-free) angularities \cite{Berger:2003iw,Almeida:2008yp,Ellis:2010rwa,Larkoski:2014uqa} or energy 
correlation functions with arbitrary angular exponent.  This would enable us to extend our results 
to the case with $\alpha\neq 2$.}
We thereby achieve full NNLL resummation for the soft-drop groomed jet mass.\footnote{The jet mass has been calculated at NNLL using other 
methods~\cite{Chien:2012ur,Jouttenus:2013hs,Dasgupta:2012hg}
as has 2-subjettiness~\cite{Feige:2012vc}. However, without grooming the jets, 
there are non-global logarithms which are not resummed (and which may or may not be quantitatively important)
and uncontrollable sensitivity to pileup (which is very quantitatively important).}

It is straightforward to generalize from $e^+e^-$ to $pp$ collisions, since the distribution is determined by collinear physics within the jet, independent of the initial state.
The main new ingredient in $pp$ collisions is that  jets may be initiated by quarks or gluons.  
As we will show, soft-drop grooming the jet enables an infrared and collinear safe definition of the jet flavor at leading power in $\ecf{2}{\alpha}$ and $\zcut$ by simply summing the flavors of partons in the groomed jet.
Using this procedure, we are able to match our NNLL resummed distribution of soft-drop groomed jet mass to  
fixed order results for $pp\to Z+j$ events (including relative $\mathcal{O}(\alpha_s^2)$ corrections to the Born process).

The outline of this paper is as follows.  In \Sec{sec:obs}, we review the definition of the soft drop grooming algorithm and the energy correlation functions.  In \Sec{sec:factth}, we present the factorization theorem for soft-drop groomed energy correlation functions in $e^+e^-\to $ dijets events.  In this section, we will also present a detailed power-counting analysis of soft-dropped observables to determine the range of validity of the factorization theorem.  Our factorization theorem has many non-trivial consequences, which we review in \Sec{sec:factcons}.  These include absence of non-global logarithms, process independence, and small hadronization corrections.  In \Sec{sec:nnll}, we describe and present the ingredients necessary for NNLL resummation.  Here, we also describe our method for extracting anomalous dimensions from \eventtwo.  
We then match our NNLL results with fixed-order calculations for $e^+e^-$ collisions in \Sec{sec:eematch} and for $pp\to Z+j$ events in \Sec{eq:ppmatch}, comparing 
with Monte Carlo simulations in each case.  In \Sec{sec:conc}, we summarize and conclude.  The calculational details for NNLL resummation are collected in appendices.

\section{Observables}\label{sec:obs}

In this section, we review the soft drop grooming algorithm and the energy correlation functions.
Although previous work has focused on   jets produced in $pp$ collisions, we will provide definitions
for both lepton and hadron collider environments.
\subsection{Soft Drop Grooming Algorithm}

Given a set of constituents of a jet with radius $R$,
the soft drop grooming algorithm \cite{Larkoski:2014wba} proceeds in the following way:
\begin{enumerate}
\item Recluster the jet with a sequential $k_T$-type \cite{Catani:1991hj,Catani:1993hr,Ellis:1993tq} jet algorithm.  This produces an infrared and collinear (IRC) safe branching history of the jet.  The $k_T$ clustering metric for jets in $e^+e^-$ collisions is
\begin{equation}
d^{e^+e^-}_{ij}=\min\left[
E_i^{2p},E_j^{2p}
\right](1-\cos\theta_{ij})\,,
\end{equation}
where $E_i, E_j$ are the energies of particles $i$ and $j$ and $\theta_{ij}$ is their relative angle.  $p$ is a real number that defines the particular jet algorithm.  For jets produced in $pp$ collisions, the $k_T$ clustering metric is
\begin{equation}
d^{pp}_{ij}=\min\left[
p_{Ti}^{2p},p_{Tj}^{2p}
\right]R^2_{ij}\,,
\end{equation}
where $p_{Ti},p_{Tj}$ are the transverse momenta of particles $i$ and $j$ with respect to the beam and $R^2_{ij}$ is their relative angle in the pseudorapidity-azimuth angle plane. 

While the original implementation of soft drop was restricted to reclustering with the Cambridge/Aachen algorithm ($p=0$) \cite{Dokshitzer:1997in,Wobisch:1998wt,Wobisch:2000dk}, we will also briefly consider reclustering with the anti-$k_T$ algorithm ($p=-1$) \cite{Cacciari:2008gp} in \Sec{sec:antikt}.

\item Sequentially step through the branching history of the reclustered jet.
At each branching, check the soft drop criterion.  
For $e^+e^-$ collisions, we require
\begin{equation}
\frac{\min[E_i,E_j]}{E_i+E_j}>\zcut \left(\sqrt{2}\frac{\sin\frac{\theta_{ij}}{2}}{\sin\frac{R}{2}}\right)^{\beta}\,.
\end{equation}
This is known as {\bf the soft drop criterion}.  If the branching fails this requirement, then the softer of the two daughter branches is removed from the 
jet.  The soft drop groomer then continues to the next branching in the remaining clustering history.  For $pp$ collisions, the soft drop criterion is
\begin{equation}
\frac{\min[p_{Ti},p_{Tj}]}{p_{Ti}+p_{Tj}}>\zcut \left(\frac{R_{ij}}{R}\right)^{\beta}\,.
\end{equation}

\item The procedure continues until the soft drop criterion is satisfied.  At that point, soft drop terminates, and returns the jet groomed of the branches that failed the soft drop criterion.

\end{enumerate}

Once the jet has been groomed, any observable can be measured on its remaining constituents.

\subsection{Energy Correlation Functions}\label{sec:e2defs}

On jets that have been groomed by soft drop, we measure the two-point energy correlation functions \cite{Banfi:2004yd,Jankowiak:2011qa,Larkoski:2013eya}.  We do this mainly for concreteness; the general properties of the factorized formula we will present apply for a much broader class of observables.  For jets in $e^+e^-$ collisions, the two-point energy correlation function $\ecf{2}{\alpha}$ is
\begin{equation}\label{eq:eee2}
\left.\ecf{2}{\alpha}\right|_{e^+e^-}=\frac{1}{E_J^2}\sum_{i<j\in J} E_i E_j\left(
\frac{2p_i\cdot p_j}{E_i E_j}
\right)^{\alpha/2}\,,
\end{equation}
where $E_J$ is the sum of the energies of particles in the jet, the sum runs over distinct pairs $i,j$ of particles in the jet, $p_i$ is the four-vector momentum of particle $i$, and the angular exponent $\alpha$ is required to be greater than 0 for IRC safety.  For $\alpha=2$ and a jet that has massless constituents, the two-point energy correlation function reduces to the normalized, squared jet mass:
\begin{equation}
\left.\ecf{2}{2}\right|_{e^+e^-}=\frac{m_J^2}{E_J^2}\,.
\end{equation}
The energy correlation functions have the nice property that they are insensitive to recoil effects \cite{Banfi:2004yd,Larkoski:2014uqa} and do not include explicit axes in their definition.

For jets produced in $pp$ collisions, the energy correlation functions are appropriately modified by replacing spherical coordinates with cylindrical coordinates:
\begin{equation}\label{eq:ppe2}
\left.\ecf{2}{\alpha}\right|_{pp}=\frac{1}{p_{TJ}^2}\sum_{i<j\in J} p_{Ti} p_{Tj}R_{ij}^\alpha\,,
\end{equation}
where $p_{TJ}$ is the transverse momentum of the jet and $R_{ij}$ is the separation of particles $i$ and $j$ in the pseudorapidity-azimuthal angle plane.  For jets at central rapidities and in the limit that all particles in the jet are collinear, \Eq{eq:ppe2} reduces to  \Eq{eq:eee2}.  This property in particular will enable us to recycle results calculated in $e^+e^-$ collisions to the case of $pp$ collisions.

\section{Factorization Theorem}\label{sec:factth}

In this section, we derive the factorization formula for energy correlation functions measured on soft-drop groomed jets in the region of phase space where $\ecf{2}{\alpha}\ll\zcut\ll 1$ using SCET \cite{Bauer:2000yr,Bauer:2001ct,Bauer:2001yt,Bauer:2002nz}.  We begin with a power counting analysis based on the scales $\ecf{2}{\alpha}$ and $\zcut$ relevant to soft-drop groomed jets. This enables us to identify all modes and their momentum scalings that contribute at leading power.  Using these scales and the associated modes, we derive the factorization formula.
We then show that the jet function in the factorization formula can be re-factorized due to a collinear-soft mode
which decouples from the collinear-but-not-soft modes as a result of soft drop.

\subsection{Power Counting and Modes} 

\begin{figure}
\begin{center}
\includegraphics[width=10cm]{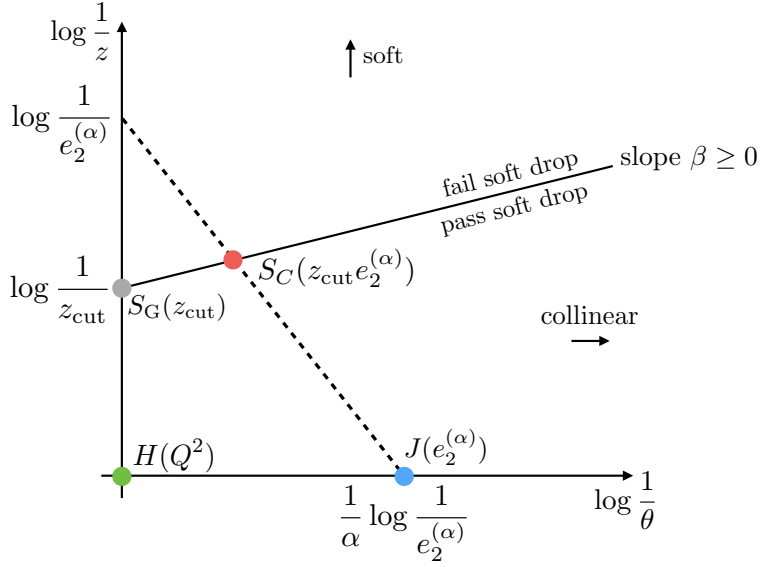}
\end{center}
\caption{
Location of modes appearing in the soft drop factorization theorem in the plane defined by energy fraction $z$ and splitting angle $\theta$ of emissions in the jet.  The solid diagonal line separates the regions of phase space where emissions pass and fail soft drop. All emissions along the dashed line that pass soft drop contribute at leading power to the measured value of $\ecf{2}{\alpha}$.
}
\label{fig:modesplot}
\end{figure}

For jets on which the soft drop groomer is applied and the energy correlation functions are measured, there are three relevant dimensonless scales: the jet radius $R$, the soft drop parameter $\zcut$, and $\ecf{2}{\alpha}$.  
Typically, jet radii are $R\sim 1$ . We are interested in the singular region $\ecf{2}{\alpha} \to 0$ for a fixed value
of $\zcut$. Thus we can assume $\ecf{2}{\alpha} \ll \zcut$. We will also assume $\zcut \ll 1$ to refactorize
the jet function.
The limits $R\ll1$ or $\zcut \sim 1$ could be considered as well, but  are beyond the scope of our analysis.

 We will use scaling arguments to 
 identify the regions of phase space that are present at leading power and then take the limit where each region becomes a separate sector, that no longer interacts with the other regions.

For a jet to have $\ecf{2}{\alpha}\ll1$, all particles must be either soft or collinear to the jet axis.  In particular, a particle with energy $E= zE_J$ at an angle $\theta$ from the jet axis must satisfy
\begin{equation}
\label{eq:eascale} z\theta^\alpha  \lesssim \ecf{2}{\alpha} \,.
\end{equation}
This is a line in the $\log(1/z)$-$\log(1/\theta)$ plane, as shown in \Fig{fig:modesplot}. Anything below
the dashed line in this figure is too hard to be consistent with a given value of $\ecf{2}{\alpha}$. 
The soft drop criterion is that 
\begin{equation}
\label{eq:zcutscale} \zcut \lesssim z\theta^{-\beta} \,,\\
\end{equation}
This is the region below the solid line in \Fig{fig:modesplot}.

 To find the relevant modes for the factorized expression, we need to
identify the distinct characteristic momentum scalings that approach the
singular regions of phase space in the limit $\ecf{2}{\alpha} \ll \zcut \ll 1$. For a particular
scaling, the constraints in \Eqs{eq:eascale}{eq:zcutscale}
will either remain relevant or decouple.  We can characterize the
relevant regions by their scalings in light-cone coordinates. Defining
$n^\mu$ as the jet direction and $\bar{n}^\mu$ as the direction
backwards to the jet, then light-cone coordinates are triplets
$p=(p^-,p^+,p_\perp)$ where $p^- = \bar{n}\cdot p$, $p^+ =n \cdot p$ and
$p_\perp$ are the components transverse to $n$.  On-shell massless
particles have $p^+ p^- = p_\perp^2$.  The energy fraction is $z= p^0/Q
= \frac{1}{2}(p^+ + p^-)/Q$ and the angle to the jet axis in the
collinear limit is $\theta = {p_\perp}/{p^0}$.

We start with the {\bf soft modes}, emitted at large angles $\theta\sim 1$, but still within the jet. 
If such radiation were to pass soft drop, with energy fraction greater than $\zcut$,
it would set $\ecf{2}{\alpha} \gtrsim \zcut$; this contradicts our assumed hierarchy $\ecf{2}{\alpha} \ll \zcut$. 
Therefore, soft wide-angle radiation is removed by soft drop and is not constrained by $\ecf{2}{\alpha}$. These modes
thus have momenta that scale like
\begin{equation}\label{eq:softmom}
p_s\sim \zcut Q(1,1,1)\,.
\end{equation}
They  contribute only to the normalization of the distribution, not to its shape. 

Next consider the collinear radiation, emitted at small angles $\theta \ll 1$. All collinear radiation
has $p^- \gg p^+$. Then, from \Eq{eq:eascale}, we find
\begin{equation}
\label{eq:scalinge}
\ecf{2}{\alpha} \sim \frac{(p^+)^{\alpha/2} (p^-)^{1-\alpha/2}}{Q}
\end{equation}
Collinear modes can either have $z \sim 1$ or be parameterically soft $z \ll 1$.
 
For modes with $z \sim 1$, we have $z \gg \zcut$. Thus $p^- \sim Q$ and
$p^+ \sim Q (\ecf{2}{\alpha})^{2/\alpha}$ independent of $\zcut$. Their scaling is
\begin{equation}\label{eq:collmom}
p_c\sim Q\left(1, (\ecf{2}{\alpha})^{2/\alpha},(\ecf{2}{\alpha})^{1/\alpha} \right)\,.
\end{equation}
We call these modes {\bf collinear modes}, although strictly they are not-soft collinear modes. 

Collinear radiation that can have $z \sim \zcut \ll 1$ we call {\bf collinear-soft}. 
In this case, $p^- \sim z Q$ and $ p^+ \sim \theta^2 z Q$.
These modes are simultaneously compatible with  \Eqs{eq:eascale}{eq:zcutscale}.
Their scaling is determined by saturating these parametric relationships, which leads to
\begin{align}\label{eq:csoftmom}
p_{cs}\sim  (\zcut)^{\frac{\alpha}{\alpha+\beta}}(\ecf{2}{\alpha})^{\frac{\beta}{\alpha+\beta}}Q\left(1,\left(
\frac{\ecf{2}{\alpha}}{\zcut}
\right)^{\frac{2}{\alpha+\beta}},\left(
\frac{\ecf{2}{\alpha}}{\zcut}
\right)^{\frac{1}{\alpha+\beta}}\right)\,.
\end{align}
This is the point in phase space labeled $S_C(\zcut \ecf{2}{\alpha})$ in \Fig{fig:modesplot}.

\subsection{Factorization and refactorization}
\begin{figure}
\begin{center}
\includegraphics[width=10cm]{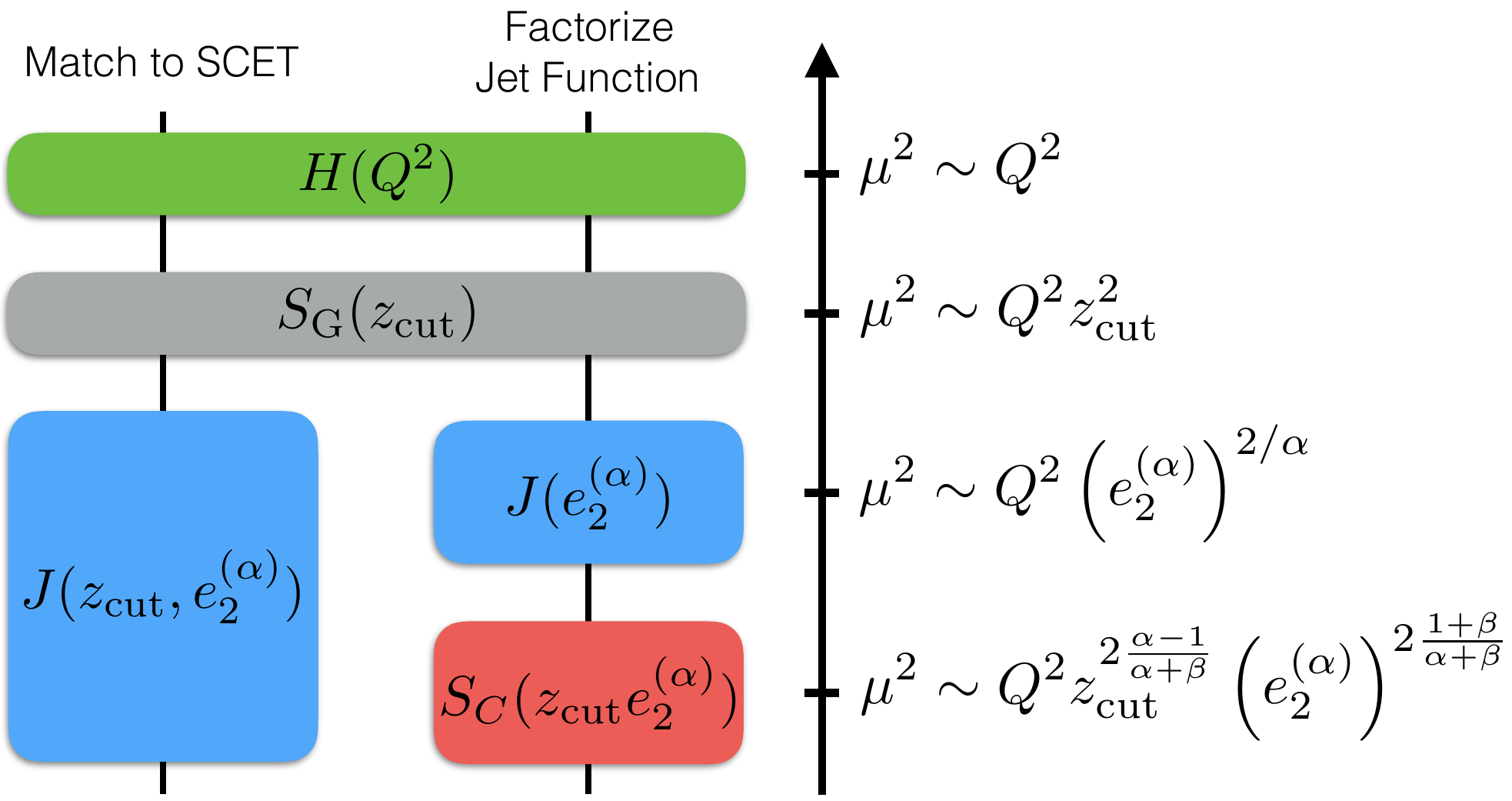}
\end{center}
\caption{Illustration of the multi-stage matching procedure to derive the soft drop factorization theorem.  As discussed in the text, we first match QCD to SCET, then factorize the jet function into collinear and collinear-soft modes.  Canonical scales of all modes in the factorization theorem are shown on the right, ordered in virtuality where we assume that $\alpha > 1$ and $\beta \geq0$.
}
\label{fig:scalemodes}
\end{figure}

With the relevant scalings identified, we proceed to derive the factorization formula. 
For simplicity, we focus on the case of $e^+e^- \to $ hemisphere jets, 
with $\ecf{2}{\alpha}$ measured on each hemisphere.
Jets at hadron colliders can be treated similarly, as we discuss in \Sec{sec:factcons}.  
\Fig{fig:scalemodes} illustrates the relevant modes and their scales.

We begin with the usual SCET factorization formula, in the absence of soft drop grooming. 
The hard, collinear and soft modes are separated in the limit of small observables.
This leads to~\cite{Fleming:2007qr,Schwartz:2007ib,Ellis:2010rwa}
\begin{equation}\label{eq:zerostepfact}
\frac{d^2\sigma}{d\ecf{2,L}{\alpha}\, d\ecf{2,R}{\alpha}} = H(Q^2) \times S\left(\ecf{2,L}{\alpha} ,  \ecf{2,R}{\alpha}\right) 
\otimes J(\ecf{2,L}{\alpha})\otimes J(\ecf{2,R}{\alpha})
\end{equation}
for the ungroomed hemispheres in $e^+e^- \to$ dijets events, provided $\ecf{2,L}{\alpha},\ecf{2,R}{\alpha} \ll 1$. 
Here, $\otimes$ denotes convolution in $\ecf{2,L}{\alpha}$ or $\ecf{2,R}{\alpha}$ appropriately.  
To get to this equation, one can match to full QCD to get the hard function, then decouple the soft and collinear degrees of freedom
to pull the jet and soft functions apart~\cite{Bauer:2000yr,Bauer:2001ct,Bauer:2001yt,Bauer:2002nz}. Alternatively, one can use the method of regions approach~\cite{Beneke:1997zp,Becher:2014oda}, or the on-shell phase space approach~\cite{Feige:2013zla,Feige:2014wja,Feige:2015rea}. 
Importantly, $\ecf{2}{\alpha}$ is insensitive to recoil effects from soft emissions that displace the jet axis from the direction of hard, collinear particles \cite{Banfi:2004yd,Larkoski:2014uqa}, and so the jet and soft functions are completely decoupled. 

Next we write down the hard-soft-jet factorization formula in the presence of soft drop grooming, assuming the hierarchy $\ecf{2}{\alpha} \ll \zcut \ll 1$.
 With this assumption, soft radiation emitted at large angles must necessarily fail the soft drop criterion.
Thus, all wide angle soft radiation in the jets (in this case, the hemisphere jets) is groomed and cannot contribute to the observable. 
All that remains of the global soft function is a $\zcut$-dependent normalization
factor $S_\glob(\zcut)$. This leads to 
\begin{equation}\label{eq:onestepfact}
\frac{d^2\sigma}{d\ecf{2,L}{\alpha}\, d\ecf{2,R}{\alpha}} = H(Q^2) \times S_\glob(\zcut) \times J_{\ze}\left(\zcut,\ecf{2,L}{\alpha}\right) \times J_{\ze}\left(\zcut,\ecf{2,R}{\alpha}\right)\,.
\end{equation}
$S_\glob(\zcut)$ gives the cross section for the radiation from a set of Wilson lines that 
fails the soft drop criterion.                                                                                       
An explicit calculation of $S_\glob$ for hemisphere jets at one-loop is given in Appendix \ref{app:softfct}.
With the collinear and soft modes decoupled, we can lower the virtuality of the collinear modes without further matching.

The jet function $J_{\ze}$ still depends on multiple scales, so to resum all the large logarithms it must be re-factorized.
To see that it refactorizes, note first that in addition to being collinear,  
radiation in the jet function that is sensitive to the scale set by $\zcut$ must also be soft, 
by the assumption that $\zcut \ll 1$.  
Equivalently, emissions with order-1 energy fractions are not constrained by the scale $\zcut$. 
We can thus factorize the jet function into two pieces depending on their energy fraction:
\begin{equation}\label{eq:softfctzero}
J_{\ze}(\zcut,\ecf{2}{\alpha}) = J(\ecf{2}{\alpha})\otimes S_C(\zcut\ecf{2}{\alpha})\,.
\end{equation}
Here, 
 $J(\ecf{2}{\alpha})$ is the jet function that only depends on $\ecf{2}{\alpha}$ and only receives contributions from emissions with order-1 energy fraction.  $S_C(\zcut\ecf{2}{\alpha})$ is the soft limit of the unfactorized jet function $J_{\ze}(\zcut,\ecf{2}{\alpha})$.
The scaling of the collinear and collinear-soft modes are given in \Eqs{eq:collmom}{eq:csoftmom} as discussed above. 
  Note that, importantly, because the collinear-soft mode arises from refactorization of a jet function, it is a color singlet and only depends on two back-to-back directions.  Because the jet function only depends on $\ecf{2}{\alpha}$, it is sensitive to a single infrared scale.  

The step in \Eq{eq:softfctzero} is the most unusual and important in the derivation. That the collinear-soft function
depends on only a single combination of $\zcut$ and $\ecf{2}{\alpha}$ is absolutely critical to being able to resum all the logs of $\ecf{2}{\alpha}$. We therefore devote \Sec{sec:singcs} to showing explicitly that the collinear-soft function depends on a unique combination of $\zcut$ and $\ecf{2}{\alpha}$ as determined by the parametric scaling of the modes of \Eq{eq:csoftmom}, and so is also only sensitive to a single infrared scale.

 Inserting \Eq{eq:softfctzero} into \Eq{eq:onestepfact} results in the 
factorization formula for soft drop energy correlation functions:
\begin{equation}\label{eq:factdefnew}
\boxed{
\frac{d^2\sigma}{d\ecf{2,L}{\alpha}\, d\ecf{2,R}{\alpha}} = H(Q^2) S_\glob(\zcut) \left[S_{C}\left(\zcut\ecf{2,L}{\alpha}\right)\otimes J(\ecf{2,L}{\alpha})\right]\left[S_{C}\left(\zcut\ecf{2,R}{\alpha}\right)\otimes J(\ecf{2,R}{\alpha})\right]
}
\end{equation}
The pieces of the factorization theorem are:
\begin{itemize}
\item $H(Q^2)$ is the hard function for production of an $e^+e^-\to$ dijets event.

\item $S_\glob(\zcut)$ is the global soft function. It integrates the radiation coming from Wilson lines in the jet directions
that fails the soft drop criterion. Its modes fail soft drop and have momenta that scale as determined in \Eq{eq:softmom}.

\item $J(\ecf{2}{\alpha})$ is the jet function describing the emission of collinear radiation from a jet.  Its modes parametrically pass soft drop and have momenta that scale as determined in \Eq{eq:collmom}.

\item $S_{C}\left(\zcut\ecf{2}{\alpha}\right)$ is the collinear-soft function describing the emission of soft radiation boosted along the direction of a jet.  Its modes may or may not pass soft drop and have momenta that scale as determined in \Eq{eq:csoftmom}.  We denote the single scale that the collinear-soft function depends on as $\zcut\ecf{2,L}{\alpha}$ for brevity; it is shorthand for \Eq{abbrev}.
\end{itemize}
We present the operator definitions and explicit one-loop results for all of these functions in the appendices.

The appearance of collinear-soft modes in this factorization theorem has some similarities and differences with respect to the identification of other collinear-soft modes in the literature \cite{Bauer:2011uc,Procura:2014cba,Larkoski:2015zka,Larkoski:2015kga,Becher:2015hka,Chien:2015cka}.  The original construction of a collinear-soft mode in \Ref{Bauer:2011uc} followed from boosting two jets in an event far from their center-of-mass frame in an effective theory of collinear dijets called SCET$_+$.  The collinear-soft mode in SCET$_+$ is sensitive to three Wilson line directions: the two from the collinear jets and the backward direction from boosting all other jets in the event.  This collinear-soft mode was also exploited in \Ref{Larkoski:2015kga} in the resummation of jet observables that are sensitive to multi-prong substructure.

The collinear-soft mode in the factorization theorem presented here, however, is more similar to modes identified from the measurement of multiple observables on a jet, each of which is only sensitive to radiation about a single hard core \cite{Procura:2014cba,Larkoski:2015zka,Larkoski:2015kga,Becher:2015hka,Chien:2015cka}.  For example, \Ref{Procura:2014cba} presented a factorization theorem for jets on which two angularities \cite{Berger:2003iw,Almeida:2008yp,Ellis:2010rwa,Larkoski:2014uqa} are measured.  At leading power, angularities are only sensitive to the hard jet core, and so the collinear-soft modes only know about two Wilson line directions: the jet axis and the backward direction.  
More recently, collinear-soft modes of this type have been used to resum NGLs \cite{Larkoski:2015zka,Becher:2015hka} and logarithms of the jet radius \cite{Chien:2015cka}.

\subsection{The Single Scale of the Collinear-Soft Function}\label{sec:singcs}
To demonstrate explicity that the collinear-soft function only depends on a single scale, we can make the following scaling argument.  The collinear-soft function has the following form:
\begin{align}
S_C\left(\zcut,\ecf{2}{\alpha}\right) = \sum_n \mu^{2n\epsilon} \int d\Pi_n\, |\mathcal{M}_n|^2\,\Theta_\text{SD}\,\delta_{\ecf{2}{\alpha}}\,.
\end{align}
Here, $n$ is the number of final state collinear-soft particles, $d\Pi_n$ is on-shell Lorentz-invariant phase space in $d=4-2\epsilon$ dimensions:
\begin{align}
d\Pi_n = \prod_{i=1}^n \frac{d^d k_i}{(2\pi)^d} 2\pi\delta(k_i^2)\Theta(k_i^0)\,,
\end{align}
$\mu$ is the renormalization scale, and $\mathcal{M}_n$ is the amplitude for the production of the final state.  $\Theta_\text{SD}$ represents the soft drop grooming algorithm, which applies constraints on the final state and $\delta_{\ecf{2}{\alpha}}$ represents the measurement of $\ecf{2}{\alpha}$ on the final state:
\begin{equation}
\delta_{\ecf{2}{\alpha}} = \delta\left(
\ecf{2}{\alpha} - \frac{2^\alpha}{Q}\sum_i (k_i^-)^{1-\alpha/2}(k_i^+)^{\alpha/2}
\right)\,,
\end{equation}
where the sum runs over the set of final state particles $\{i\}$ that remain in the jet after grooming.  To write this expression, we have used the definition of $\ecf{2}{\alpha}$ from \Sec{sec:e2defs} and expanded in the collinear-soft limit, as in \Eq{eq:scalinge}.

Now, we rescale the momenta in light-cone coordinates that appear in the phase space integral in the following way:
\begin{align}\label{eq:scalings}
k^- & \to  (\zcut)^{\frac{\alpha}{\alpha+\beta}}(\ecf{2}{\alpha})^{\frac{\beta}{\alpha+\beta}} k^-\,,\\
k^+ & \to  (\zcut)^{\frac{\alpha-2}{\alpha+\beta}}(\ecf{2}{\alpha})^{\frac{2+\beta}{\alpha+\beta}}k^+\,,\nonumber\\
k_\perp & \to  (\zcut)^{\frac{\alpha-1}{\alpha+\beta}}(\ecf{2}{\alpha})^{\frac{1+\beta}{\alpha+\beta}} k_\perp\,.\nonumber 
\end{align}
At leading power in exactly $d=4$, the phase space measure $d\Pi_n$ and the squared matrix element $|\mathcal{M}_n|^2$ scale exactly inversely.  Therefore, in $d$ dimensions, under this rescaling, we have
\begin{equation}
d\Pi_n\, |\mathcal{M}_n|^2 \to \left(
(\zcut)^{\frac{\alpha-1}{\alpha+\beta}}(\ecf{2}{\alpha})^{\frac{1+\beta}{\alpha+\beta}}
\right)^{-2n\epsilon}d\Pi_n\, |\mathcal{M}_n|^2\,.
\end{equation}

Next, look at how the measurement functions $\Theta_\text{SD}$ and $\delta_{\ecf{2}{\alpha}}$ change under the rescaling of \Eq{eq:scalings}.  First, consider the soft drop groomer $\Theta_\text{SD}$.  This consists of two parts: one, the reclustering with the Cambridge/Aachen algorithm and the second, the energy requirement on the clustered particles.  The clustering metric of the Cambridge/Aachen algorithm is just the pairwise angle
\begin{equation}
d_{ij}^{\text{C/A}}=\theta_{ij}^2\,,
\end{equation}
and a pair $\{i,j\}$ of particles in the jet are clustered if they have the smallest $d_{ij}^{\text{C/A}}$.  Importantly, the reclustering of the jet with soft drop is completely inclusive: all particles in the jet are clustered with no jet radius parameter.  Therefore, for collinear-soft modes, there are only three types of clustering constraints that can be enforced, depending on what $d_{ij}^{\text{C/A}}$'s are being compared.  If in the clustering history we compare the angles between two collinear-soft particles $i$ and $j$ to the jet axis, this corresponds to the constraint
\begin{equation}
\Theta\left(
\frac{k_i^+}{k_i^-}-
\frac{k_j^+}{k_j^-}
\right)\,.
\end{equation}
This is invariant under the rescalings of \Eq{eq:scalings}.  If in the clustering history we compare the angle between a collinear-soft particle $i$ to the jet axis and the angle between two collinear-soft particles $j$ and $k$, we have the constraint
\begin{equation}
\Theta\left(
\frac{k_i^+}{k_i^-}-
\frac{k_j\cdot k_k}{k_j^-k_k^-}
\right)\,,
\end{equation}
which is also invariant under the rescalings of \Eq{eq:scalings}.  Finally, we can compare the angle between a pair of collinear-soft particles $i$ and $j$ to the angle between another pair of collinear-soft functions $k$ and $l$, this leads to
\begin{equation}
\Theta\left(
\frac{k_i\cdot k_j}{k_i^-k_j^-}-
\frac{k_k\cdot k_l}{k_k^-k_l^-}
\right)\,.
\end{equation}
This too is invariant under \Eq{eq:scalings}.  
Therefore, for all possible clustering structures, the Cambridge/Aachen algorithm is invariant under the rescalings of \Eq{eq:scalings}.

The soft drop energy requirement on any number of particles that have been reclustered takes the form:
\begin{align}
\Theta\left(
\sum_i z_i - \zcut \theta^\beta 
\right)\,,
\end{align}
where $z_i$ is the energy fraction of particle $i$ and $\theta$ is the angle that the cluster of particles $\{i\}$ makes with the jet axis.  In terms of light-cone coordinates, this can be written as:
\begin{align}
\Theta\left(
\sum_i z_i - \zcut \theta^\beta 
\right) = \Theta\left(
k^- - \zcut Q\left(\frac{k_\perp}{k^-}\right)^\beta 
\right)\,,
\end{align}
where
\begin{align}
k_\perp & = \left|\sum_i \vec k_{\perp,i} \right|\,,\\
k^- & = \sum_i k_i^- \,.
\end{align}
Applying the rescalings of \Eq{eq:scalings}, this constraint becomes
\begin{align}
\Theta\left(
\sum_i z_i - \zcut \theta^\beta 
\right) \to 
\Theta\left(
\sum_i z_i -  \theta^\beta 
\right)
\,.
\end{align}
Note that the low scale $\zcut$ has been removed from this constraint.
  
Under the rescaling, the measurement constraint $\delta_{\ecf{2}{\alpha}}$ becomes
\begin{align}
 \delta\left(
\ecf{2}{\alpha} - \frac{2^\alpha}{Q}\sum_i (k_i^-)^{1-\alpha/2}(k_i^+)^{\alpha/2}
\right) \to  \frac{1}{\ecf{2}{\alpha}}\delta\left(
1 - \frac{2^\alpha}{Q}\sum_i (k_i^-)^{1-\alpha/2}(k_i^+)^{\alpha/2}
\right)\,.
\end{align}
Therefore, the low scale $\ecf{2}{\alpha}$ has been removed from this constraint.

Putting this all together, 
the collinear-soft function can be rewritten as
\begin{equation}
S_C\left(\zcut,\ecf{2}{\alpha}\right) = \sum_n \mu^{2n\epsilon}\left(
(\zcut)^{\frac{\alpha-1}{\alpha+\beta}}(\ecf{2}{\alpha})^{\frac{1+\beta}{\alpha+\beta}}
\right)^{-2n\epsilon} \frac{1}{\ecf{2}{\alpha}}\int d\Pi_n\, |\mathcal{M}_n|^2\,\Theta_\text{SD}^{\zcut = 1}\,\delta_{\ecf{2}{\alpha}=1}\,.
\end{equation}
We have used the notation that $\Theta_\text{SD}^{\zcut = 1}$ is the soft drop grooming algorithm with $\zcut = 1$ and $\delta_{\ecf{2}{\alpha}=1}$ is the measurement with $\ecf{2}{\alpha} = 1$.  All low scales have been explicitly removed from the phase space integral. 
This function is now seen to be a function only of the single scale
\begin{equation}
\label{abbrev}
\text{``}\zcut \ecf{2}{\alpha}{}\text{''}= (\zcut)^{\frac{\alpha-1}{\alpha+\beta}}(\ecf{2}{\alpha})^{\frac{1+\beta}{\alpha+\beta}}.
\end{equation}
The quotes just mean that the left-hand side is our abbreviation for the unwieldly quantity on the right-hand side. 

This proves that, to all orders, the collinear-soft function has dependence on only a single infrared scale, defined by this combination of $\zcut$ and $\ecf{2}{\alpha}$.
Notice that the proof relied on the choice of Cambridge/Aachen reclustering in the soft drop grooming algorithm.

This completes the derivation of factorization and re-factorization for soft-drop groomed hemispheres in $e^+e^-$ collisions, on which two-point energy correlation functions have been measured.
All functions in the factorized cross section in \Eq{eq:factdefnew} are sensitive to a single infrared scale and so all large logarithms can be resummed with the renormalization group.

\section{Consequences of Factorization Theorem}\label{sec:factcons}

Before we use the factorization theorem of \Eq{eq:factdefnew} to make predictions for the cross section, 
we discuss consequences of this formula in some detail. 
Because the factorization theorem was derived without respect to any fixed order, these results hold to all orders.   

 Many of these consequences follow from the fact that soft wide angle radiation does not contribute to the shape of 
the soft-drop groomed $\ecf{2}{\alpha}$ distribution for $\ecf{2}{\alpha}\ll\zcut\ll1$, a property that persists even for jets at hadron colliders.
For example, it follows immediately from this fact that the shape of such a distribution is insensitive to 
contamination from pile-up and underlying event.
 
In this section, we will furthermore prove that at leading power, there are no NGLs that affect the shape of the soft-drop groomed $\ecf{2}{\alpha}$ distribution in this regime.
This was explicitly shown at ${\cal O}(\alpha_s^2)$ in \Refs{Dasgupta:2013ihk,Dasgupta:2013via,Larkoski:2014wba} and plausibility arguments were presented for all orders, but this is the first proof.  
The factorization theorem also exhibits sample independence to a large degree, because the shape of the distribution is only sensitive to collinear physics.  
We will also demonstrate that soft-drop groomed energy correlation functions are less sensitive to hadronization than their ungroomed counterparts.

\subsection{Absence of Non-Global Logarithms to All Orders}

NGLs in cross sections of observables measured on individual hemispheres in $e^+e^-$ collisions arise from a parametric separation of the scales in the hemispheres.  Their leading effects are exclusively non-Abelian and quantify the correlation between the two hemispheres.  Clearly, for a correlation to be present, there must be correlated radiation emitted into both hemispheres.  If we measure the energy correlation functions $\ecf{2}{\alpha}$ on both hemispheres and demand that $\ecf{2}{\alpha}\ll 1$, then the radiation in the event must be soft wide-angle or collinear.  At leading power, it is not possible to have correlations between different collinear directions (beyond total momentum conservation) as this would violate the collinear factorization of gauge theory amplitudes.  Therefore, correlations and NGLs can only arise from soft, wide-angle radiation in the event with these assumptions.

The factorization theorem for ungroomed hemisphere energy correlation functions is~\cite{Fleming:2007qr,Schwartz:2007ib,Ellis:2010rwa}
\begin{equation}
\left.\frac{d^2\sigma}{d\ecf{2,L}{\alpha}\, d\ecf{2,R}{\alpha}}\right|_\text{ug}=H(Q^2)S(\ecf{2,L}{\alpha},\ecf{2,R}{\alpha})\otimes J(\ecf{2,L}{\alpha})\otimes J(\ecf{2,R}{\alpha})\,,
\end{equation}
where ``ug'' denotes ungroomed.  The cross section explicitly depends on soft wide-angle radiation through the soft function $S(\ecf{2,L}{\alpha},\ecf{2,R}{\alpha})$, and so if either $\ecf{2,L}{\alpha}\ll\ecf{2,R}{\alpha}$ or $\ecf{2,R}{\alpha}\ll\ecf{2,L}{\alpha}$, NGLs will be present in this factorization theorem.  Because the soft function depends on two scales, all of the singular dependence cannot be determined by renormalization group invariance.  More generally, non-global structure present in the soft function has been studied at ${\cal O}(\alpha_s^2)$ \cite{Hornig:2011iu,Hornig:2011tg,Kelley:2011ng} and beyond~\cite{Schwartz:2014wha,Khelifa-Kerfa:2015mma} and recently, methods have been developed to control all-orders behavior \cite{Caron-Huot:2015bja,Larkoski:2015zka,Neill:2015nya,Becher:2015hka}.  However, NGLs represent an obstruction to resummation of the cross section to NLL and beyond.

For groomed hemisphere energy correlation functions, our factorization theorem instead takes the form of \Eq{eq:factdefnew}:
\begin{equation}
\hspace{-0.28cm}\frac{d^2\sigma}{d\ecf{2,L}{\alpha}\, d\ecf{2,R}{\alpha}} = H(Q^2) \, S_\glob(\zcut) \left[S_{C}(\zcut\ecf{2,L}{\alpha})\otimes J(\ecf{2,L}{\alpha})\right]\left[S_{C}(\zcut\ecf{2,R}{\alpha})\otimes J(\ecf{2,R}{\alpha})\right]\,.
\end{equation}
All soft, wide angle radiation throughout the event is described by $S_\glob(\zcut)$, which is sensitive only to the single scale $\zcut$.  Therefore, there are no NGLs present in this factorization theorem.  Even with a hierarchy between $\ecf{2,L}{\alpha}$ and $\ecf{2,R}{\alpha}$, these observables are completely decoupled at leading power in $\ecf{2}{\alpha}$ and $\zcut$.  Additionally, the shape of the distribution is also independent of jet radius effects and the precise way in which the hemispheres are defined.

When we discuss soft-drop groomed jets in $pp$ collisions in \Sec{eq:ppmatch}, we will place no constraint on global soft radiation throughout the event, unlike the case of $e^+e^-\to$ hemisphere jets.  Nevertheless, the shape of the soft-drop groomed $\ecf{2}{\alpha}$ distribution will still have no NGLs, jet radius effects, etc., due to universality of the collinear limit of QCD amplitudes.  The normalization, however, will in general be sensitive to scales both in the jet (set by $\zcut$ and the jet radius $R$) and scales outside of the jet (set by the partonic collision energy).  To eliminate these effects in $pp$ collisions, we can normalize the cross section, say, to integrate to unity.

\subsection{Process Independence}

Strictly speaking, the factorization theorem of \Eq{eq:factdefnew} depends on the process.  It includes the hard function, which is process dependent, and a soft function, that knows about all hard jet directions.  Nevertheless, there is a sense in which the factorization theorem is process independent.  Normalizing the cross section completely removes the hard and soft function dependence.  Then, by the universal collinear factorization of QCD amplitudes, if we are completely inclusive over the right hemisphere, then the differential cross section of the soft-drop groomed energy correlation function in the left hemisphere is given by
\begin{equation}
\frac{d\sigma}{d\ecf{2,L}{\alpha}} = {\cal N} S_{C}(\zcut\ecf{2,L}{\alpha})\otimes J(\ecf{2,L}{\alpha})\,,
\end{equation}
where we assume that $\ecf{2,L}{\alpha}\ll \zcut \ll1$ and ${\cal N}$ is some normalization factor.  
That is, in the deep infrared where $\ecf{2,L}{\alpha}\ll \zcut \ll1$, all radiation in the groomed jet is constrained to be collinear.  
Therefore, in this limit and for a fixed jet energy, the shape of the distribution for quark jets
is independent of the process that created the quark jets,
due to the universality of QCD matrix elements in the collinear limit.

This collinear factorization property of soft-drop groomed observables can be exploited for jets in $pp$ collisions.  
Unlike the dominant case in $e^+e^-$ collisions, jets at a $pp$ collider can be either quark or gluon.  
Of course, on a jet-by-jet level, we cannot determine whether a jet was initiated by a quark or gluon.  
However, for a given process, we can determine the relative fraction of quark and gluon jets in the sample.  
For jets produced at a $pp$ collider, the process independence manifests itself in the cross section as
\begin{equation}
\frac{d\sigma^{pp}}{d\ecf{2}{\alpha}} = D_q S_{C,q}(\zcut\ecf{2}{\alpha})\otimes J_q(\ecf{2}{\alpha})+D_g S_{C,g}(\zcut\ecf{2}{\alpha})\otimes J_g(\ecf{2}{\alpha})\,,
\end{equation}
where $D_q$ ($D_g$) is proportional to the fraction of quark (gluon) jets in the sample.  
The relative fraction of quark and gluon jets can be determined from fixed-order calculations, 
using a simple algorithm for determining the flavor of a groomed jet.   
We will describe this in detail in \Sec{eq:ppmatch} when we match our resummed distribution to fixed order in $pp\to Z+j$ events.

\subsection{Hadronization Corrections}\label{sec:hadest}

With a factorization formula, one can estimate the size and importance of non-perturbative corrections to the cross section.  We will only consider non-perturbative corrections to the shape,
 as the normalization can be set by hand.  Therefore, non-perturbative corrections can only enter into our factorization theorem via the jet or collinear-soft functions.

From \Eqs{eq:collmom}{eq:csoftmom}, the scales appearing in the jet and collinear-soft functions are
\begin{align}
\mu_{J}&=Q\left(\ecf{2}{\alpha}\right)^{1/\alpha}\,,\\
\mu_{S_C}&=Q\zcut^{\frac{\alpha-1}{\alpha+\beta}}\left(\ecf{2}{\alpha}\right)^{\frac{1+\beta}{\alpha+\beta}}\,.
\end{align}
  If either of the scales approaches $\Lambda_\text{QCD}$, then we expect there to be large corrections to the perturbative cross section due to non-perturbative physics.  We can estimate when non-perturbative corrections become large by setting these scales to be $\Lambda_\text{QCD}$.  For $\alpha>1$ and $\beta \geq 0$, the collinear-soft mode has a lower virtuality than the collinear mode, so it will probe the non-perturbative region of phase space first.  The value of $\ecf{2}{\alpha}$ at which the collinear-soft mode becomes non-perturbative is
\begin{equation}\label{eq:npest}
\mu_{S_C}=\Lambda_\text{QCD} \quad\Rightarrow\quad \left.\ecf{2}{\alpha}\right|_\text{NP}\simeq  \left(
\frac{\Lambda_\text{QCD}}{\zcut Q}
\right)^{\frac{\alpha-1}{1+\beta}}\cdot  \frac{\Lambda_\text{QCD}}{Q}\,.
\end{equation}
This estimate can be compared with the Monte Carlo analysis of hadronization corrections to the soft-drop groomed energy correlation functions from \Ref{Larkoski:2014wba}.  In particular, the estimate of \Eq{eq:npest} of when non-perturbative corrections become important for $\alpha=2$ as a function of $\beta$ agrees exceptionally well with Fig.~10(a) of \Ref{Larkoski:2014wba}.

For $\beta < \infty$, the soft drop groomer reduces the effect of non-perturbative corrections with respect to the ungroomed observable.  This can be simply seen from \Eq{eq:npest}, in which the prefactor 
\begin{equation}
\left(
\frac{\Lambda_\text{QCD}}{\zcut Q}
\right)^{\frac{\alpha-1}{1+\beta}}
\end{equation} 
approaches unity as the grooming is removed ($\beta \to \infty$). 
This factor is less than 1 for $\beta < \infty$, provided $\alpha > 1$ and $\Lambda_\text{QCD}<\zcut Q$.  For high energy jets, this suppression can be substantial.  For example, for $\alpha = 2$ (corresponding to jet mass) and $\beta = 0$ (corresponding to mMDT groomer) non-perturbative effects become important at
\begin{equation}
\left.\ecf{2}{2}\right|^{\beta = 0}_\text{NP}\simeq   \frac{\Lambda_\text{QCD}^2}{\zcut Q^2}\,.
\end{equation}
This agrees with the estimate of the size of nonperturbative corrections for the mMDT groomer from \Ref{Dasgupta:2013ihk}.

\section{Achieving NNLL Accuracy}\label{sec:nnll}

In this section, we determine the anomalous dimensions necessary to resum the large logarithms of soft-dropped energy correlation functions through NNLL accuracy. The practical details of how one assembles these ingredients, in the framework of SCET, to construct a resummed cross section are given in \App{app:resum}. 
We will discuss matching to fixed-order and demonstrate our ability to make phenomenological predictions in subsequent sections.

Resummation in SCET is accomplished with renormalization group evolution.  Solving the renormalization group equations to a given logarithmic accuracy requires anomalous dimensions to a particular fixed order.
The anomalous dimensions of the functions in the factorization theorem must sum to zero, 
because the cross section is independent of the renormalization scale.

\begin{table}
\begin{center}
\begin{tabular}{c| c c c c c c}
 & $\Gamma_\text{cusp}$ & $\gamma$ & $\beta$ & $\tilde F(\partial_\omega)$ & $c_F$ & Matching\\
 \hline
 LL& $\alpha_s$ & -  & $\alpha_s$ & - &  - & - \\
 NLL & $\alpha_s^2$ & $\alpha_s$  & $\alpha_s^2$ & $\alpha_s$ & - & $\alpha_s$ \\
 NNLL & $\alpha_s^3$ & $\alpha_s^2$ & $\alpha_s^3$ & $\alpha_s^2$ & $\alpha_s$ & $\alpha_s^2$
\end{tabular}
\end{center}
\caption{
$\alpha_s$-order of ingredients needed for resummation to the accuracy given.  $\Gamma_\text{cusp}$ is the cusp anomalous dimension, $\gamma$ is the non-cusp anomalous dimension, and $\beta$ is the QCD $\beta$-function.  $\tilde F(\partial_\omega)$ are the logarithms in the low-scale matrix elements that have been Laplace transformed and $c_F$ are constants in the low-scale matrix elements.  The final column shows the relative order to which the resummed cross section can be matched to fixed-order.
}\label{tab:logtab}
\end{table}

Recall that, for $e^+e^- \to$ hemisphere jets, the factorization theorem for soft-drop groomed energy correlation functions is
\begin{equation}\label{eq:factagain}
\hspace{-0.28cm}\frac{d^2\sigma}{d\ecf{2,L}{\alpha}\, d\ecf{2,R}{\alpha}} = H(Q^2) S_\glob(\zcut) \left[S_{C}(\zcut\ecf{2,L}{\alpha})\otimes J(\ecf{2,L}{\alpha})\right]\left[S_{C}(\zcut\ecf{2,R}{\alpha})\otimes J(\ecf{2,R}{\alpha})\right]\,.
\end{equation}
\Tab{tab:logtab} presents the order to which anomalous dimensions and constants of the functions in this factorization theorem must be computed for particular logarithmic accuracy (see, e.g., \Ref{Almeida:2014uva}).  
The cusp anomalous dimension $\Gamma_\text{cusp}$ and the QCD $\beta$-function are known through three-loop order \cite{Korchemsky:1987wg,Vogt:2000ci,Berger:2002sv,Moch:2005tm,Tarasov:1980au,Larin:1993tp} and we present them in \App{app:cab}.  The hard function $H(Q^2)$ for $e^+e^-\to q\bar q$ is known to high orders and its non-cusp anomalous dimension $\gamma_H$ is known at three-loop order \cite{vanNeerven:1985xr,Matsuura:1988sm};  we present the relevant pieces in \App{app:hard}.  For arbitrary angular exponents $\alpha$ and $\beta$, little else in the factorization theorem is known at sufficiently high accuracy to resum to NNLL.

The goal of this section is to fill in the rest of the table, to achieve full NNLL accuracy.
We start in \Sec{sec:nnll20} restricting to $\alpha = 2$ (jet mass) and $\beta = 0$ (mMDT groomer).  For this case, all of the missing ingredients can be determined by recycling results from the literature, up to calculable clustering effects from the soft drop algorithm.  In \Sec{sec:nnlla2ball}, we consider $\alpha=2$ and $\beta \geq 0$ and 
demonstrate that one can 
extract unknown two-loop non-cusp anomalous dimensions with \eventtwo.  It is possible to extend our analysis to angular exponents for the energy correlation functions beyond $\alpha = 2$, but we do not do it in this 
paper.\footnote{The two-loop non-cusp anomalous dimension of the soft function for event-wide (recoil-free) angularities \cite{Berger:2003iw,Almeida:2008yp,Ellis:2010rwa,Larkoski:2014uqa} as a function of the angular exponent has been calculated in \Ref{Bell:2015lsf}.  Recoil-free angularities and two-point energy correlation functions have identical anomalous dimensions \cite{Larkoski:2014uqa} and could be used in the same way as the calculation for $\alpha = 2$.}

\subsection{NNLL for $\alpha = 2$, $\beta = 0$}\label{sec:nnll20}

We first consider angular exponents $\alpha = 2$ and $\beta = 0$.  In this case, the soft drop requirement enforced at every branching reduces to an energy cut
\begin{equation}
\min[E_i,E_j] > \zcut (E_i+E_j)\,.
\end{equation}
On the soft-drop groomed jets we then measure
\begin{equation}
\ecf{2}{2} = \frac{m_g^2}{E_{g}^2}\,,
\end{equation}
where the subscript $g$ denotes that the mass and energy are measured on the groomed jet.  The jet functions in the factorization theorem 
are independent of the soft drop groomer, so we are able to use results from the literature for these.  The inclusive jet function has
been calculated to two loops~\cite{Bauer:2003pi,Bosch:2004th,Becher:2010pd,Becher:2006qw} and the non-cusp anomalous dimension of the inclusive jet function is known to three loops \cite{Neubert:2004dd,Becher:2006mr}.  We present the relevant expressions in \App{app:jetfunc}.

This leaves the soft function $S_\glob(\zcut)$ and the collinear-soft function $S_C(\zcut\ecf{2}{2})$ to be determined.  Their one-loop expressions are easily calculable, and we present the results in \App{app:softfct} and \App{app:csoftfct}.  To determine their two-loop non-cusp anomalous dimensions, we exploit the renormalization group consistency of the factorization theorem.  The sum of the anomalous dimensions must vanish at each order:
\begin{equation}\label{eq:scanom}
0=\gamma_H+\gamma_S+2\gamma_J+2\gamma_{S_C}\,,
\end{equation}
where $\gamma_F$ denotes the anomalous dimension of function $F$ in the factorization theorem, and we have used the symmetry of the left and right hemispheres of the event.  Therefore, only one unknown anomalous dimension remains, which we take to be $\gamma_S$.

\subsubsection{Two-Loop Soft Function}\label{sec:sfanomb0}

To calculate the two-loop non-cusp anomalous dimension $\gamma_S$ we need to calculate the soft function $S_\glob(\zcut)$ with two real emissions.  The two-loop expression for the soft function is
\begin{equation}\label{eq:sd2loop}
\left.S_\glob(\zcut)\right|_{\alpha_s^2}=\int [d^dk_1]_+[d^dk_2]_+ \,|{\cal M}(k_1,k_2)|^2 \Theta_\text{SD}\,.
\end{equation}
Here, $[d^dk_1]_+$ is the positive-energy on-shell phase space measure in $d=4-2\epsilon$ dimensions:
\begin{equation}
[d^dk_1]_+=\frac{d^dk_1}{(2\pi)^d}2\pi\delta(k_1^2)\Theta(k_1^0)\,,
\end{equation}
and $|{\cal M}(k_1,k_2)|^2$ is the squared matrix element for two soft emissions from a $q\bar q$ dipole.  The explicit expression for $|{\cal M}(k_1,k_2)|^2$ can be found in \Ref{Catani:1999ss}.  $\Theta_\text{SD}$ is the phase space constraint imposed by the soft drop groomer. Recall that, for consistency with the assumed hierarchy $\ecf{2}{\alpha} \ll \zcut$, soft modes must fail soft drop.

If the particles in the hemispheres are reclustered using the Cambridge/Aachen algorithm, $\Theta_\text{SD}$ can be written as
\begin{align}\label{eq:sdb0ps}
\Theta_\text{SD} 
&= 
\Theta(-\eta_1\eta_2)\,
\Theta\left(\zcut\frac{Q}{2}-k_1^0\right)
\Theta\left(\zcut\frac{Q}{2}-k_2^0\right) \\
&
+
\Theta(\eta_1\eta_2)
\left[
\Theta(\theta_{1J}-\theta_{12})
\Theta(\theta_{2J}-\theta_{12})
\Theta\left(\zcut\frac{Q}{2}-k_1^0-k_2^0\right)
\right.\nonumber\\
&\hspace{1cm}\left.
+
\left[1-\Theta(\theta_{1J}-\theta_{12})\Theta(\theta_{2J}-\theta_{12})\right]
\Theta\left(\zcut\frac{Q}{2}-k_1^0\right)\Theta\left(\zcut\frac{Q}{2}-k_2^0\right)
\right]
\nonumber\,.
\end{align}
The first line of \Eq{eq:sdb0ps} corresponds to particles 1 and 2 lying in different hemispheres 
(opposite rapidity with respect to the $q\bar q$ dipole), and so each particle individually must fail soft drop.  
$Q$ is the center of mass energy and so $Q/2$ is the energy in one hemisphere.  
The second and third lines correspond to the configuration where both particles lie in the same hemisphere.  
$\theta_{12}$ is the angle between the particles and $\theta_{iJ}$ (for $i=1,2$)
is the angle particle $i$ makes with that hemisphere's axis.
If $\theta_{12}$ is less than both $\theta_{1J}$ and $\theta_{2J}$ then,  according to the Cambridge/Aachen algorithm,
 the soft particles are clustered first.  
Therefore, the sum of the energies of particles 1 and 2 must fail soft drop.  
If instead one of the particles is closer to the jet axis, then they are clustered separately and must individually 
fail soft drop.  

To proceed, we separate the squared matrix element into Abelian and non-Abelian pieces, according to their color coefficient.  At this order, the squared matrix element takes the form
\begin{align}
|{\cal M}(k_1,k_2)|^2 = |{\cal M}_\text{n-A}(k_1,k_2)|^2 + \frac{1}{2!}|{\cal M}(k_1)|^2|{\cal M}(k_2)|^2\,,
\end{align}
Here, ``n-A'' denotes the non-Abelian component of the squared matrix element, which includes the $C_F C_A$ and $C_F n_f T_R$ color channels.  The Abelian contribution is just the symmetrized product of the one-loop result, with a color factor of $C_F^2$.  We will consider these two pieces separately, starting with the non-Abelian term.

\subsubsection*{Non-Abelian Clustering Effects}

Note that except for the effects from Cambridge/Aachen clustering, soft drop is just imposing a soft energy veto 
on each hemisphere.  The two-loop soft function with a soft energy veto was calculated in \Ref{vonManteuffel:2013vja}.  That calculation showed that the two-loop Abelian piece (proportional to $C_F^2$) to the energy vetoed soft function satisfies non-Abelian exponentiation.  The two-loop non-cusp anomalous dimension for a hemisphere energy vetoed soft function is then purely non-Abelian and was extracted in \Ref{Chien:2015cka}.  The non-Abelian part of the soft function with an energy veto at two-loops is
\begin{equation}\label{eq:veto2loop}
\left.S_\text{veto}\right|_{\text{n-A,}\alpha_s^2}=\int [d^dk_1]_+[d^dk_2]_+ \,|{\cal M}_\text{n-A}(k_1,k_2)|^2 \Theta_\text{veto}\,.
\end{equation}
The phase space cut $\Theta_\text{veto}$ is
\begin{align}
\Theta_\text{veto}&=
\Theta\left(
\Lambda-k_1^0-k_2^0
\right)\,, \nonumber
\end{align}
where $\Lambda$ is the veto scale.
We can then write the two-loop soft function for soft drop as
\begin{align}\label{eq:sdveto}
&\left.S_\glob(\zcut)\right|_{\text{n-A,}\alpha_s^2}
=
\left.S_\text{veto}\right|_{\text{n-A,}\alpha_s^2}+\int [d^dk_1]_+[d^dk_2]_+ \,|{\cal M}_\text{n-A}(k_1,k_2)|^2\left[ \Theta_\text{SD}- \Theta_\text{veto}\right]\,,
\end{align}
where the veto scale is set to $\Lambda=\zcut Q/2$.  The difference between the soft drop and energy veto phase space constraints 
is purely a clustering effect, given by
\begin{align}\label{eq:clustsd}
\Theta_\text{SD}- \Theta_\text{veto}&=
\big\{\Theta(\eta_1\eta_2)
\left[
1-\Theta(\theta_{1J}-\theta_{12})\Theta(\theta_{2J}-\theta_{12})
\right]+\Theta(-\eta_1\eta_2)\big\}\\
&
\times\Theta\left(\zcut\frac{Q}{2}-k_1^0\right)\Theta\left(\zcut\frac{Q}{2}-k_2^0\right)\Theta\left(k_1^0+k_2^0-\zcut\frac{Q}{2}\right)
\nonumber\,.
\end{align}

\Eq{eq:sdveto} enables us to calculate much more simply the two-loop non-cusp anomalous dimension of the soft function.  The anomalous dimension can then be written as
\begin{equation}
\gamma_S = \gamma_{\text{veto}}+\gamma_\text{C/A}\,.
\end{equation}
$\gamma_{\text{veto}}$ is the two-loop non-cusp 
anomalous dimension of $S_\text{veto}$ extracted in \Ref{Chien:2015cka}:
\begin{equation}\label{eq:ssanom}
\gamma_{\text{veto}}^{\alpha_s^2}=\left(\frac{\alpha_s}{4\pi}\right)^2C_F\left[
\left(
\frac{1616}{27}-56\zeta_3
\right)C_A-\frac{448}{27}n_f T_R-\frac{2\pi^2}{3}\beta_0
\right]\,,
\end{equation}
where $\beta_0$ is the one-loop $\beta$-function coefficient:
\begin{equation}
\beta_0=\frac{11}{3}C_A-\frac{4}{3}n_f T_R\,.
\end{equation}
Then, we only need to determine the contribution to the anomalous dimension from residual Cambridge/Aachen clustering effects, $\gamma_\text{C/A}$.

The non-Abelian clustering effects are contained in
\begin{align}
&\left.S_\glob(\zcut)\right|^\text{C/A}_{\text{n-A},\alpha_s^2}=\int [d^dk_1]_+[d^dk_2]_+ \,|{\cal M}_\text{n-A}(k_1,k_2)|^2\left[ \Theta_\text{SD}- \Theta_\text{veto}\right]\,.
\end{align}
The squared non-Abelian matrix element does not have collinear singularities when the angle of the particles from the jet axis is strongly ordered.  Therefore, in this integral there is only a collinear divergence when the two emissions become collinear to the jet axis in a non-strongly ordered way.  The coefficient of this divergence is proportional to the correction to the two-loop anomalous dimension due to clustering effects in the non-Abelian color channel.  The divergence can be extracted with the standard plus-function prescription and the correction to the anomalous dimension can be found.  While we were unable to find an analytic expression, its approximate numerical value is\footnote{This anomalous dimension does not seem to be a linear combination of the usual transcendental numbers appearing in other two-loop anomalous dimensions.}
\begin{align}
&\left.S_\glob(\zcut)\right|^\text{C/A}_{\text{n-A},\alpha_s^2}=\left(
\frac{\alpha_s}{4\pi}
\right)^2C_F\left[
-9.31 C_A-14.04 n_f T_R
\right]\left(
\frac{4\mu^2}{\zcut^2Q^2}
\right)^{2\epsilon}\frac{1}{4\epsilon}+{\cal O}(\epsilon^0)\,.
\end{align}
The contribution to the anomalous dimension is then
\begin{equation}\label{eq:snaanom}
\gamma_\text{C/A}^{\text{n-A,}\alpha_s^2}=\left(
\frac{\alpha_s}{4\pi}
\right)^2C_F\left[
-9.31 C_A-14.04 n_f T_R
\right]\,.
\end{equation}

\subsubsection*{Abelian Clustering Effects}

The Abelian contribution can be calculated similarly.  However, unlike the non-Abelian contribution, the exponentiation of the one-loop result will describe at least some of the two-loop Abelian piece.  If the square of the one-loop result does not account for all of the two-loop result, then non-Abelian exponentiation breaks down.  This does not mean that exponentiation breaks down or that the cross section cannot be resummed, just that the anomalous dimension of the purely Abelian piece will need to be corrected at every logarithmic order.  So, for the two-loop non-cusp anomalous dimension, we need to determine the part of the soft function that is not accounted for by non-Abelian exponentiation.

To do this, we start from the full expression for the Abelian term at two-loops:
\begin{equation}
\left.S_\glob(\zcut)\right|_{\text{A,}\alpha_s^2}=\frac{1}{2!}\int [d^dk_1]_+[d^dk_2]_+ \,|{\cal M}(k_1)|^2|{\cal M}(k_2)|^2 \Theta_\text{SD}\,.
\end{equation}
We then add and subtract the one-loop phase space constraints:
\begin{align}\label{eq:Sabelian}
&\left.S_\glob(\zcut)\right|_{\text{A,}\alpha_s^2}\\
&
\hspace{0.5cm}
=\frac{1}{2!}\int [d^dk_1]_+[d^dk_2]_+ \,|{\cal M}(k_1)|^2|{\cal M}(k_2)|^2 \Theta\left(
\zcut\frac{Q}{2}-k_1^0
\right)\Theta\left(
\zcut\frac{Q}{2}-k_2^0
\right)
\nonumber \\
&
\hspace{1cm}
+\frac{1}{2!}\int [d^dk_1]_+[d^dk_2]_+ \,|{\cal M}(k_1)|^2|{\cal M}(k_2)|^2 \left[\Theta_\text{SD}-\Theta\left(
\zcut\frac{Q}{2}-k_1^0
\right)\Theta\left(
\zcut\frac{Q}{2}-k_2^0
\right)\right]\,. \nonumber
\end{align}
The difference between the phase space constraints is a clustering effect, given by
\begin{align}
\Theta_\text{SD}&-\Theta\left(
\zcut\frac{Q}{2}-k_1^0
\right)\Theta\left(
\zcut\frac{Q}{2}-k_2^0
\right)\\
&
=-\Theta(\eta_1\eta_2)\,
\Theta(\theta_{1J}-\theta_{12})\Theta(\theta_{2J}-\theta_{12})\nonumber\\
&\hspace{0.5cm}
\times\Theta\left(\zcut\frac{Q}{2}-k_1^0\right)\Theta\left(\zcut\frac{Q}{2}-k_2^0\right)\Theta\left(k_1^0+k_2^0-\zcut\frac{Q}{2}\right)
\nonumber\,.
\end{align}
As with the non-Abelian term, this phase space constraint completely removes all soft divergences and the strongly-ordered collinear limit.  The remaining divergence can be isolated by standard plus-function techniques.  For the two-loop Abelian Cambridge/Aachen clustering term, we find the numerical result
\begin{align}
\left.S_\glob(\zcut)\right|^\text{C/A}_{\text{A,}\alpha_s^2}&=
\left(
\frac{\alpha_s}{4\pi}
\right)^2  34.01 \,C_F^2
\left(
\frac{4\mu^2}{\zcut^2Q^2}
\right)^{2\epsilon}\frac{1}{4\epsilon}+{\cal O}(\epsilon^0)\,,
\end{align}
for the second integral in \Eq{eq:Sabelian}.
The contribution to the anomalous dimension is then
\begin{equation}\label{eq:saanom}
\gamma_\text{C/A}^{\text{A,}\alpha_s^2}=\left(
\frac{\alpha_s}{4\pi}
\right)^2  34.01\,C_F^2\,.
\end{equation}

\subsubsection{Two-Loop Anomalous Dimension and Comparison with \eventtwo}

Combining Eqs.~\eqref{eq:ssanom}, \eqref{eq:snaanom} and \eqref{eq:saanom}, the total two-loop non-cusp anomalous dimension for the soft function is
\begin{equation}
\gamma_S^{\alpha_s^2}=\left(\frac{\alpha_s}{4\pi}\right)^2C_F\left[34.01\,C_F+
\left(
\frac{1616}{27}-56\zeta_3-9.31
\right)C_A-\left(\frac{448}{27}+14.04\right)n_f T_R-\frac{2\pi^2}{3}\beta_0
\right]
\,.
\label{eq:softnoncusp}
\end{equation}
The two-loop non-cusp anomalous dimension for the collinear-soft function is found by consistency using \Eq{eq:scanom}.  Note that this anomalous dimension has no $\log\zcut$ terms.  Therefore, the anomalous dimensions of no functions in the factorization theorem have $\log\zcut$ dependence.  
This is a consequence of the fact that each function of our factorization theorem in \Eq{eq:factdefnew} depends on a single infrared scale, 
allowing NNLL resummation of all logarithms of $\zcut$ and $\ecf{2}{2}$ alike. 
As we discuss in \Secs{sec:singcs}{sec:antikt}, this result relies on the choice of Cambridge/Aachen reclustering in the soft drop algorithm.

We can verify this result by comparing the resummed distribution, truncated at $\mathcal{O}(\alpha_s^2)$, 
with the singular region of the full QCD result, computed to the same fixed order.
For the full QCD result,
we have implemented soft drop into \eventtwo~\cite{Catani:1996vz}, 
a Monte Carlo code that generates fixed-order results up to $\mathcal{O}(\alpha_s^2)$ in $e^+e^-$ collisions.  Our specific implementation is as follows.  We generate $e^+e^-$ collisions at 1 TeV center of mass energy and identify event hemispheres with the exclusive $k_T$ algorithm \cite{Catani:1991hj}.  We then recluster each hemisphere using the Cambridge/Aachen algorithm and apply soft drop with $\beta = 0$.  On each of the soft-drop groomed hemispheres, we then measure the energy correlation function $\ecf{2}{2}$ and record the larger of the two values, which we denote by $\ecf{2,H}{2}$ and refer to as the heavy groomed mass.  This is simply related to the cross section of our factorization theorem:
\begin{align}
&\frac{d\sigma}{d\ecf{2,H}{2}}
=\int d\ecf{2,L}{2}\,d\ecf{2,R}{2}
\frac{d^2\sigma}{d\ecf{2,L}{2}\,d\ecf{2,R}{2}}
\left[
\Theta\left(
\ecf{2,L}{2}-\ecf{2,R}{2}
\right)\delta\left(
\ecf{2,H}{2}-\ecf{2,L}{2}
\right)+
(L \leftrightarrow R)
\right]\,.
\end{align}

\begin{figure}[t]
\centering
\subfloat[]{\label{fig:ev2a2b0a2l2}
\includegraphics[width=0.45\linewidth]{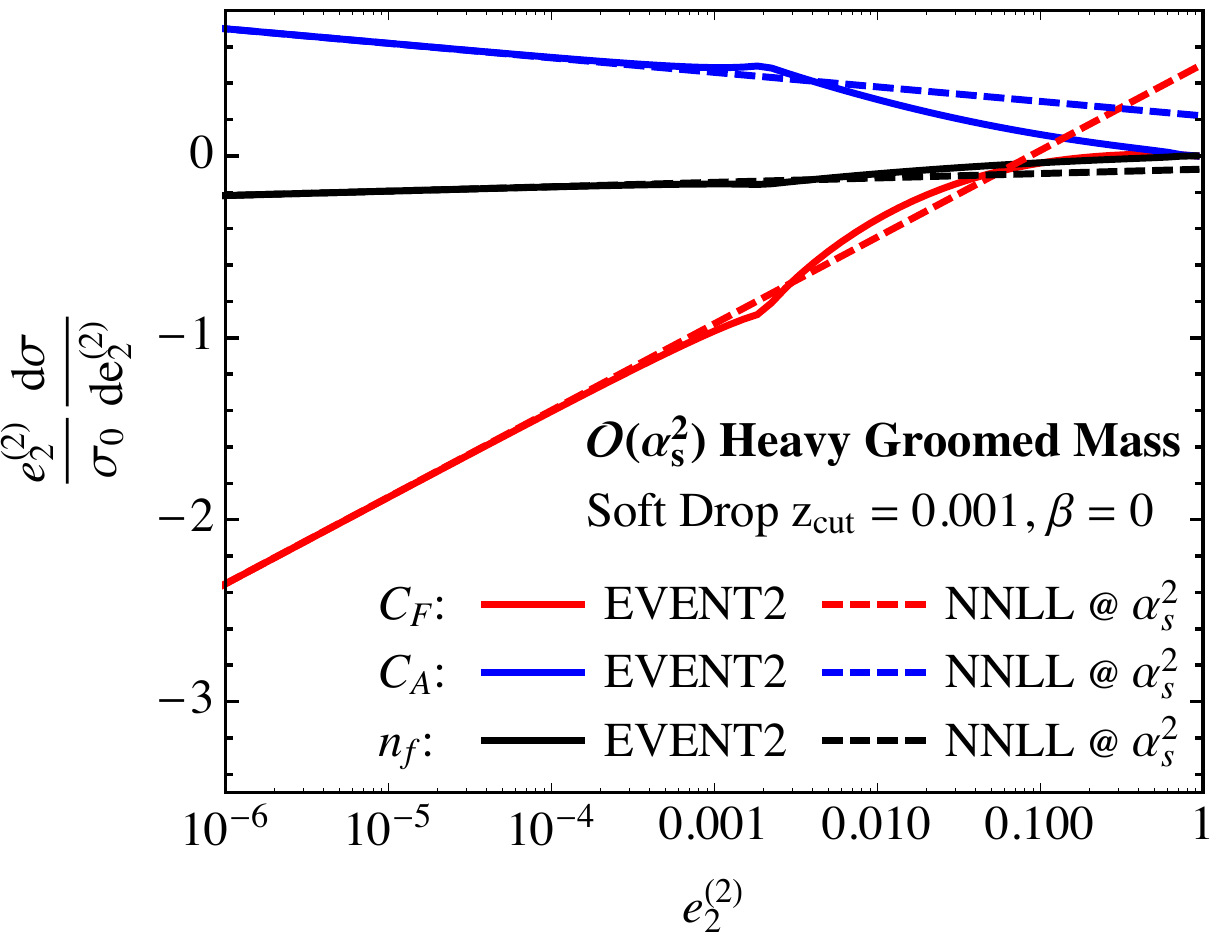}
}\quad
\subfloat[]{\label{fig:ev2a2b0a2l1}
\includegraphics[width=0.45\linewidth]{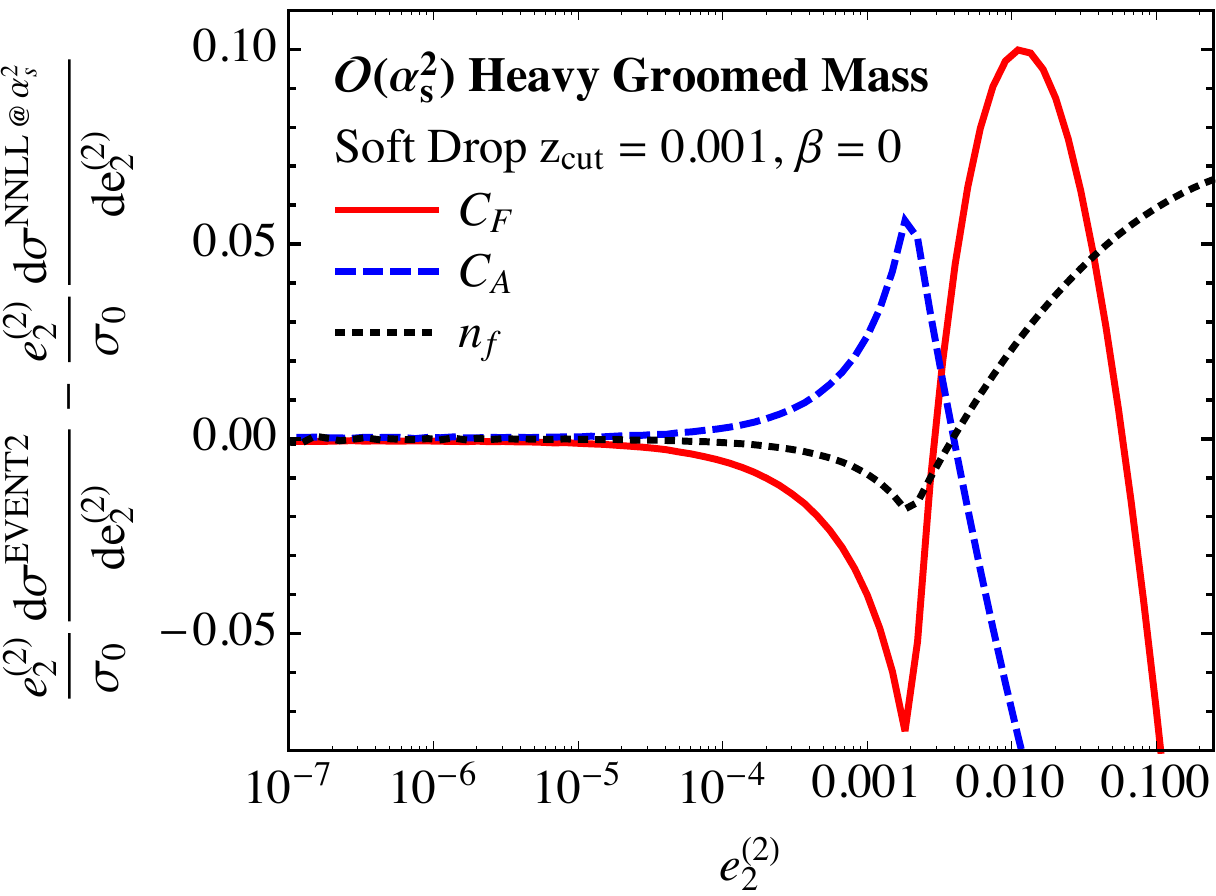}
}\\
\caption{
Verification of our factorization theorem at ${\cal O}(\alpha_s^2)$ for soft-drop grooming with $\zcut = 0.001$ and $\beta=0$.  Solid curves are numerical results from \eventtwo, and dashed curves are ${\mathcal O}(\alpha_s^2)$ terms in our NNLL distribution, 
plotted in the three color channels $C_F^2$, $C_FC_A$, and $C_Fn_fT_R$. (a) shows a direct comparison and (b) the difference. 
}
\label{fig:ev2a2b0}
\end{figure}

In \Fig{fig:ev2a2b0}, we compare \eventtwo~results to the prediction of the factorized expression at NNLL expanded to ${\cal O}(\alpha_s^2)$.  For soft drop with $\beta = 0$, soft logarithms are removed, which means that at ${\cal O}(\alpha_s^2)$, the cross section has the schematic form
\begin{equation}
\ecf{2,H}{2}\frac{d\sigma^{\alpha_s^2}}{d\ecf{2,H}{2}} \sim  \alpha_s^2 C_0 \log \ecf{2,H}{2}+\alpha_s^2 C_1\,,
\end{equation}
where $C_0$ and $C_1$ are constants.  We plot the cross section separated into the three color channels ($C_F^2$, $C_F C_A$, and $C_Fn_f T_R$).  We set $\zcut=0.001$ to suppress power corrections of $\zcut$.  Excellent agreement between our factorization theorem and \eventtwo~is observed at small $\ecf{2}{2}$, demonstrating that we have captured all singular terms of the full QCD result in our factorization theorem to ${\cal O}(\alpha_s^2)$.

\subsection{Reclustering with anti-$k_T$}\label{sec:antikt}
It is illuminating to study the clustering effects in the soft function in more detail.  In this section, we re-calculate the clustering effects with the anti-$k_T$ algorithm, instead of the standard Cambridge/Aachen algorithm.  
We find that the clustering effects with the anti-$k_T$ algorithm are intimately related to the corresponding effects calculated in jet veto calculations.  
This can be understood relatively simply by re-expressing the clustering conditions in a form analogous to the clustering metric of the longitudinally-invariant $k_T$ algorithm.

To calculate the two-loop soft function for soft drop defined with anti-$k_T$ reclustering, we only need to calculate the clustering effects unique to this algorithm.  
We will denote the phase space constraints for the anti-$k_T$ reclustering as $\Theta^{\text{a}k_T}_\text{SD}$, but we will not explicitly present them here.  
The two-loop soft function is
\begin{align}
&\left.S^{\text{a}k_T}(\zcut)\right|_{\alpha_s^2}\\
&
\hspace{1cm}
=\int [d^dk_1]_+[d^dk_2]_+ \,|{\cal M}(k_1,k_2)|^2 \Theta_\text{veto}+\int [d^dk_1]_+[d^dk_2]_+ \,|{\cal M}(k_1,k_2)|^2\left[ \Theta^{\text{a}k_T}_\text{SD}- \Theta_\text{veto}\right].
\nonumber
\end{align}
The relevant phase space constraints can be written as
\begin{align}
\Theta^{\text{a}k_T}_\text{SD}-\Theta_\text{veto}
&=\Theta(\eta_1\eta_2)
\left[
1-\Theta\left(\max[k_1^0,k_2^0]\theta_{1J}-\frac{Q}{2}\theta_{12}\right)\Theta\left(\max[k_1^0,k_2^0]\theta_{2J}-\frac{Q}{2}\theta_{12}\right)
\right]\nonumber\\
&
\times\Theta\left(\zcut\frac{Q}{2}-k_1^0\right)\Theta\left(\zcut\frac{Q}{2}-k_2^0\right)\Theta\left(
k_1^0+k_2^0-\zcut\frac{Q}{2}
\right)\,.
\end{align}
With this, we can calculate the divergent part of the two-loop soft function from clustering effects and extract the anomalous dimension.   As with Cambridge/Aachen, we can write the two-loop non-cusp anomalous dimension as
\begin{equation}
\gamma_S^{\text{a}k_T}=\gamma_{\text{veto}}+\gamma_{\text{a}k_T}\,,
\end{equation}
where $\gamma_{\text{a}k_T}$ is the part of the anomalous dimension purely from clustering effects.  We find
\begin{align}
\gamma_{\text{a}k_T}&=-8\left(
\frac{\alpha_s}{4\pi}
\right)^2C_F\left\{
\left[
\left(
\frac{131}{9}-\frac{4}{3}\pi^2-\frac{44}{3}\log 2
\right)C_A+\left(
-\frac{46}{9}+\frac{16}{3}\log 2
\right)n_f T_R
\right]\log \zcut
\right.\nonumber\\
&
\hspace{2cm}
+\left(
-\frac{269}{6}+\frac{7}{2}\zeta_3+\frac{274}{9}\log 2+\frac{11\pi^2}{9}+\frac{44}{3}\log^2 2
\right)C_A\nonumber \\
&
\hspace{2cm}
\left.
+\left(
\frac{53}{3}-\frac{4\pi^2}{9}-\frac{116}{9}\log 2-\frac{16}{3}\log^2 2
\right)n_f T_R
\right\}  \,.
\label{eq:anom-akt}
\end{align}

This anomalous dimension is fascinating.  
 First, note that there is no $C_F^2$ term, implying that non-Abelian exponentiation holds for anti-$k_T$ reclustering,
in contrast to what we found for the Cambridge/Aachen algorithm. That is, all logarithms at $\mathcal{O}(\alpha_s^2)$
with color factor $C_F^2$ are accounted for by exponentiating the one-loop result. This is to be expected: Since the anti-$k_T$ algorithm
clusters soft gluons (with energy fractions of order $\zcut$) one-by-one with the hard jet core unless two soft gluons have angular separation 
$\Delta R \lesssim \zcut$, clustering effects are merely a power correction for Abelian gluons.

Also, unlike the case for Cambridge/Aachen reclustering, there is explicit $\log\zcut$ dependence in the anomalous dimension of \Eq{eq:anom-akt}.  
This shows that we do not resum logarithms of $\zcut$ to full NNLL accuracy when anti-$k_T$ clustering is used in soft drop.  
The coefficient of the $\log\zcut$ term is identical to the coefficient of the logarithm of the jet radius $R$ found from 
clustering effects in jet veto calculations \cite{Banfi:2012yh,Becher:2012qa,Banfi:2012jm,Becher:2013xia,Stewart:2013faa}.  
This connection between soft drop and jet veto calculations can be made clearer by a simple rewriting of the clustering metric.

The $k_T$ class of clustering metrics for $e^+e^-$ collisions can be written as
\begin{equation}
d_{ij}=\min\left[E_i^{2p},E_j^{2p}\right]\theta_{ij}^2\,,
\end{equation}
for particles $i$ and $j$, with $p$ an integer that defines the jet algorithm.  In the soft function, soft particles are either clustered with each other or with the jet axis.  For $\beta = 0$, these soft particles have characteristic energy fraction $\zcut\ll 1$.  In terms of energy fractions, the clustering metric of two soft particles is
\begin{equation}
d_{ij}=\min\left[z_i^{2p},z_j^{2p}\right]Q^{2p}\theta_{ij}^2\sim \zcut^{2p}Q^{2p}\theta_{ij}^2\,,
\end{equation}
and for a soft particle $i$ with the jet axis it is
\begin{equation}
d_i=\min[1,\zcut^{2p}]Q^{2p}\theta_i^2\,,
\end{equation}
where $\theta_i$ is the angle between particle $i$ and the jet axis.

Consider $p<0$.  In this case, the two soft gluons are 
(parametrically) clustered together when
\begin{equation}
\zcut^{p}\theta_{ij} < \min[\theta_i,\theta_j]\,,
\end{equation}
or equivalently, when
\begin{equation}
\theta_{ij} <\zcut^{|p|}\min[\theta_i,\theta_j]\,.
\end{equation}
The effective clustering metric in this case is then
\begin{equation}
d_{ij}^{\text{eff}}=\min\left[z_i^{2p},z_j^{2p}\right]\frac{\theta_{ij}^2}{\zcut^{2|p|}\min[\theta_i^2,\theta_j^2]}\,, \qquad d_i^{\text{eff}}=z_i^{2p}\,.
\end{equation}
With $p=-1$, this is the clustering metric for the inclusive anti-$k_T$ algorithm with effective jet radius $R = \zcut \ll1$.  There will now be logarithms of the jet radius that arise.  
The $\log\zcut$ term in the anomalous dimension has the identical coefficient as the $\log R$ term in jet veto calculations because $\zcut$ and $R$ act as the angular scale for collinear splittings in the respective soft functions.

In summary, while we could use anti-$k_T$ to recluster the jet for soft drop grooming, we could not resum all large logarithms to the same precision without a different factorization theorem.  Therefore, reclustering in soft drop with the Cambridge/Aachen algorithm is preferred from a theory perspective.

\subsection{NNLL for $\alpha = 2$, $\beta \geq0$}\label{sec:nnlla2ball}

For soft drop with angular exponent $\beta > 0$, we cannot recycle results from the literature
to reach NNLL precision. Instead, a completely new two-loop calculation of either the soft
or collinear-soft function is needed. But without such a calculation, we can perform NNLL resummation 
for particular values of $\beta>0$, using numerical simulations to estimate the ingredients we lack.
We will demonstrate this explicitly in the case of $\beta=1$, and the result will allow us to study
features of NNLL distributions for energy correlation functions with less aggressive grooming.

The same method we used to validate anomalous dimensions for $\beta = 0$ can be used to extract the anomalous dimension for $\beta>0$. 
This method relies on the fact that all ingredients necessary for NNLL resummation with $\alpha=2$, $\beta>0$
are known except the two-loop non-cusp anomalous dimensions of the soft and collinear-soft functions. 
As mentioned above, renormalization group invariance determines one of these, say $\gamma_{S_C}^{(1)}$, in terms of the
other anomalous dimensions. So only one unkown, $\gamma_S^{(1)}$, remains and we can extract it at 
fixed order.

To do this for a given $\beta>0$, we can use \eventtwo~to obtain numerical results at $\mathcal{O}(\alpha_s^2)$
for the groomed $\ecf{2,H}{2}$ distribution with several moderately small values of $\zcut$. 
From each of these distributions, we can subtract the known terms, which we get by expanding the NNLL distribution
to fixed order. This leaves a term proportional to the unknown $\gamma_S^{(1)}$, as well as power corrections
suppressed by $\ecf{2,H}{2}$ or $\zcut$. By computing the distribution down to very small $\ecf{2,H}{2}$, 
we can ignore the $\ecf{2,H}{2}$ power corrections.  Reducing power corrections from $\zcut$ is limited by the numerical precision of \eventtwo~because 
our factorization theorem only applies for $\ecf{2,H}{2} \ll \zcut$.   Instead, we can
fit the $\zcut$ power corrections to linear combinations of $\zcut \log^n(\zcut) \log^m(\ecf{2,H}{2})$.
At $\mathcal{O}(\alpha_s^2)$ it is appropriate to use $0 \leq n+m \leq 3$, though in practice we found
terms with $m \geq 2$ to be difficult to fit. 
With the non-negligible power corrections thus removed, we can then extract the remaining anomalous dimension.

While the procedure outlined above is straightforward, an explicit calculation of $\gamma_S^{(1)}$ or 
$\gamma_{S_C}^{(1)}$ for $\beta>0$ is of course desirable. On practical time scales, numerical 
extractions are limited to rough approximations, due to 
inadequate numerical precision in the deep infrared.
Nevertheless, an estimate is sufficient for our purposes here, which are to demonstrate the advantages of 
resumming jet substructure observables to NNLL, and to examine various levels of grooming. 
Thus, we will test the above procedure on $\beta=0$, and learn about the associated
uncertainties by comparing with our direct calculation, \Eq{eq:softnoncusp}. 
Then we will move to $\beta=1$, and extract $\gamma_S^{(1)}$ in that case.

\begin{figure}[t]
\centering
\subfloat[]{\label{fig:extractL}
\includegraphics[height=0.35\linewidth]{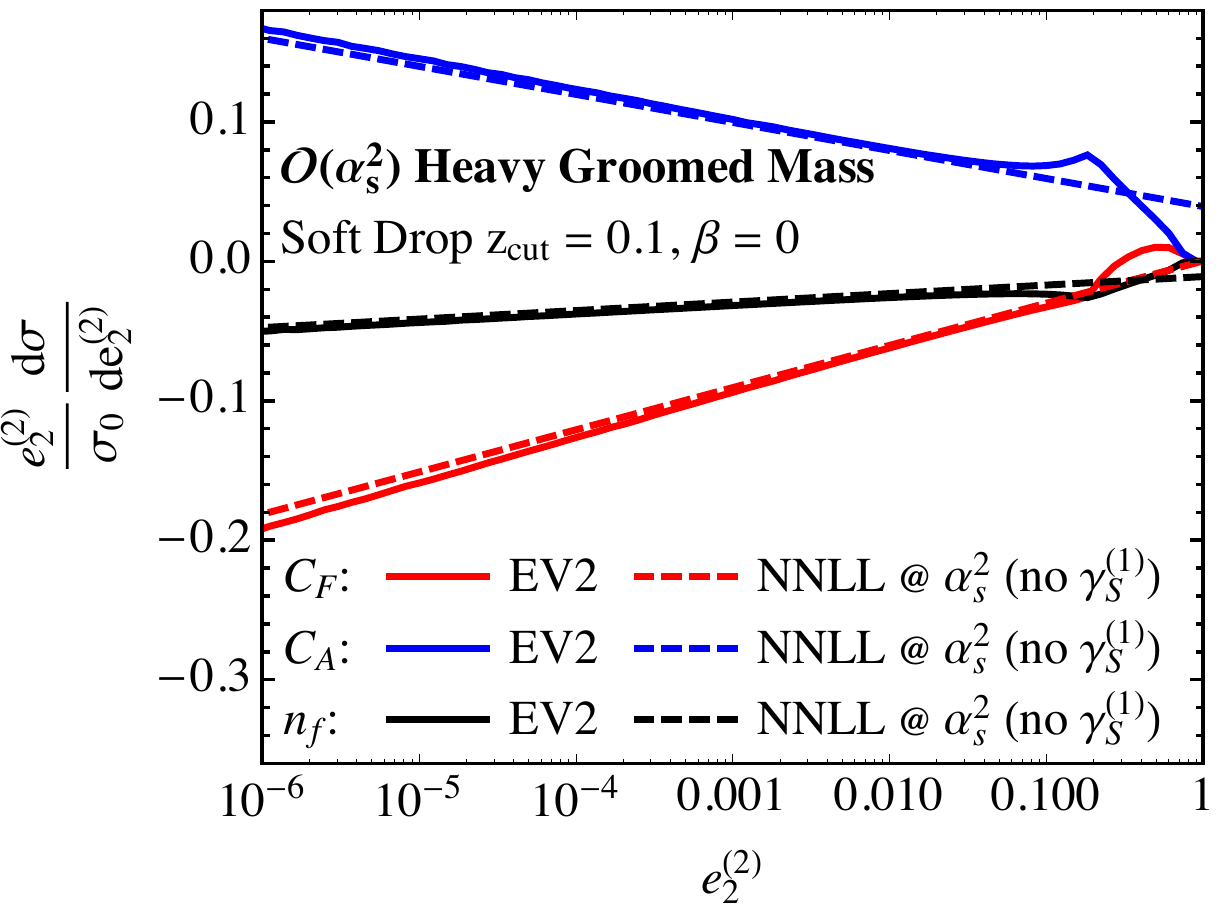}
}\quad
\subfloat[]{\label{fig:extractR}
\includegraphics[height=0.35\linewidth]{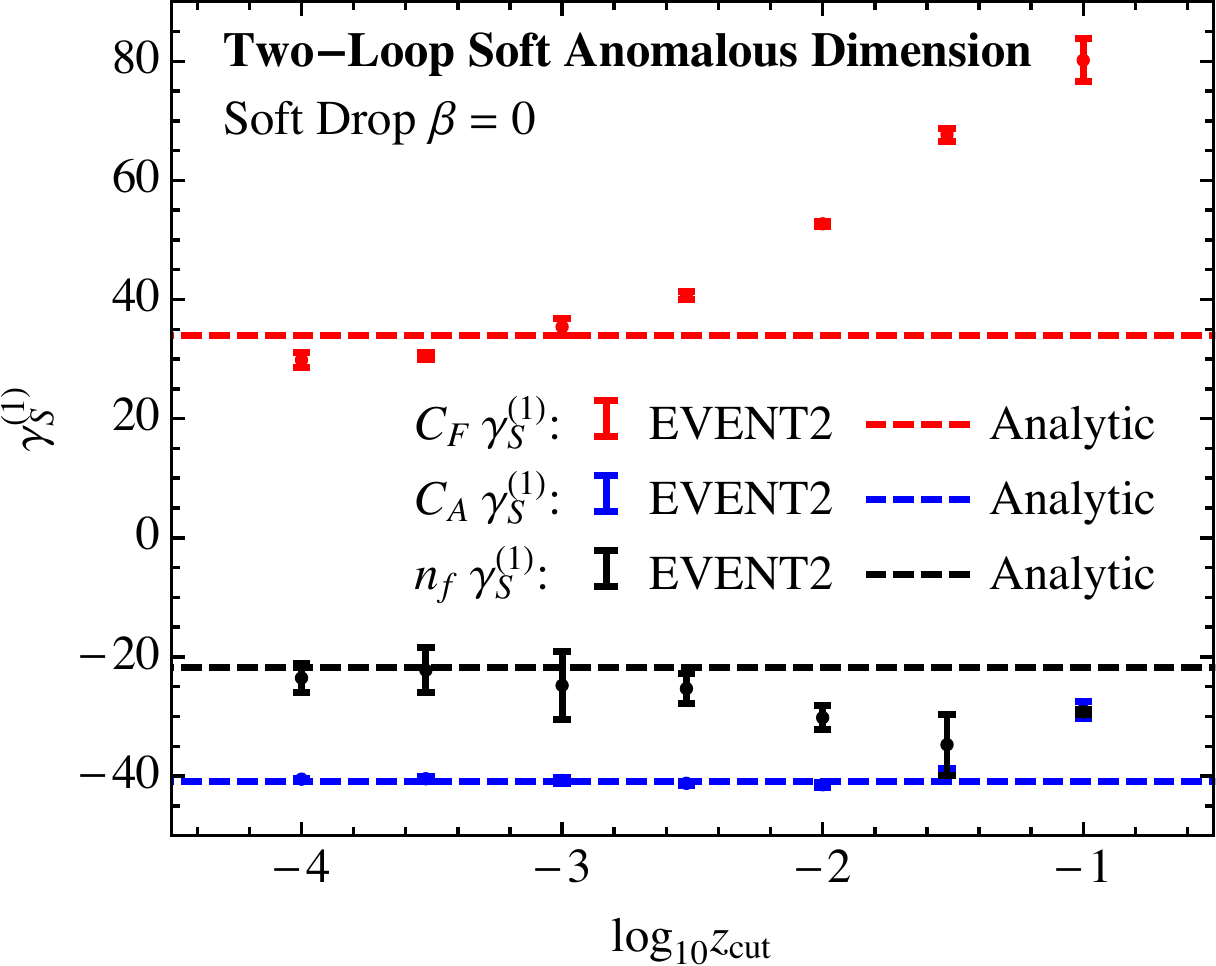}
}\\
\caption{
Demonstration of non-cusp anomalous dimension extraction in \eventtwo. 
(a) Solid curves are numerical results from \eventtwo~at 
${\cal O}(\alpha_s^2)$ with $\beta=0$ and $\zcut=0.1$.  
Dashed curves are ${\cal O}(\alpha_s^2)$ terms in NNLL distribution, without the term
proportional to $\gamma^{(1)}_S$. 
Discrepancy results from $\zcut$ power corrections in solid curves and missing $\gamma^{(1)}_S$
in dashed curves. Subtracting $\zcut \log^n(\zcut) \log(\ecf{2,H}{2})$ power corrections
from dashed curves, we extract the remaining offsets.
(b) As $\zcut \to 0$, remaining offsets allow extraction of $\gamma_S^{(1)}$ in rough agreement
with \Eq{eq:softnoncusp}.
}
\label{fig:extract}
\end{figure}

In \Fig{fig:extractL} we show numerical results at ${\cal O}(\alpha_s^2)$ from \eventtwo~with 
$\beta=0$ and $\zcut=0.1$. 
Also shown is the NNLL distribution, expanded to fixed order, but without 
the $\gamma^{(1)}_S$ term. The discrepancy between the curves is thus due to the missing $\gamma^{(1)}_S$
term and $\zcut$ power corrections. Using several distributions like this
one, with values of $\zcut$ between $10^{-4}$ and $10^{-1}$, we fit the $\zcut$ power corrections. 
\Fig{fig:extractR} shows the remaining offsets between our analytical curves
and the results of \eventtwo, after $\zcut \log^n(\zcut)\log(\ecf{2,H}{2})$ power corrections  
have been subtracted. On each point in this plot, the error bar represents the standard deviation 
in \eventtwo~output, across $\ecf{2,H}{2}$ bins.
The offset that remains as $\zcut \to 0$ is the $\gamma^{(1)}_S$ we would extract using this method.
One can see from \Fig{fig:extractR} that agreement with our analytical calculation, \Eq{eq:softnoncusp}, 
is quite good, with some discrepancy in the $C_F$ channel. 

\begin{table}
\begin{center}
\begin{tabular}{c|ccc}
 Soft Drop $\gamma_S^{(1)}$& $C_F$ & $C_A$ & $n_f$ \\
 \hline
 $\beta=0$ extraction & $28 \pm 1.5$ & $-40 \pm 1$ & $-23 \pm 3$ \\
 \hline
 $\beta=0$ calculation & $34.01$ & $-40.90$ & $-21.86$ \\
 \hline \hline
 $\beta=1$ extraction & $6 \pm 12$ & $-9.5 \pm 2$ & $-8 \pm 7$
\end{tabular}
\end{center}
\caption{
Extraction of two-loop non-cusp anomalous dimension $\gamma_S^{(1)}$ of wide-angle soft function in the three different color channels.
For $\beta=0$ comparison with our direct calculation is possible.
See text for discussion of uncertainties.
}
\label{tab:extraction}
\end{table}

\Tab{tab:extraction} lists the numerical results of $\gamma_S^{(1)}$ using this 
method.\footnote{In carrying out the procedure just described, we tuned 
\eventtwo~parameters to favor the infrared. In particular, 
we use of order 1 trillion events, with $\texttt{CUTOFF}=10^{-15}$ and phase-space 
sampling exponents $\texttt{NPOW1}=\texttt{NPOW2}=5$.  This procedure corresponded to centuries of CPU time.
}
The uncertainties quoted for the $\beta = 0$ extraction in the 
table come from the standard deviation in \eventtwo~output across $\ecf{2,H}{2}$ bins,
which introduces an error in the identification of constant offsets. 
These should be compared with our direct calculation in the second line
of the table. The discrepancy in the $C_F$ channel gives us a sense of additional numerical uncertainties,
which are significant. Similar disagreements have been encountered 
before, e.g.~in \Ref{Chien:2010kc}, in the context of $C_F$ channel extractions from \eventtwo, and to
resolve it might require significantly longer run times.

As stated above, a rough estimate of $\gamma^{(1)}_S$ for $\beta>0$ is sufficient for our purposes,
so we have applied the method described above to the case of $\beta=1$. See the third line of 
\Tab{tab:extraction} for the results of the extraction. Uncertainties quoted in this line of the table
have two sources: (i) variance in \eventtwo~output, and (ii) additional numerical precision issues,
which we took to be the difference (both absolute and relative) between extraction and direct 
calculation in the $\beta=0$ 
test. In each color channel, we took the maximum of these uncertainties and inflated it by a 
factor of 2.
 
The estimate in \Tab{tab:extraction} allows us to 
study NNLL distributions of $\ecf{2,H}{2}$ groomed with $\beta=1$. 
In the resulting distributions, the uncertainties associated with the
imperfect extraction are relatively small; e.g.~see \Fig{fig:eematchnnll} below.
Still, a direct calculation of either $\gamma^{(1)}_S$ or $\gamma^{(1)}_{S_C}$ for $\beta>0$ 
would of course be preferred, but we leave this to future work.

\section{Matching NNLL to Fixed Order in $e^+e^-\to$ dijets}\label{sec:eematch}

Using the results calculated in the previous sections, here we match our resummed differential cross section for 
soft-drop groomed energy correlation functions to fixed-order for hemisphere jets produced in $e^+e^-$ collisions.  
We first match resummed results at NLL and NNLL to ${\mathcal O}(\alpha_s)$ and ${\mathcal O}(\alpha_s^2)$, respectively, using \eventtwo~and demonstrate that 
theoretical uncertainties are greatly reduced at NNLL.  We then compare several Monte Carlo parton shower simulations 
to our matched NNLL results.  We compare both parton and hadron level Monte Carlo to our perturbative analytic results, 
and include a simple model of hadronization in our calculation.  We leave a detailed understanding and justification of 
incorporating hadronization into the resummed and matched cross section to future work.

\subsection{Matching Resummation to Fixed-Order}

With the explicitly calculated and extracted two-loop non-cusp anomalous dimensions of the soft function in the soft drop factorization theorem \Eq{eq:factdefnew}, we are able to resum the differential cross section through NNLL accuracy in the region where $\ecf{2}{2}\ll\zcut \ll1$.  Anomalous dimensions of all functions are collected in the appendices and we present the explicit form of the resummed cross section in \App{app:resum}.  This resummed cross section is only valid in the region where $\ecf{2}{2}\ll\zcut \ll1$, and will not provide an accurate description of the cross section outside this region.  To accurately describe the cross section throughout the full phase space requires matching the resummed result to fixed-order.

While there are many ways to do this at various levels of sophistication, we choose to use simple additive matching. 
That is, we construct matched distributions according to
\begin{equation}\label{eq:addmatch}
\frac{d\sigma_\text{match}}{d\ecf{2}{2}}=\frac{d\sigma_\text{resum}}{d\ecf{2}{2}}+\frac{d\sigma_\text{FO}}{d\ecf{2}{2}}-\frac{d\sigma_\text{resum,FO}}{d\ecf{2}{2}}\,.
\end{equation}
Here, $d\sigma_\text{resum}$ is the resummed cross section, calculated to the appropriate logarithmic accuracy.  $d\sigma_\text{FO}$ is the fixed-order differential cross section calculated to a particular order in $\alpha_s$.  $d\sigma_\text{resum,FO}$ is the resummed cross section truncated at the same accuracy as the fixed-order cross section.  In the infrared phase space region, this term will exactly cancel the singularities in the fixed-order cross section, only leaving the resummed cross section plus power corrections.  In the hard phase space region, this term cancels the resummed cross section, up to higher orders in $\alpha_s$.

The logarithmic accuracy of the resummed cross section was defined in \Sec{sec:nnll}, 
and here we specify the fixed orders that we use in the matching procedure.
We additively match the analytic NLL distributions to ${\mathcal O}(\alpha_s)$ fixed order results,
which include one real emission from the $q \bar q$ dipole. We match NNLL distributions to ${\mathcal O}(\alpha_s^2)$ results,
which include up to two real emissions. 
\eventtwo~is able to generate $e^+e^-$ collisions through ${\mathcal O}(\alpha_s^2)$, except for the two-loop virtual contribution.  
The two-loop virtual term only contributes at $\ecf{2}{2}=0$, so our differential distributions are unaffected by this omission.

\begin{figure}[t]
\centering
\subfloat[]{\label{fig:eematchnll}
\includegraphics[width=0.45\linewidth]{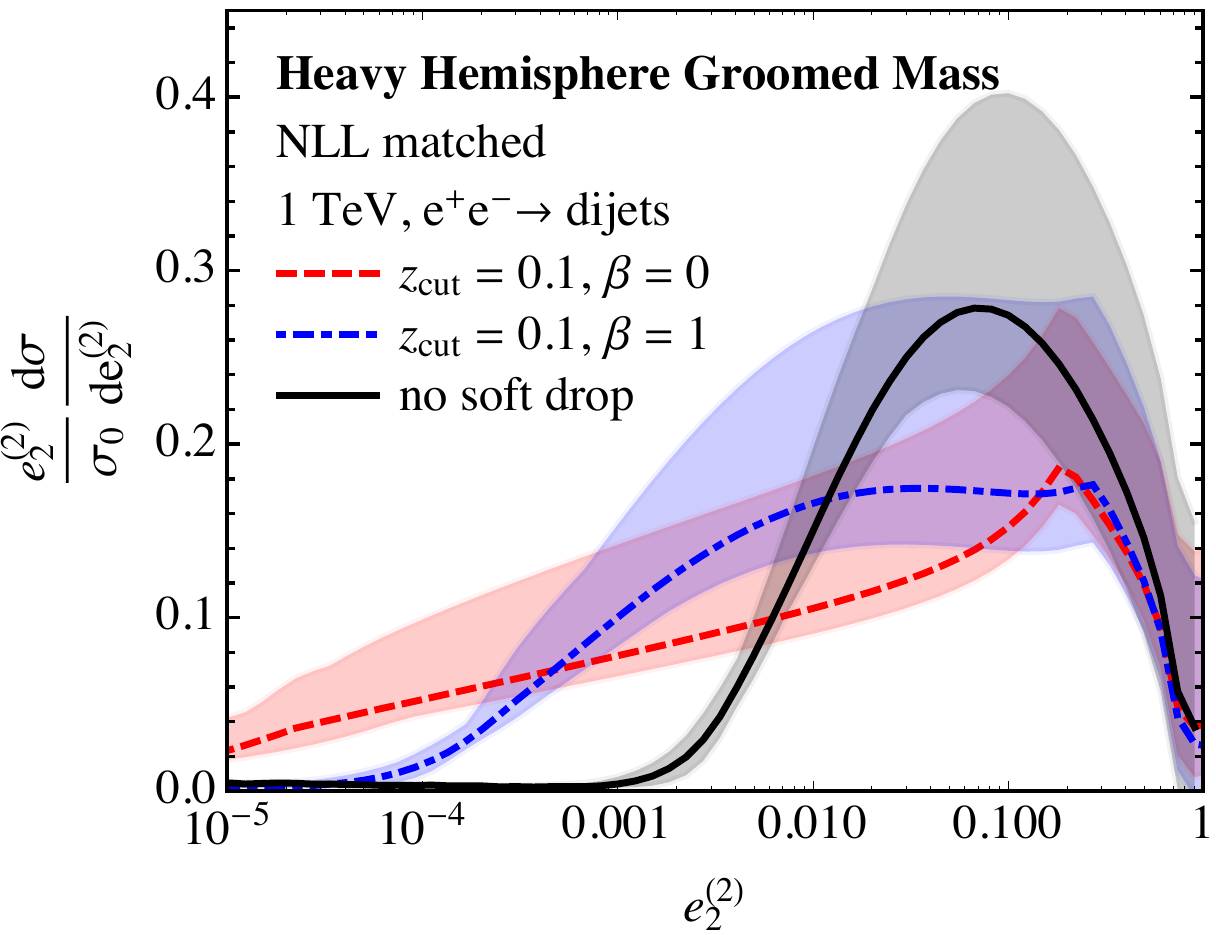}
}\quad
\subfloat[]{\label{fig:eematchnnll}
\includegraphics[width=0.45\linewidth]{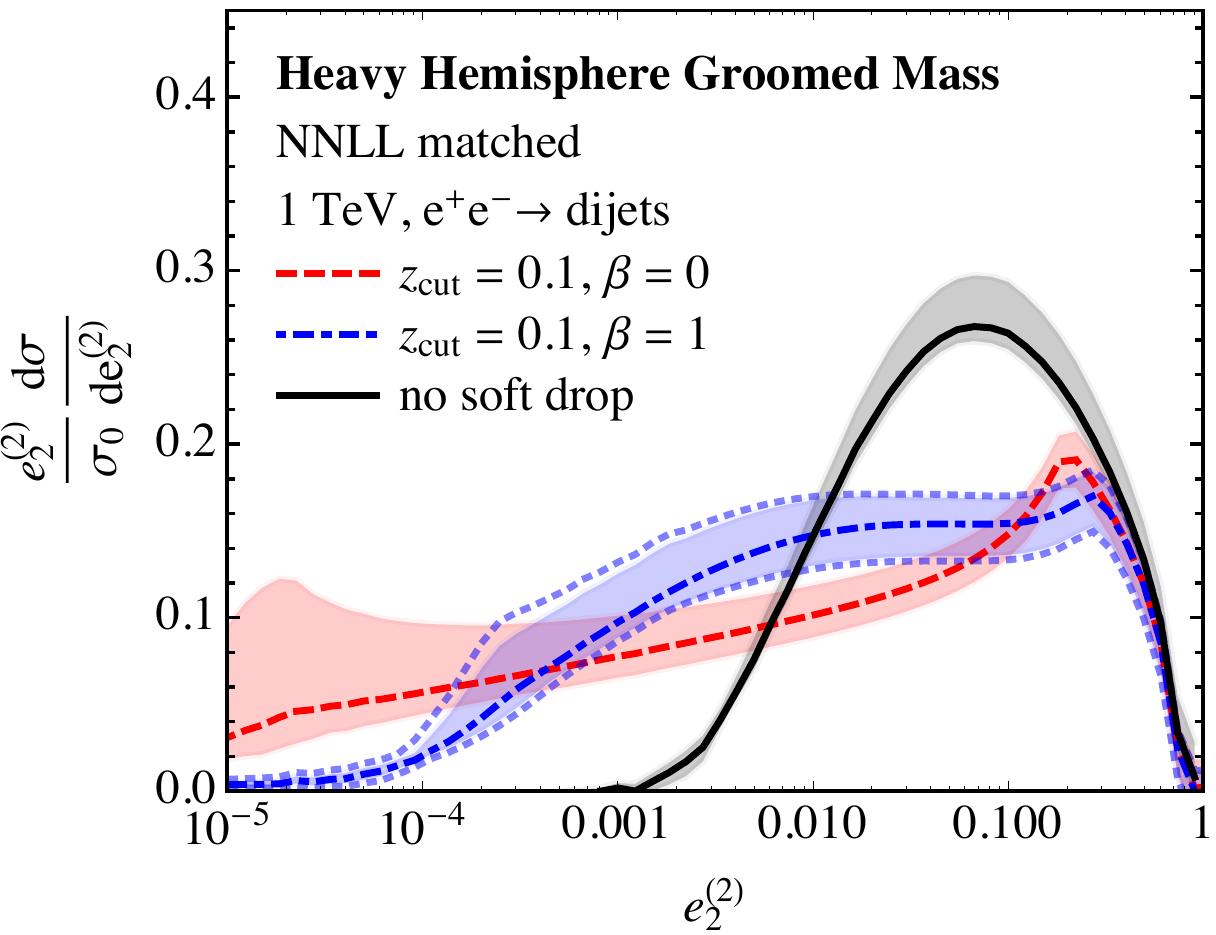}
}
\caption{
(a) NLL matched distributions for heavy hemisphere $\ecf{2}{2}$ in $e^+e^-$ collisions with soft drop grooming $\zcut = 0.1$ and $\beta = 0$, $\beta=1$, and without soft drop.  
Estimates of theoretical uncertainties are represented by the shaded bands. (b) The corresponding matched distributions at NNLL.  
For soft drop with $\beta = 1$, the dotted lines represent the extent of the theoretical uncertainties when the variation of the two-loop non-cusp anomalous dimension is included. 
Note the significant reduction in uncertainties at NNLL.
}
\label{fig:eematch}
\end{figure}

In \Fig{fig:eematch}, we plot the resummed and matched differential cross sections for the larger $\ecf{2}{2}$ of the two hemispheres at NLL and NNLL with various levels of soft drop grooming.  
Here, we consider dijet production in $e^+e^-$ collisions at 1 TeV center-of-mass energy and identify hemispheres with the exclusive $k_T$ algorithm \cite{Catani:1991hj}.  The parameters of soft drop are $\zcut = 0.1$ and we show both $\beta = 0$ and $\beta = 1$.  We also show the ungroomed heavy hemisphere $\ecf{2}{2}$ distribution.  In these plots, we include estimates of theoretical uncertainties represented by the lighter bands about the central curve.  While more sophisticated methods for estimating uncertainties exist, we simply vary the natural scales that appear in the functions of the factorization theorem up and down by a factor of two.  We then take the envelope of these scale variations as an estimate of theoretical uncertainties.  
This simple prescription is sufficient for our main purpose in showing uncertainty bands: 
to demonstrate the reduction in theoretical uncertainty in moving from NLL to NNLL.

Included in these uncertainty estimates is a variation in our treatment of the 
Landau pole of the strong coupling $\alpha_s$.
For scales $\mu > 1$ GeV, $\alpha_s$ is evaluated according to its perturbative running.  For $\mu < 1$ GeV, we freeze $\alpha_s$ to its value at the scale $\mu = 1$ GeV.  This is not intended to be a model for hadronization or non-perturbative physics, but is just intended to maintain finite cross section predictions at small $\ecf{2}{2}$ values.  To estimate the sensitivity of our results to the scale at which we freeze the coupling, we vary this 1 GeV scale by a factor of two, and include the effect in the uncertainty bands of \Fig{fig:eematch} as well.

Finally, we have shown the uncertainty bands around the $\beta=1$ curves at NNLL with and without 
the uncertainty in our estimate of the two-loop non-cusp anomalous dimension of the soft function. One can 
see from the figure that this imperfect extraction has only a relatively small effect on the overall uncertainty
at this order. 

Importantly, we allow the normalization of the cross section to change under these scale variations.  
That is, the curves in \Fig{fig:eematch} are constructed according to \Eq{eq:addmatch}. The normalization 
of each distribution displayed is meaningful, since we resum all large logs in both the shape and the normalization. 
While the central value curves don't change much in going from NLL to NNLL, the uncertainties are dramatically reduced, and this is 
partly due to the increased accuracy in the normalization.

\subsection{Comparison to Monte Carlo}

\begin{figure}[t]
\centering
\subfloat[]{\label{fig:her_sd0}
\includegraphics[width=0.45\linewidth]{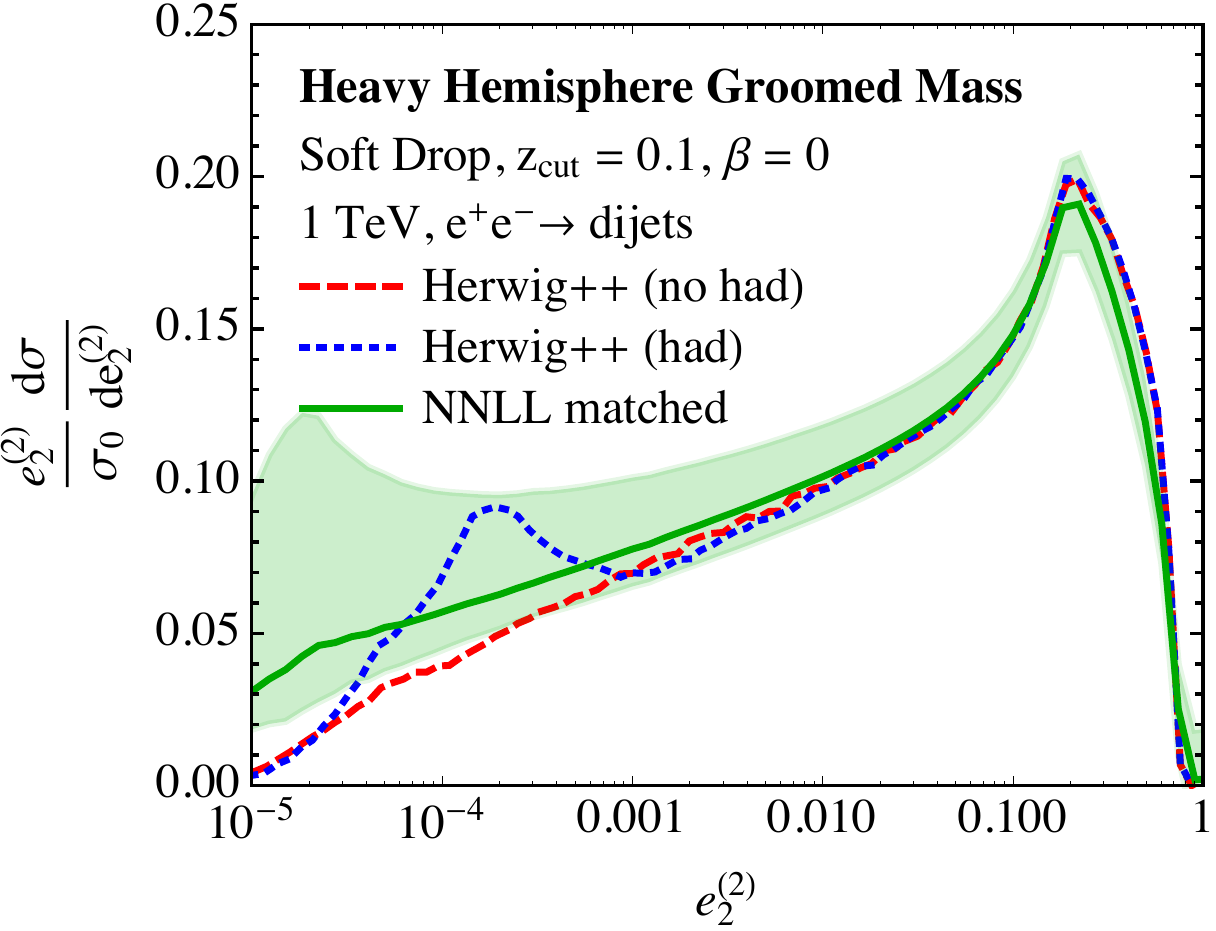}
}\quad
\subfloat[]{\label{fig:py_sd0}
\includegraphics[width=0.45\linewidth]{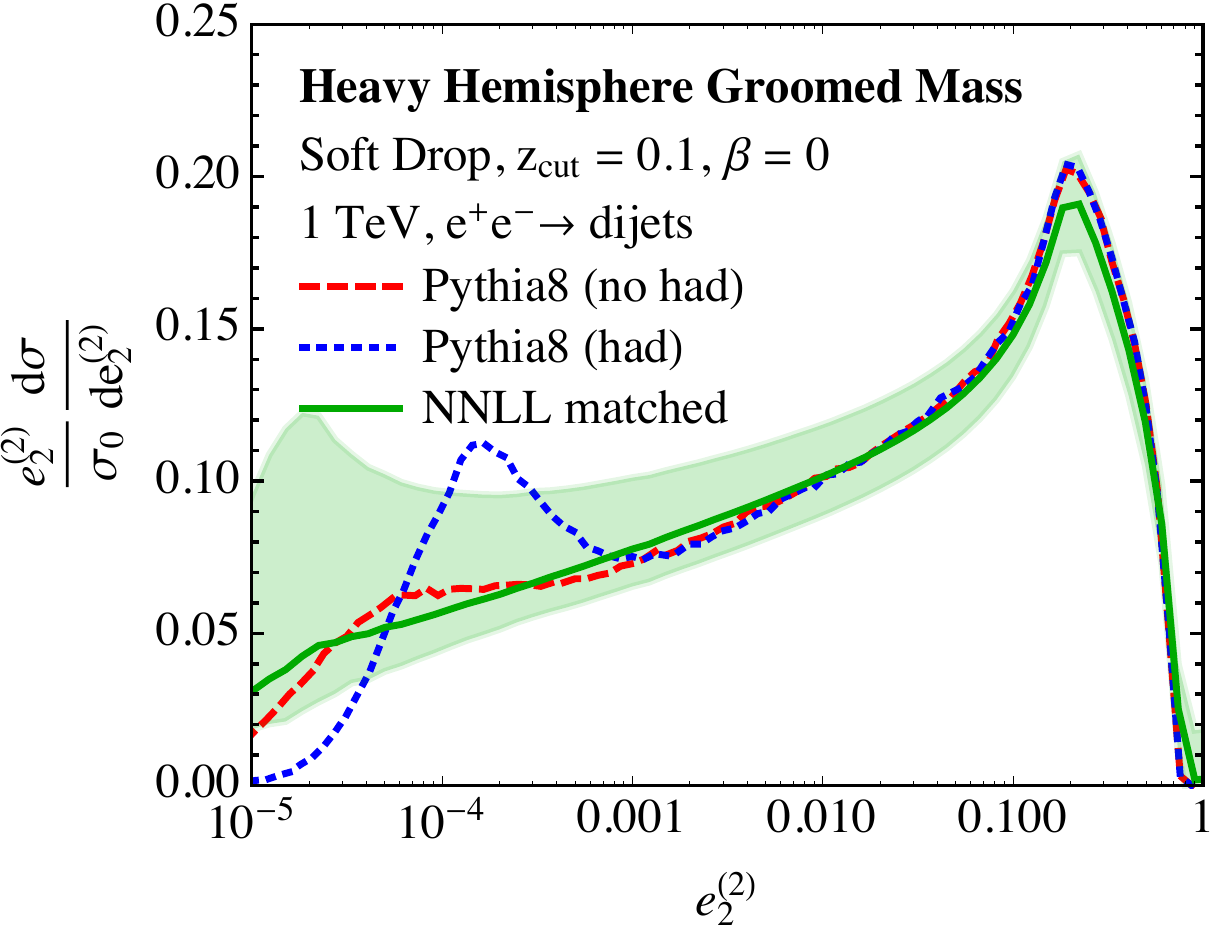}
}\\
\subfloat[]{\label{fig:vlo_sd0}
\includegraphics[width=0.45\linewidth]{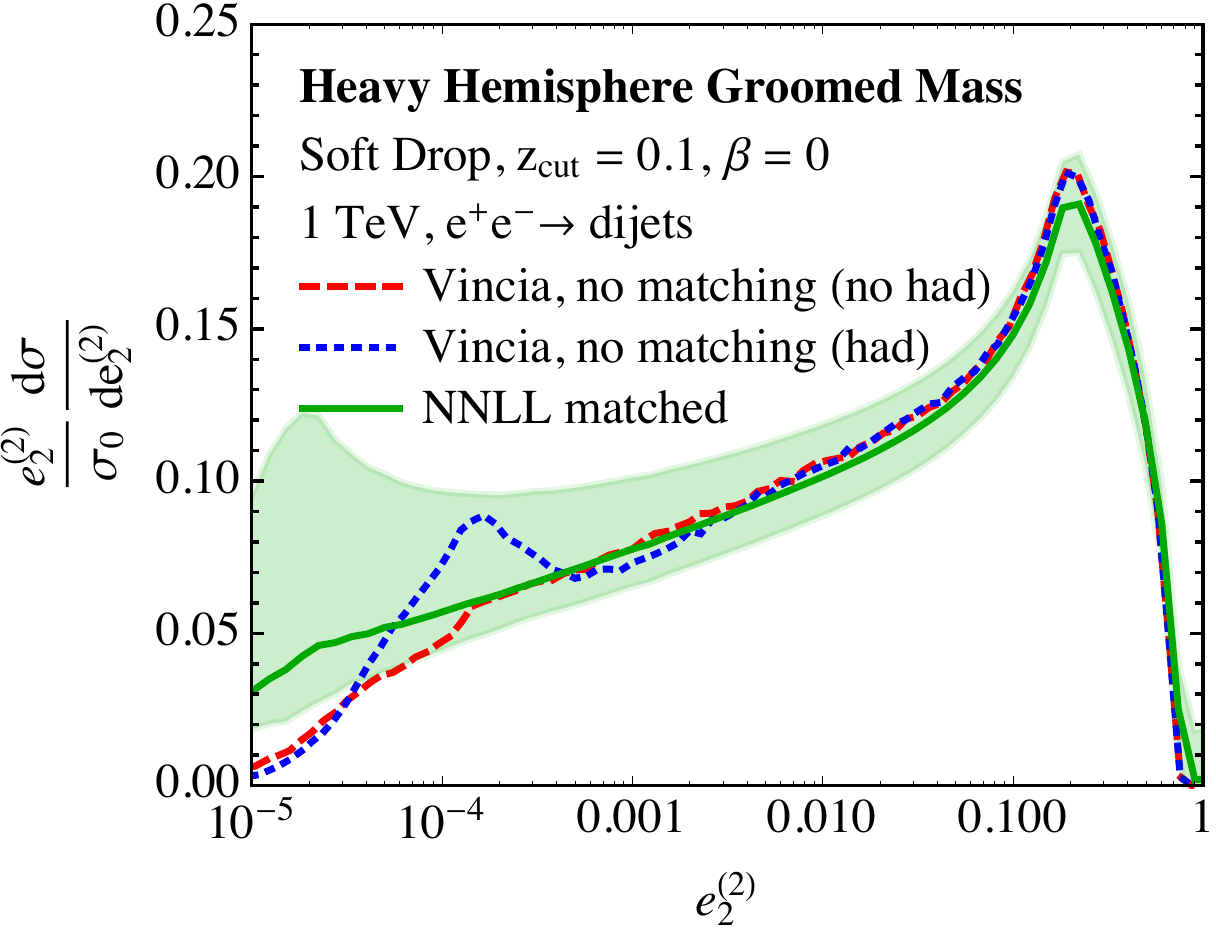}
}\quad
\subfloat[]{\label{fig:vnlo_sd0}
\includegraphics[width=0.45\linewidth]{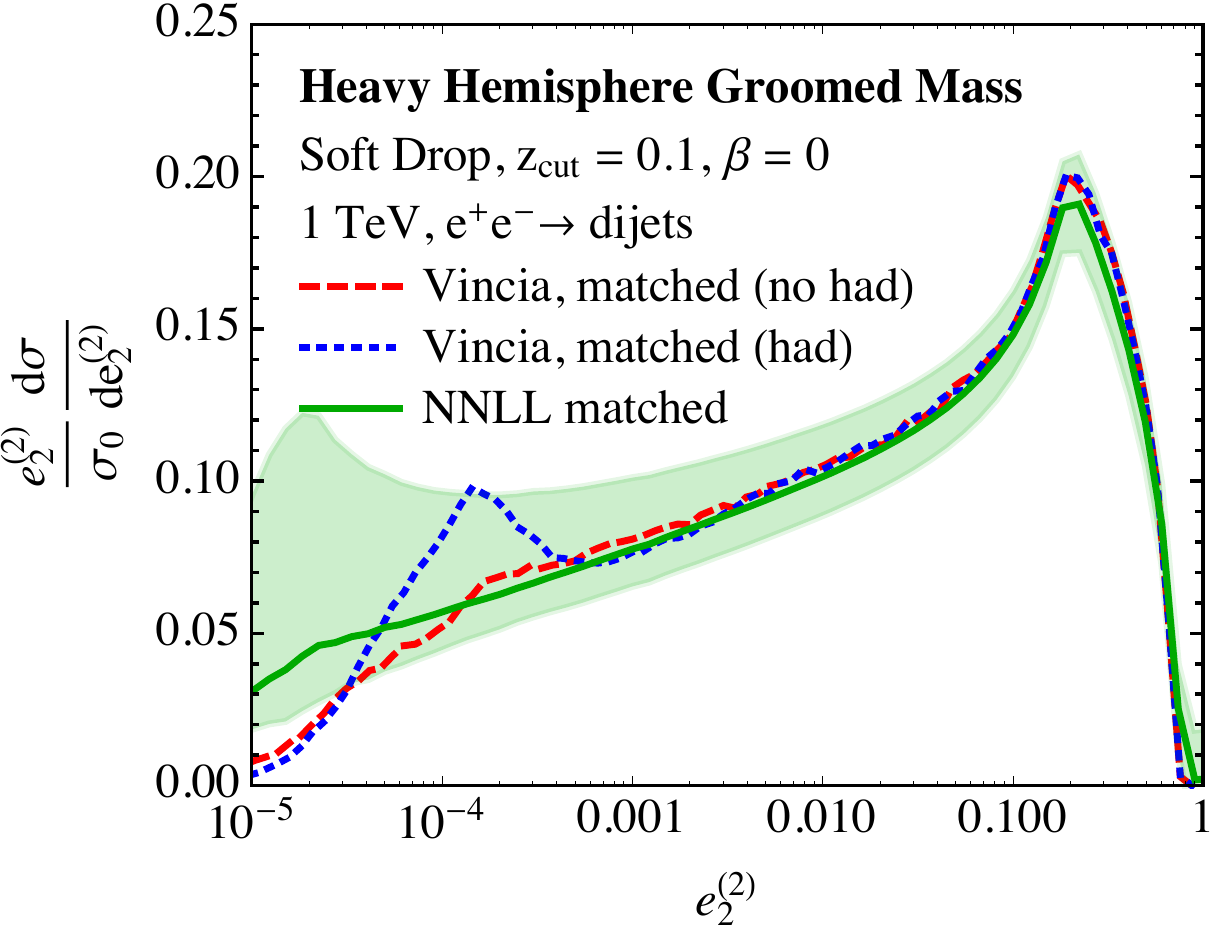}
}
\caption{
Comparison between soft-drop groomed $\ecf{2}{2}$ distributions with $\zcut = 0.1$ and $\beta = 0$ for NNLL, parton-level, and hadron-level Monte Carlo.  
All curves integrate to the same value over the range $\ecf{2}{2}\in[0.01,1]$.
}
\label{fig:ee_sd0}
\end{figure}

In this section, we compare our NNLL resummed and matched soft-drop groomed $\ecf{2}{2}$ distributions to the output of several standard Monte Carlo simulations.  We generate $e^+e^-\to $ dijets events at 1 TeV center-of-mass collision energy with \herwigpp{2.7.1} \cite{Bahr:2008pv,Bellm:2013hwb}, \pythia{8.210}  \cite{Sjostrand:2006za,Sjostrand:2014zea}, and \vincia{1.2.02} \cite{Giele:2007di,Giele:2011cb,Hartgring:2013jma,Larkoski:2013yi}.  While the \herwigpp{} and \pythia{} events are showered from the leading order process $e^+e^- \to q\bar q$, we consider \vincia{} with and without fixed-order matching included.  
The matched \vincia{} results are accurate effectively through ${\mathcal O}(\alpha_s^2)$.  
For the most direct comparison of the simulations to our NNLL matched results, we run $\alpha_s$ at two loops in all Monte Carlos (in the CMW scheme \cite{Catani:1990rr,Dokshitzer:1995ev}) and we fix $\alpha_s(m_Z)=0.118$, which is the same value used in our analytic 
calculations.  We include Monte Carlo events both at partonic level and after hadronization.  These events are then clustered into hemispheres using the exclusive $k_T$ algorithm \cite{Catani:1991hj} using \fastjet{3.1.3} \cite{Cacciari:2011ma}.  The soft drop grooming and subsequent measurement of the two-point energy correlation functions of these $e^+e^-$ events is implemented in \fastjet{} with custom code.

\begin{figure}[t]
\centering
\subfloat[]{\label{fig:her_sd1}
\includegraphics[width=0.45\linewidth]{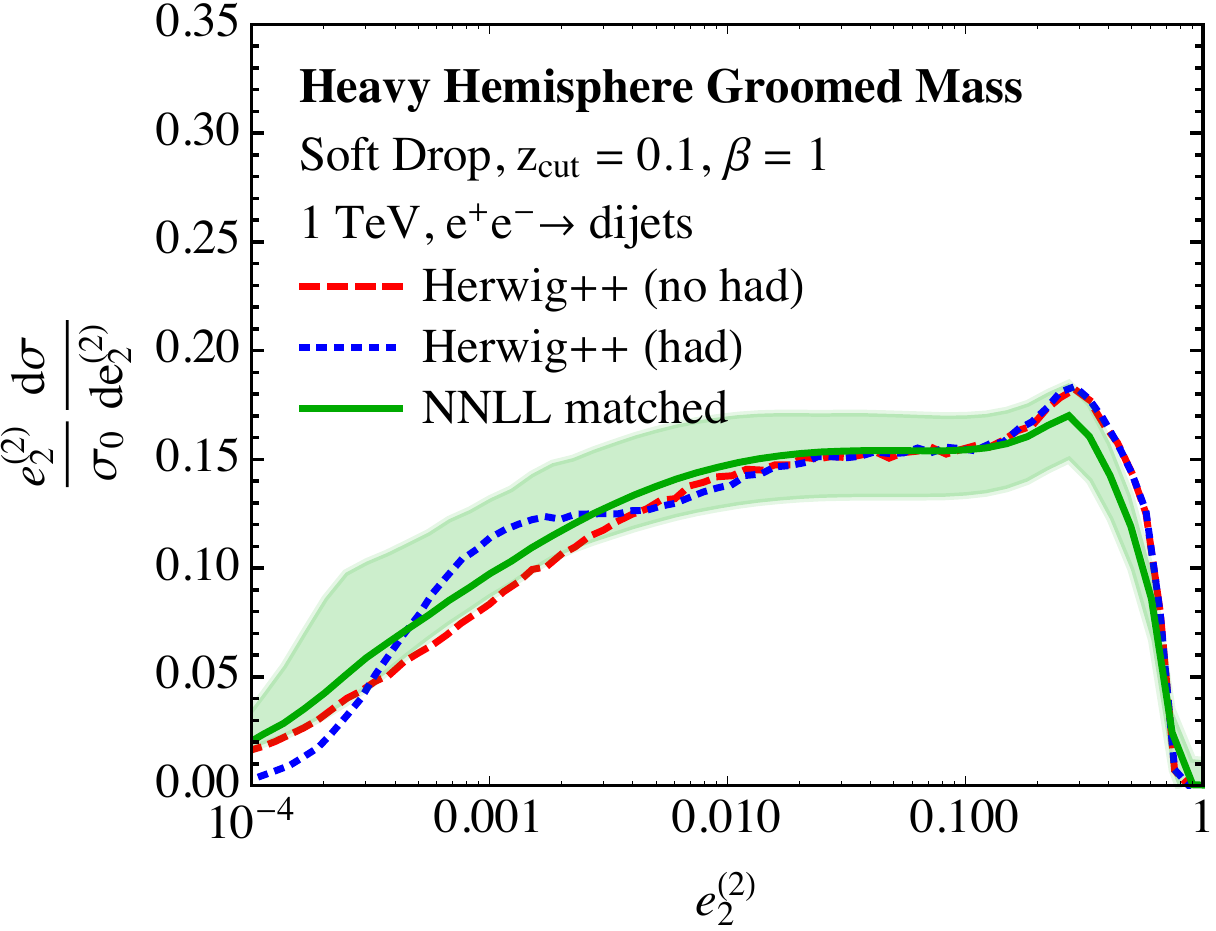}
}\quad
\subfloat[]{\label{fig:py_sd1}
\includegraphics[width=0.45\linewidth]{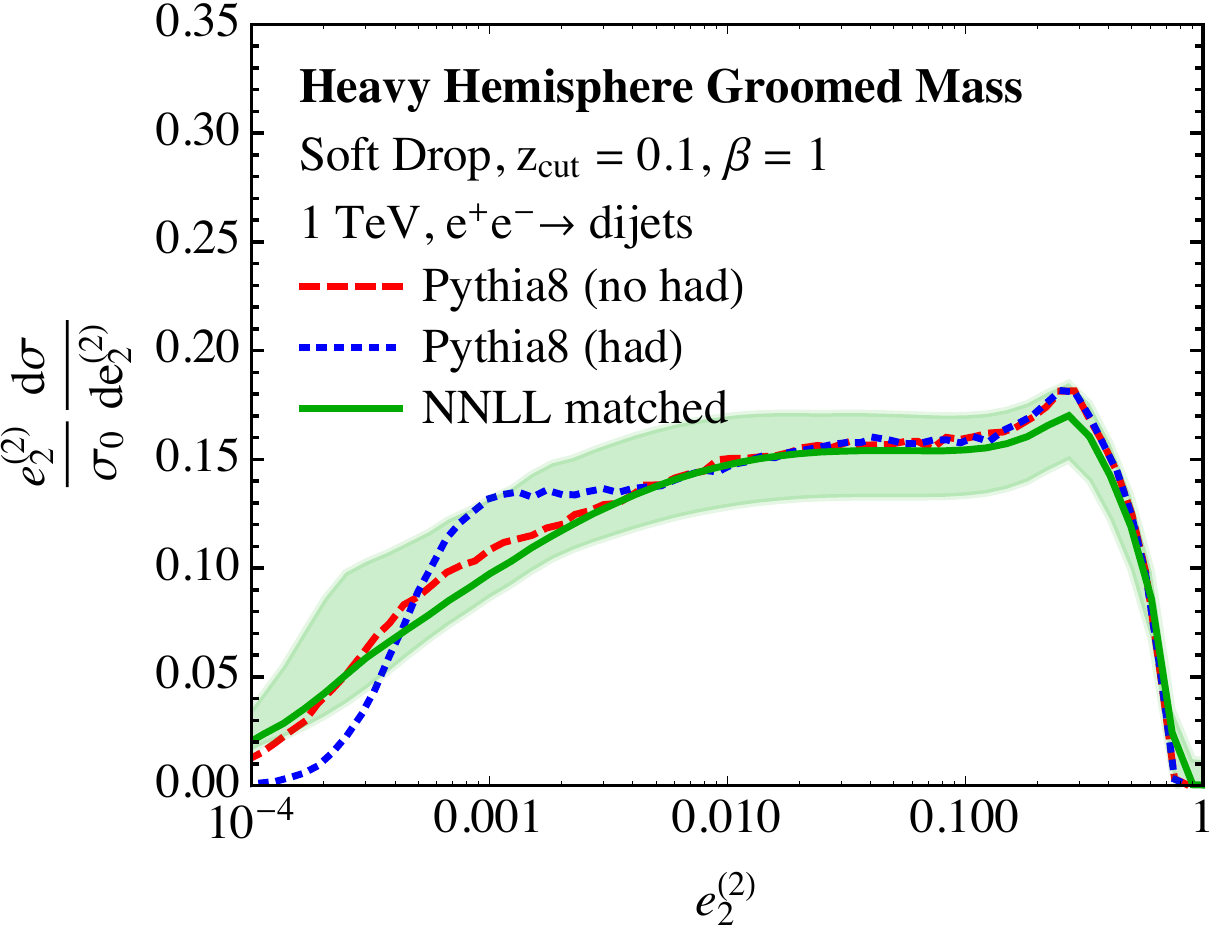}
}\\
\subfloat[]{\label{fig:vlo_sd1}
\includegraphics[width=0.45\linewidth]{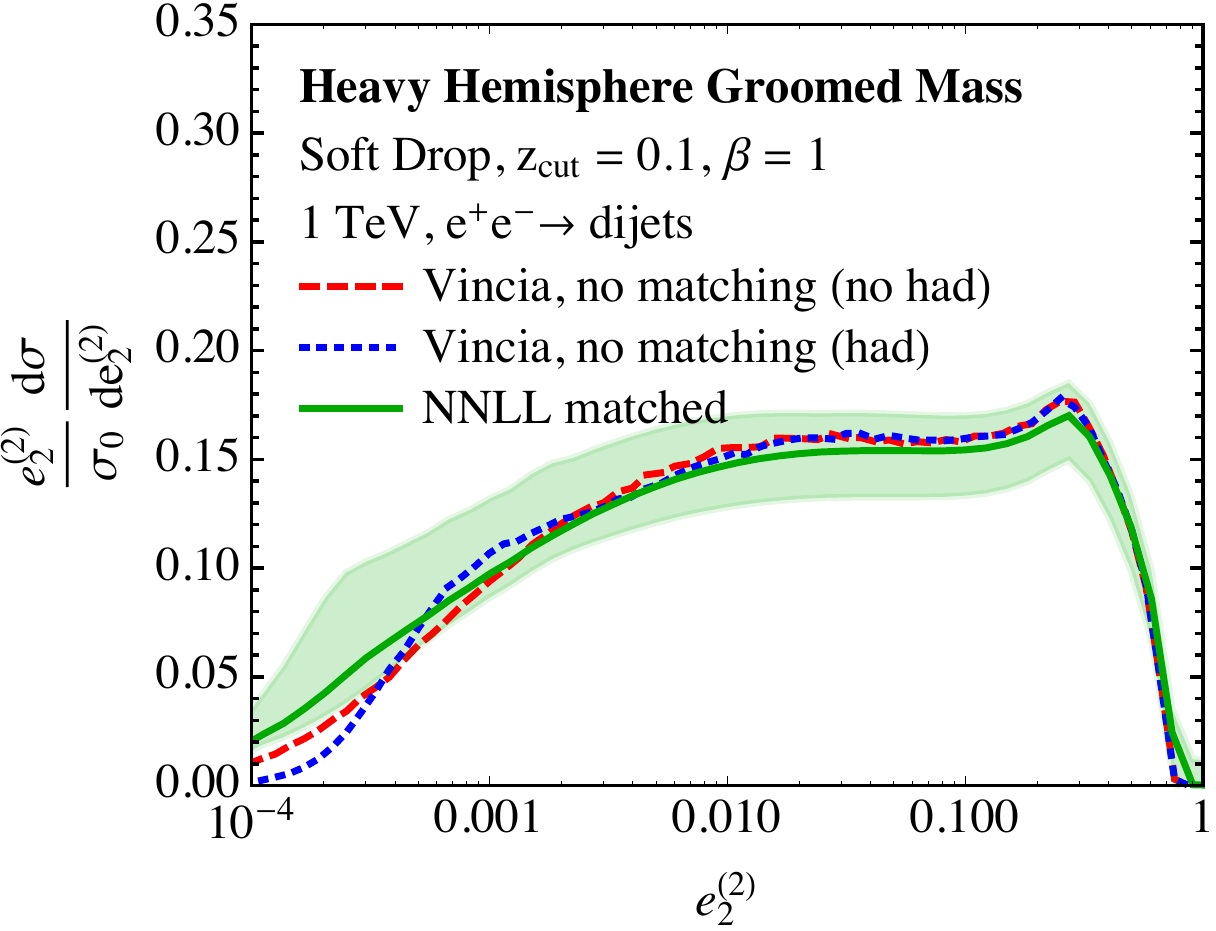}
}\quad
\subfloat[]{\label{fig:vnlo_sd1}
\includegraphics[width=0.45\linewidth]{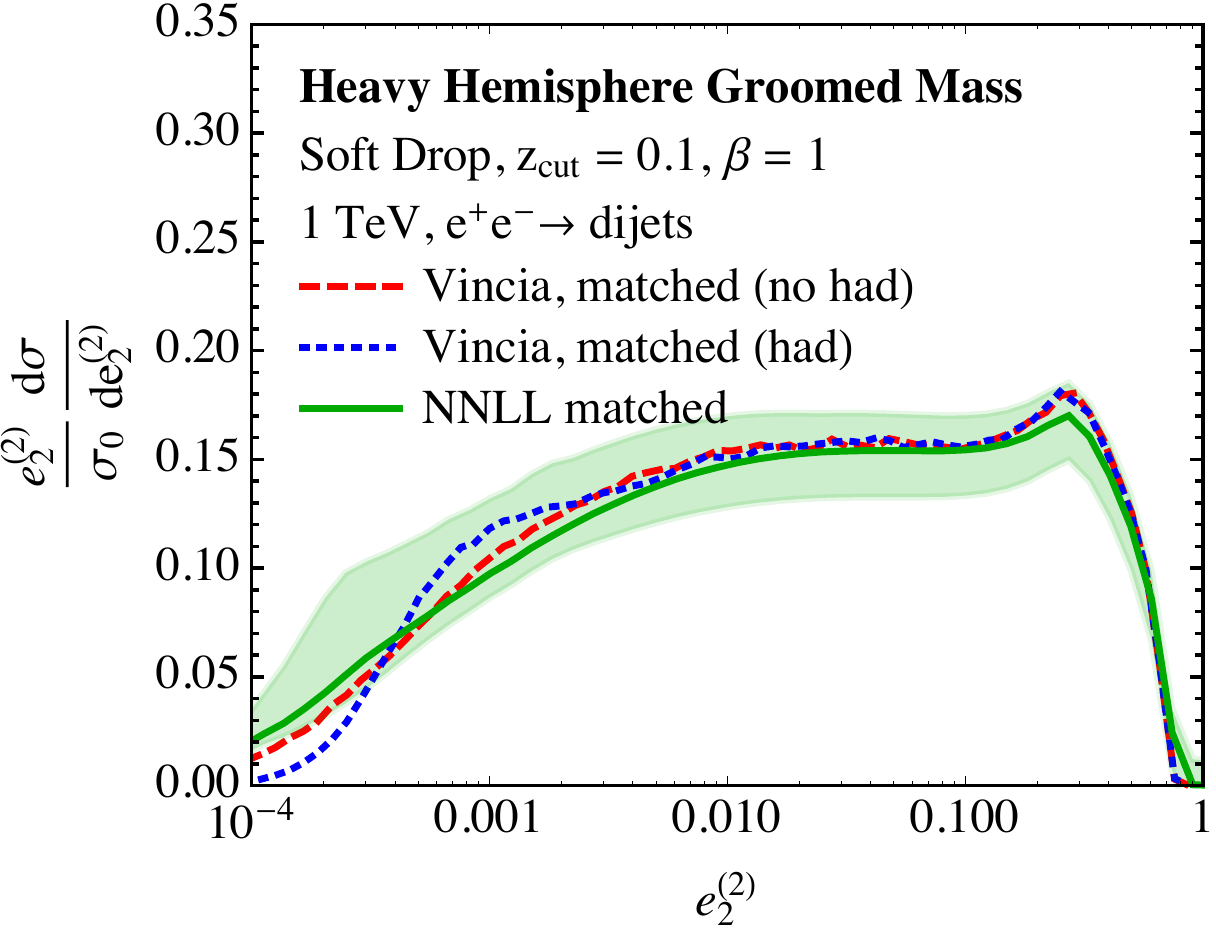}
}
\caption{
Comparison between soft-drop groomed $\ecf{2}{2}$ distributions with $\zcut = 0.1$ and $\beta = 1$ for NNLL, parton-level, and hadron-level Monte Carlo.  All curves integrate to the same value over the range $\ecf{2}{2}\in[0.01,1]$.  The uncertainty band for NNLL includes the variation of the two-loop non-cusp anomalous dimension.
}
\label{fig:ee_sd1}
\end{figure}

We compare the Monte Carlo distributions to our NNLL resummed and matched calculations in 
\Figs{fig:ee_sd0}{fig:ee_sd1}.  
In these plots, all distributions integrate to the same value over the range $\ecf{2}{2}\in [0.01,1]$.  In \Fig{fig:ee_sd0}, we compare the  
soft-drop groomed $\ecf{2}{2}$ distributions with $\zcut = 0.1$, $\beta = 0$ .  Good agreement between the Monte Carlos and our analytic calculation is observed, with (not surprisingly) the matched Monte Carlo agreeing the best.  
These distributions also show that parton- and hadron-level Monte Carlos are essentially identical for $\ecf{2}{2}\gtrsim 0.001$.  
In \Fig{fig:ee_sd1}, we compare the  
soft-drop groomed $\ecf{2}{2}$ distributions with $\zcut = 0.1$, $\beta = 1$.  Again, good agreement between the Monte Carlos and our matched NNLL result is observed, 
with the parton- and hadron-level Monte Carlos nearly identical for $\ecf{2}{2}\gtrsim 0.005$.  
The uncertainty bands for the analytic curve includes the uncertainty in the two-loop non-
cusp anomalous dimension.  

\begin{figure}[t]
\centering
\subfloat[]{\label{fig:ee_comp_sd0}
\includegraphics[width=0.45\linewidth]{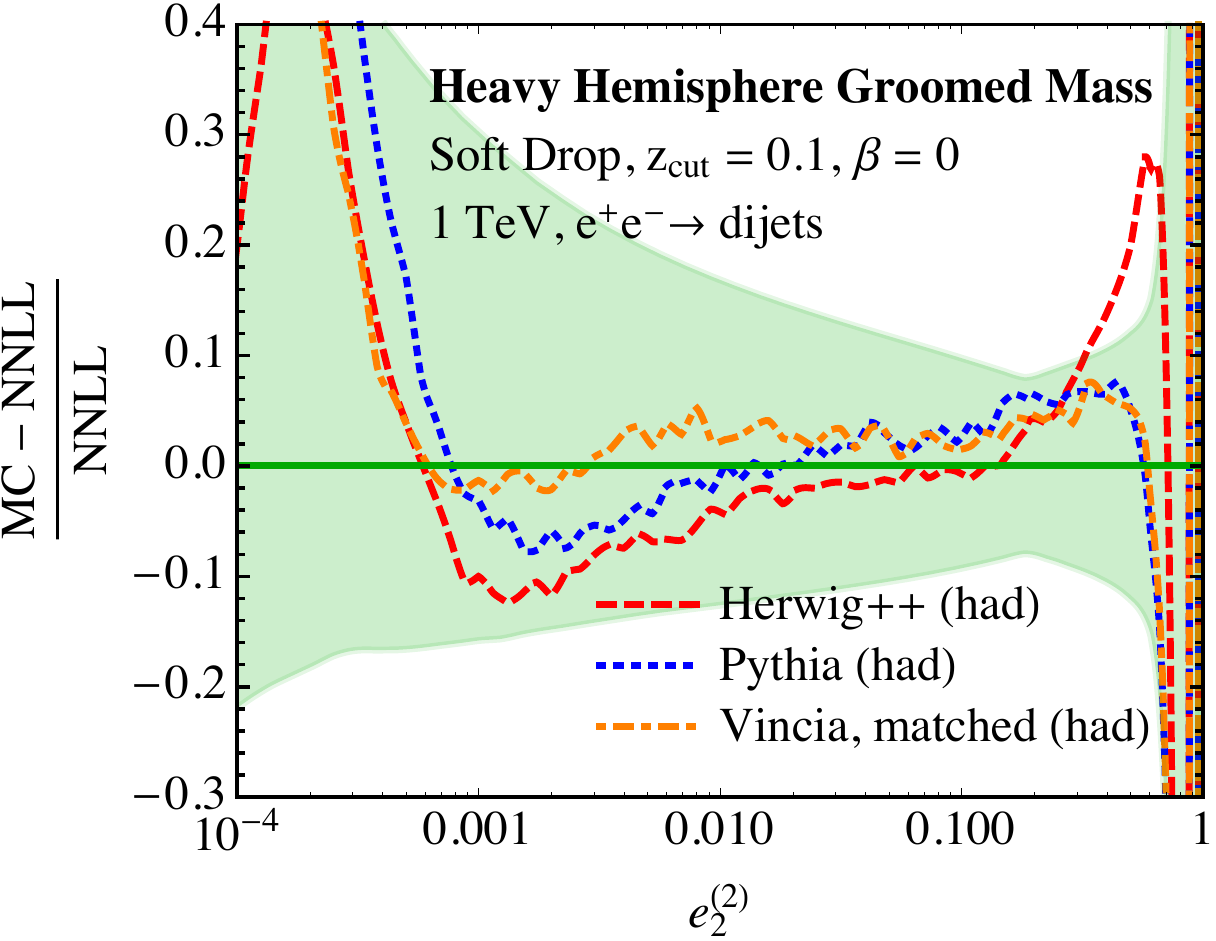}
}\quad
\subfloat[]{\label{fig:ee_comp_sd1}
\includegraphics[width=0.45\linewidth]{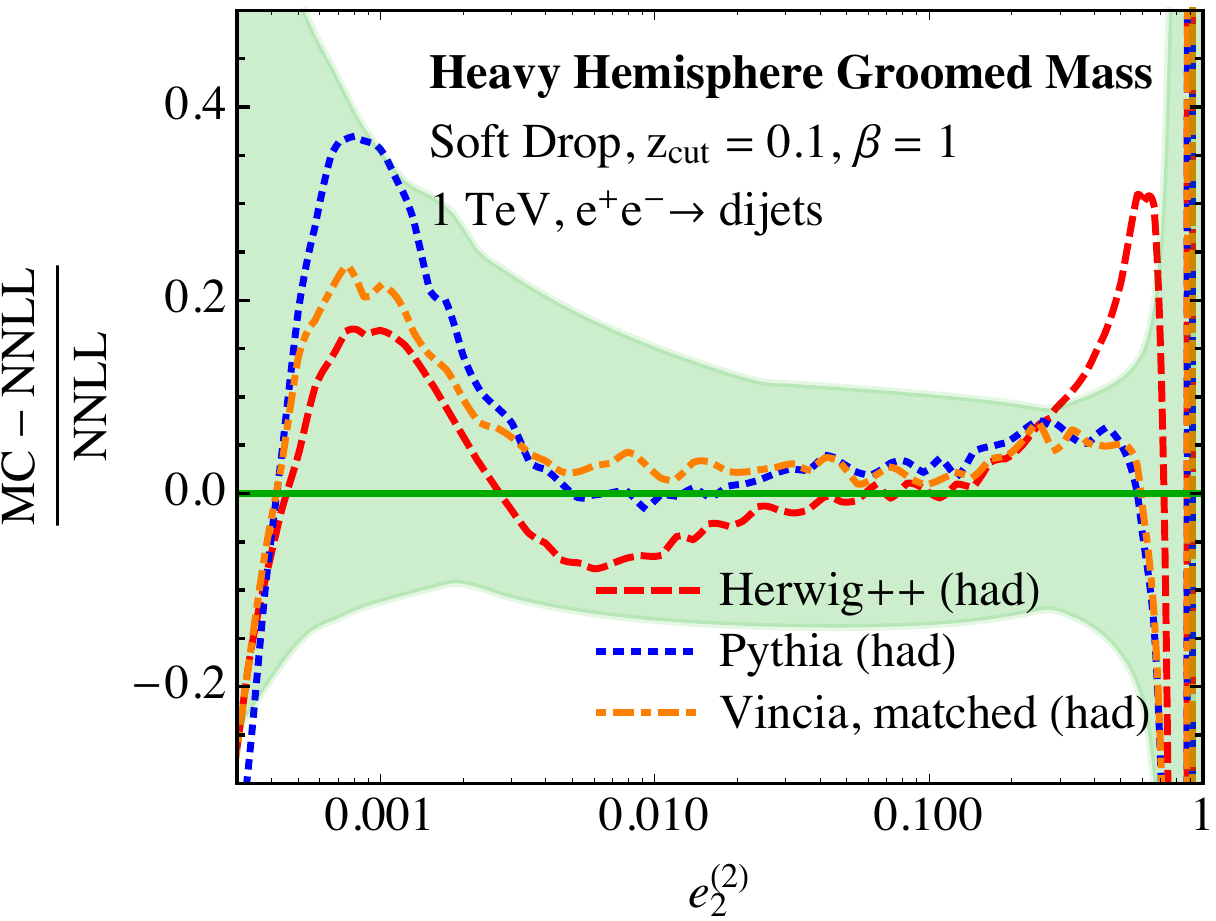}
}
\caption{
Direct comparison of hadron-level output from \herwigpp{}, \pythia{}, and \vincia{} already shown in \Figs{fig:ee_sd0}{fig:ee_sd1}. 
Soft drop is performed with $\zcut=0.1$ and both $\beta=0$ (left) and $\beta=1$ (right).
Curves are displayed as relative differences between Monte Carlo output and our matched NNLL predictions,
with theoretical uncertainties shown as a shaded band.
}
\label{fig:ee_comp}
\end{figure}

As a more direct comparison of the Monte Carlos, 
\Fig{fig:ee_comp} displays the relative difference between each of the hadron-level Monte Carlos and our matched NNLL predictions.
Again, soft drop is performed with $\zcut=0.1$, and both $\beta=0$ and $\beta=1$ are shown. 
All the Monte Carlo curves lie within our shaded band of theoretical uncertainty, but discrepancies between the different simulations are visible.

One striking feature in these plots, especially for $\beta=0$, is the presence of additional structure in the hadron-level Monte Carlo distributions at small $\ecf{2}{2}$.  It is clear that this feature is due to non-perturbative physics, and so is therefore not included in our NNLL calculation.  Nevertheless, we can include a simple model of hadronization into our calculation to see if this structure is easily explained.

For additive IRC safe observables, like thrust or jet mass, it can be shown from general principles that hadronization corrections can be incorporated in perturbative distributions by convolution with a model shape function \cite{Korchemsky:1999kt,Korchemsky:2000kp}.  In general, the energy correlation functions are additive observables, so we should be able to use shape functions to model hadronization corrections.  However, once soft drop is applied on the jet, emissions in the jet may or may not contribute to the energy correlation functions, so the observable is no longer strictly additive.  
We leave a more careful study of whether shape functions can be used to model hadronization effects in groomed 
observables to future work. Here we convolve our matched results with a simple shape function to see if qualitative 
agreement with the Monte Carlos can be achieved.

\begin{figure}[t]
\centering
\subfloat[]{\label{fig:had_sd0}
\includegraphics[width=0.45\linewidth]{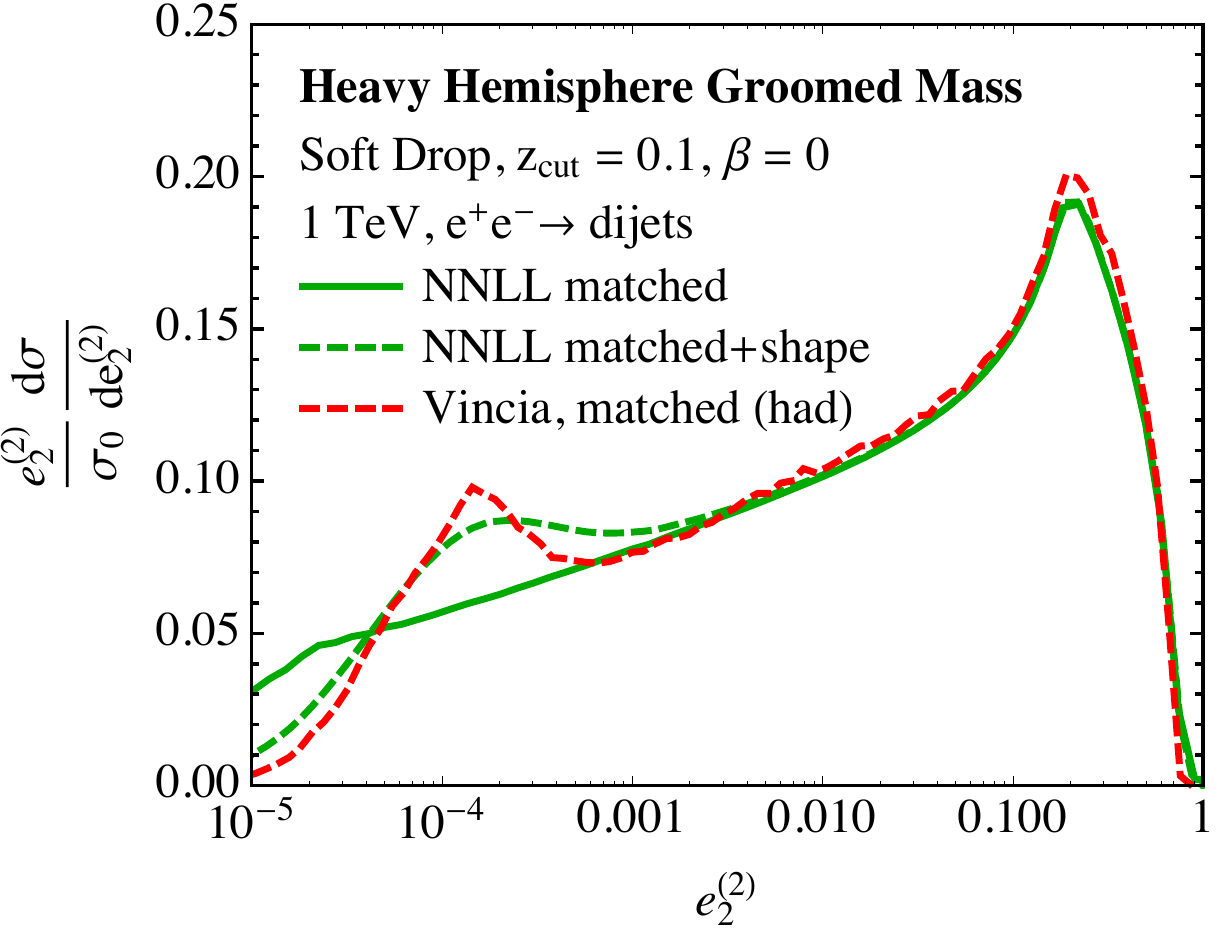}
}\quad
\subfloat[]{\label{fig:had_sd1}
\includegraphics[width=0.45\linewidth]{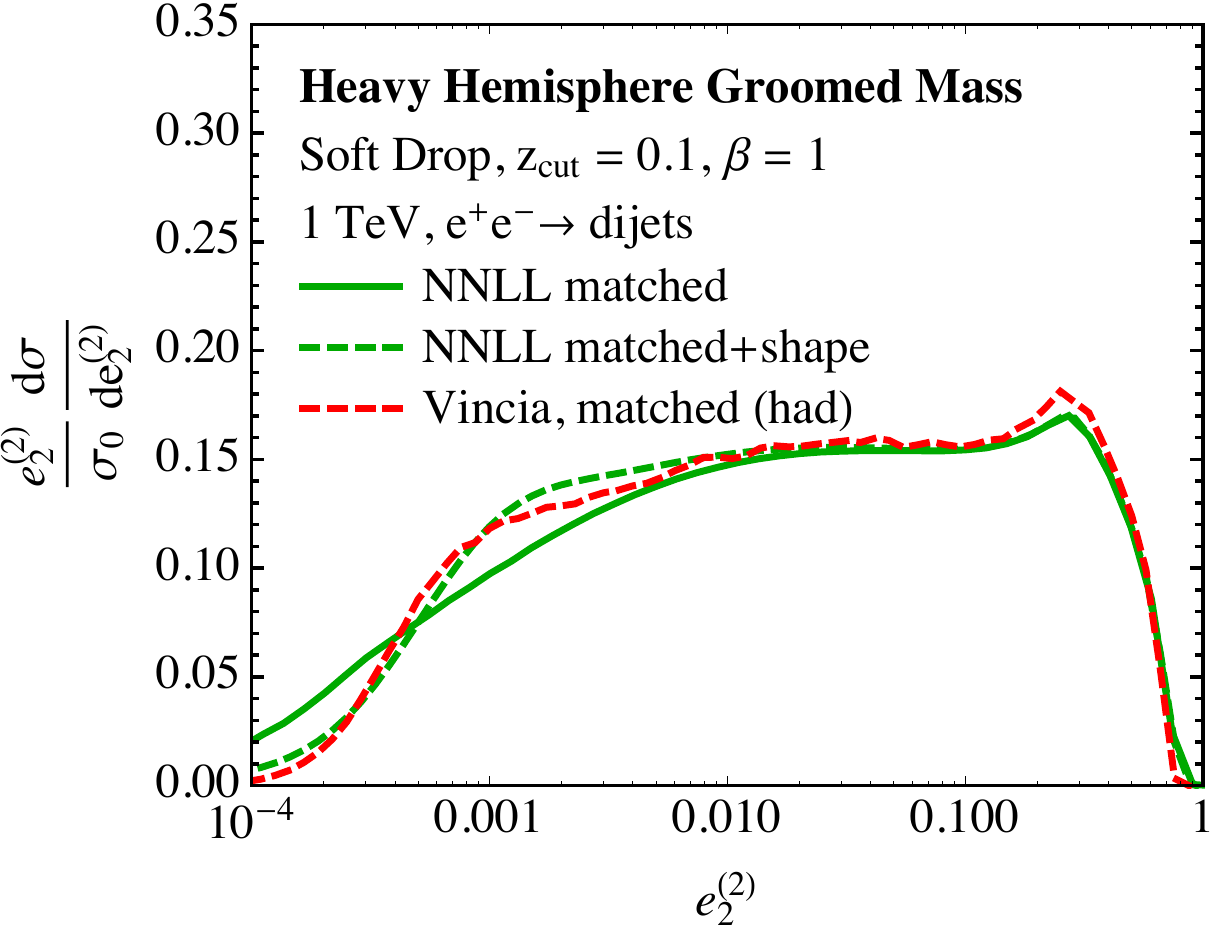}
}
\caption{
Perturbative NNLL results for soft-drop groomed $\ecf{2}{2}$ with $\zcut = 0.1$ and $\beta = 0$ (left) and $\beta  = 1$ (right), 
compared to analytic results that include the shape function of \Eq{eq:shape_func} for modeling hadronization,
and compared to hadron-level Monte Carlo.  
The parameter $\Omega = 1$ GeV.
Note that, qualitatively, the shape function produces a hadronization-bump similar to those seen in the Monte Carlos.
}
\label{fig:ee_had}
\end{figure}

Because shape functions describe non-perturbative physics, they only have support for energies comparable to $\Lambda_\text{QCD}$.  The shape function we choose is the parametrization suggested by \Ref{Stewart:2014nna}:
\begin{align}\label{eq:shape_func}
F_\text{shape}(\epsilon)=\frac{4\epsilon}{\Omega^2}e^{-2\epsilon/\Omega}\,.
\end{align}
This is normalized
\begin{equation}
\int_0^\infty d\epsilon\, F_\text{shape}(\epsilon) = 1\,,
\end{equation}
and has first moment equal to $\Omega$.  As discussed in \Sec{sec:hadest}, of all modes present in our factorization
theorem, the collinear-soft mode has the lowest virtuality, so it will have the largest sensitivity to non-perturbative
physics.  We thus convolve our perturbative distribution with the shape function, assuming non-perturbative effects
are primarily associated with the collinear-soft mode.  That is, we include hadronization corrections in the soft
drop groomed $\ecf{2}{\alpha}$ distribution according to
\begin{align}
\frac{d\sigma_\text{had}}{d\ecf{2}{\alpha}}=\int d\epsilon\, \frac{d\sigma_\text{pert}\left(\ecf{2}{\alpha}-\left(
\frac{\epsilon}{\zcut Q}
\right)^{\frac{\alpha-1}{1+\beta}} \frac{\epsilon}{Q}\right)}{d\ecf{2}{\alpha}}\, F_\text{shape}(\epsilon)\,,
\end{align}
where the argument of the perturbative distribution is shifted by the virtuality of the collinear-soft mode, \Eq{eq:npest}.

In \Fig{fig:ee_had}, we compare the matched NNLL distribution of $\ecf{2}{2}$ with and without convolution with the shape function of \Eq{eq:shape_func}, in which we set $\Omega = 1$ GeV to be comparable to the scale of hadron masses.  We show this comparison for soft drop grooming with $\beta=0$ and $\beta=1$. The peak at small $\ecf{2}{2}$ for $\beta = 0$ agrees qualitatively with the structure of the hadronized Monte Carlo distributions.  Similarly, the shape at small $\ecf{2}{2}$ for $\beta = 1$ agrees with the simulations as well.  This suggests that there might exist a shape function for describing hadronization effects in groomed jet observables, though we leave a detailed discussion and justification for such a model to future work.

\section{Matching NNLL to Fixed Order in $pp\to Z+j$}\label{eq:ppmatch}

In this section, we present predictions for soft-drop groomed $\ecf{2}{\alpha}$ distributions as measured on the 
jet in $pp\to Z+j$ events at the LHC.  The definitions of soft drop and energy correlation functions appropriate 
for jets in $pp$ collisions are given in \Sec{sec:obs}.  
As with jets from $e^+e^-$ collisions, we match our NNLL 
resummed distribution to fixed-order results that include relative ${\mathcal O}(\alpha_s^2)$ corrections to the Born process. 

There are two complications we must deal with. 
First, at $pp$ collisions, the jets will be both quark and gluon initiated. 
Second, because we only measure the observable within the jet
and do not constrain radiation throughout the rest of the event,
the simple hard-soft-jet factorization that we employed for $e^+e^- \to$ hemisphere jets will not apply here.
Nevertheless, while the normalization of the jet-observable distribution will thus be complicated and sensitive to multiple scales,
the shape of the distribution will still be controlled exclusively by collinear physics.
To address these complications, we first show how soft-drop groomed quark and gluon jets can be unambiguously defined order-by-order in perturbation theory. 
Then we discuss how the normalization of the distribution can be obtained by matching to full QCD at fixed order.
The discussion will focus on the $Z+j$ sample for concreteness, but these ideas apply equally well to any process with hard jets at a hadron collider.

\subsection{Resummed Cross Section in $pp\to Z+j$}

We define our observable on soft-drop groomed jets in $pp\to Z+j$ events in the following way.  
First, we cluster the final state according to a jet algorithm with some jet radius $R\sim 1$.  
Of the jets with pseudorapidity $|\eta_J|<\eta_{\max}$, we then identify the jet with the largest 
transverse momentum $p_{TJ}$ and require that $p_{TJ}>p_T^{\min}$.   
We groom this jet with soft drop and measure $\ecf{2}{\alpha}$ according to the definitions given in \Sec{sec:obs}
for jets in $pp$ collisions.  In this procedure, we remain inclusive over all other hadronic activity in the final state: 
we only care about the hardest jet.  

For this process, the relevant factorization formula is 
\begin{equation}\label{eq:factxsecpp}
\hspace{-0.25cm}\frac{d \sigma_\text{resum}}{ d \ecf{2}{\alpha}}=  \sum_{k = q,\bar q,g} \ppmatch_k(p_{T}^{\min},\eta_{\max},\zcut,R) S_{C,k}(\zcut\ecf{2}{\alpha}) \otimes J_k(\ecf{2}{\alpha})\,.
\end{equation}
Unlike our factorization theorem in $e^+e^-$ collisions, \Eq{eq:factxsecpp} only resums large logarithms of $\ecf{2}{\alpha}$ in the limit $\ecf{2}{\alpha} \ll \zcut \ll 1$.  There will be logarithms of $\zcut$ (and other scales in the events) in the $\ppmatch_k$ prefactor that we do not resum.  When referring to calculations of this cross section, we will specify the accuracy to which logarithms of $\ecf{2}{\alpha}$ are resummed (i.e., NLL or NNLL).
%
 We now explain the components of this formula in detail.

In \Eq{eq:factxsecpp}, $S_{C,k}(\zcut\ecf{2}{\alpha})$ and $J_k(\ecf{2}{\alpha})$ are the collinear-soft and 
jet functions for the measurement of soft-drop groomed $\ecf{2}{\alpha}$ that, by collinear factorization, 
are identical to the functions defined in $e^+e^-$ collisions.  Unlike in $e^+e^-$ collisions, however, 
these functions also have a label $k$ corresponding to the flavor of the jet, 
and a sum over the possible QCD parton flavors $k$ is included.  
The symbol $\otimes$ denotes convolution in $\ecf{2}{\alpha}$ between the collinear-soft and jet functions.

  $\ppmatch_k$ is a matching coefficient that 
can be extracted from fixed-order calculations, and it sets the normalization and relative contributions from the 
different jet flavors.  In addition to the dependence explicitly shown, $\ppmatch_k$ also depends implicitly 
on parton distributions, as different initial states produce different flavors of final state jets.

Unlike the case in $e^+e^-$ collisions, where the 
jet energy was (almost exactly) half the center-of-mass energy, due to the non-trivial parton distributions, 
the distribution of the jet $p_T$ has a finite width and depends on the cut, $p_T^{\min}$.
For a true precision prediction, we would compute the matching coefficient $D_k$ as a function of $p_{TJ}$ 
and include an integral in \Eq{eq:factxsecpp} convolving the jet and collinear-soft functions with $D_k(p_{TJ})$. 
An approach to doing this in a semi-automatic manner was discussed recently in \Refs{Becher:2014aya,Farhi:2015jca}.
But, for simplicity
we instead employ the following approximation:
we evaluate the jet and collinear-soft functions at $\overline{p}_{TJ}$, the average $p_{TJ}$.

The average jet transverse momentum $\overline{p}_{TJ}$ can be estimated by 
using 
the fact that the cross section for a jet with transverse momentum $p_{TJ}$ takes the power-law form:
\begin{equation}
\frac{1}{\sigma}\frac{d\sigma}{dp_{TJ}}\simeq \frac{n-1}{p_T^{\min}}\left(
\frac{p_T^{\min}}{p_{TJ}}
\right)^n \Theta(p_{TJ}-p_T^{\min})\,.
\end{equation}
This distribution is normalized and the mean value of $p_{TJ}$ is
\begin{align}
\overline p_{TJ} &= 
\frac{n-1}{n-2}~p_T^{\min}\,.
\end{align}
The  typical exponent is $n \sim 5$, and we take $n=5$ in our numerical computations.

The full cross section for soft-dropped $\ecf{2}{\alpha}$ (including power corrections) can be expressed as
\begin{equation}\label{eq:scratch}
  \frac{d\sigma}{d\ecf{2}{\alpha}} = \sum_{k=q,\bar q, g} D_k \,  S_{C,k}\otimes J_k 
  + \frac{d\sigma_{\text{pc}}}{d\ecf{2}{\alpha}}\,.
\end{equation}
Here, the right-most term includes all power corrections suppressed by $\ecf{2}{\alpha}$  or $\zcut$. 
The functions $S_{C,k}$ and $J_k$ should be evaluated at $\overline p_{TJ}$ but
we have suppressed their arguments for brevity.  
We will use this form of the cross section to define the matching coefficient $\ppmatch_k$ at fixed-order.
For NNLL resummation, the relative ${\mathcal O}(\alpha_s)$ corrections to $\ppmatch_k$ are required.

First, at leading order in $\alpha_s$, \Eq{eq:scratch} becomes
\begin{equation}
  \frac{d\sigma^{(0)}}{d\ecf{2}{\alpha}} = \sum_{k=q,\bar q, g} D_k^{(0)} \delta(\ecf{2}{\alpha}) 
  \,,
\end{equation}
where the superscript ${(0)}$ denotes the leading order in $\alpha_s$.  Here, we have used $J_k^{(0)} = S_{C,k}^{(0)} = \delta(\ecf{2}{\alpha})$. 
Also, since a jet has only one constituent at this order,
the distribution has no support away from $\ecf{2}{\alpha}=0$ and there are no partons to soft drop;
therefore, there are no $\ecf{2}{\alpha}$ or $\zcut$ power corrections at this order.
Integrating over all $\ecf{2}{\alpha}$, we are left with the Born-level cross section for the $k$ flavor channel 
$\sigma_k^{(0)}$, so that
\begin{equation}
\ppmatch_k^{(0)}=\sigma_k^{(0)}\,.
\end{equation}

At the next-to-leading order in $\alpha_s$, the extraction of $\ppmatch_k$ requires separating the jets by flavor. 
Since $\ppmatch_k$ is defined in each flavor channel, we need to determine the flavor of the hardest jet in each 
$pp \to Z+j$ event included in our sample.
Ordinarily, any definition of jet flavor based on the constituents of the jet is infrared-unsafe and ill-defined at leading power, because soft wide-angle emissions into a jet can 
change its flavor.\footnote{However, one infrared and collinear safe definition of jet flavor was presented in \Ref{Banfi:2006hf}.} Soft drop eliminates this problem at leading power in $\ecf{2}{\alpha}$ and $\zcut$ by removing soft wide-angle radiation from the jet.
This allows for an infrared and collinear safe definition of jet flavor at leading power in $\ecf{2}{\alpha}$ and $\zcut$.
We define the jet flavor $f_J$ as the flavor sum of the constituents of the groomed jet:
\begin{equation}
f_J=\sum_{i\in J_g}f_i\,,
\end{equation}
where $f_q=1$, $f_{\bar q}=-1$ and $f_g=0$.  The subscript on $J_g$ means that one only sums over the jet constituents 
that remain after grooming with soft drop. If $f_J=\pm 1$, then the jet is quark-type, while if $f_J=0$, it is gluon-type.  
With this jet flavor identification, we are able to determine the total fixed-order cross section for each jet flavor channel 
in $pp\to Z+j$.  We will denote the next-to-leading order term in the cross section for a jet of flavor $k$ as $\sigma_k^{(1)}$,
defined according to the phase space cuts described at the beginning of this section.

Then, at next-to-leading order in the $k$ flavor channel, \Eq{eq:scratch} becomes
\begin{equation}
 \frac{d\sigma^{(1)}_k}{d\ecf{2}{\alpha}} 
 = \ppmatch_k^{(0)} \left[ S_{C,k}^{(1)} + J_k^{(1)} \right] 
 + \ppmatch_k^{(1)}\delta(\ecf{2}{\alpha})
 + \frac{d\sigma^{(1)}_{k,\text{pc}}}{d\ecf{2}{\alpha}}\,.
\end{equation}
Here, $S_{C,k}^{(1)}$ and $J_k^{(1)}$ are the collinear-soft and jet functions at ${\mathcal O}(\alpha_s)$.
Using $\ppmatch_k^{(0)} = \sigma_k^{(0)}$, we can integrate over $\ecf{2}{\alpha}$ to find
\begin{equation}\label{eq:matchcoeffnlo}
 \ppmatch_k^{(1)} 
 = \sigma_k^{(1)} 
 - \sigma_k^{(0)} \int_0^1 d\ecf{2}{\alpha} \left[S_{C,k}^{(1)} + J_k^{(1)}\right] 
 - \sigma^{(1)}_{k,\text{pc}}\,.
\end{equation}
We computed $\sigma_k^{(1)}$ using MCFM \cite{Campbell:2002tg,Campbell:2003hd} 
with settings detailed in the next section. We computed the power corrections according to
\begin{equation}\label{eq:C1terms}
\sigma_{k,\text{pc}}^{(1)}\
  \equiv \int d\ecf{2}{\alpha} \left[\frac{d\sigma_k^{(1)}}{d\ecf{2}{\alpha}} - \sigma^{(0)}_k \left(J_k^{(1)} + S_{C,k}^{(1)}\right) \right]\,.
\end{equation}
For the first term in the integrand, we use a numerical distribution obtained with MCFM. 
Since we do not have access to this distribution at arbitrarily small values of $\ecf{2}{\alpha}$,
the integral in \Eq{eq:C1terms} extends from $\ecf{2}{\alpha}=10^{-5}$ to 1. This approximation is sufficient for 
power corrections suppressed by $\ecf{2}{\alpha}$, and the effect of dropping the $\zcut\delta(\ecf{2}{\alpha})$ term 
from the integral is negligible in comparison to the scale uncertainties shown in the next section.

This completes our extraction of the matching coefficient $D_k$ through relative ${\mathcal O}(\alpha_s)$.  With it,
the resummed cross section of \Eq{eq:factxsecpp} is complete and ready to be matched to relative ${\mathcal O}(\alpha_s^2)$ fixed-order results.

\subsection{Matching Resummation to Fixed-Order}

With the resummed differential cross section for soft-drop groomed $\ecf{2}{\alpha}$ defined in \Eq{eq:factxsecpp}, 
we next match to fixed order  for $pp \to Z+j$.
Our matching procedure will be identical to the procedure we used for $e^+e^-$ collisions;  we 
 add the difference between the exact fixed order and the expansion of the resummed distribution
 to fixed order:
\begin{equation}
\frac{d\sigma_\text{match}}{d\ecf{2}{\alpha}}=\frac{d\sigma_\text{resum}}{d\ecf{2}{\alpha}}+\frac{d\sigma_\text{FO}}{d\ecf{2}{\alpha}}-\frac{d\sigma_\text{resum,FO}}{d\ecf{2}{\alpha}}\,.
\end{equation}
We match the analytic NLL resummed distributions to fixed-order results that include the relative
${\mathcal O}(\alpha_s)$ corrections to the Born process for $pp \to Z+j$. We match NNLL distributions to 
fixed-order results including relative ${\mathcal O}(\alpha_s^2)$ corrections and up to 3 partons in the jet.

We use MCFM v.~6.8 \cite{Campbell:2002tg,Campbell:2003hd} to generate the fixed-order cross sections for soft-drop 
groomed $\ecf{2}{\alpha}$ in $pp\to Z+j$ events.  Currently, MCFM can only generate 
fixed-order corrections at ${\mathcal O}(\alpha_s)$ relative to a Born-level process, and so we will have to use some properties of the 
observable to be able to calculate to relative ${\mathcal O}(\alpha_s^2)$ accuracy.  For $\ecf{2}{\alpha}>0$, as we did in $e^+e^-$ collisions, 
we can ignore the purely two-loop virtual contribution to $pp\to Z+j$, as it has no effect on the differential 
distribution away from $\ecf{2}{\alpha}=0$.  MCFM can generate both inclusive $pp\to Z+j$ and $pp\to Z+2j$ processes 
through relative ${\mathcal O}(\alpha_s)$ accuracy.  Therefore, we can use $pp\to Z+2j$ at relative ${\mathcal O}(\alpha_s)$ in 
MCFM to calculate the relative ${\mathcal O}(\alpha_s^2)$ distribution for $pp\to Z+j$, in the region where $\ecf{2}{\alpha}>0$.

In practice, this procedure requires some care.  To define the cross section for $pp\to Z+ 2j$ in MCFM, 
we must set a minimum $p_T$ for the two jets as identified by MCFM.  This is set by the parameter {\tt ptjet\_min} 
within MCFM.  To  compute the fixed-order cross section correctly for $\ecf{2}{\alpha}$ as measured on the soft-drop 
groomed jet in $pp\to Z+j$ events, {\tt ptjet\_min} should be set to 0; this would of course produce infinity 
because $pp\to Z+2j$ lacks the virtual corrections of $pp\to Z+j$.  To regulate this divergence, we set 
{\tt ptjet\_min} $=1$ GeV and have verified that for jets with $p_{TJ}>500$ GeV, this choice has a negligible effect on the 
differential cross section of $\ecf{2}{\alpha}$ until deep in the infrared region, well beyond the point where 
resummation dominates.  Additionally, we have verified that the distribution of $\ecf{2}{\alpha}$ as measured in 
$pp\to Z+j$ at relative ${\mathcal O}(\alpha_s^2)$ is identical to that measured in $pp\to Z+2j$ at Born level with {\tt ptjet\_min} $=1$ GeV, 
up to differences deep in the infrared.  Using this procedure, we are therefore able to match to relative ${\mathcal O}(\alpha_s^2)$ with MCFM.

\begin{figure}[t]
\centering
\subfloat[]{\label{fig:ppmatchnll}
\includegraphics[width=0.45\linewidth]{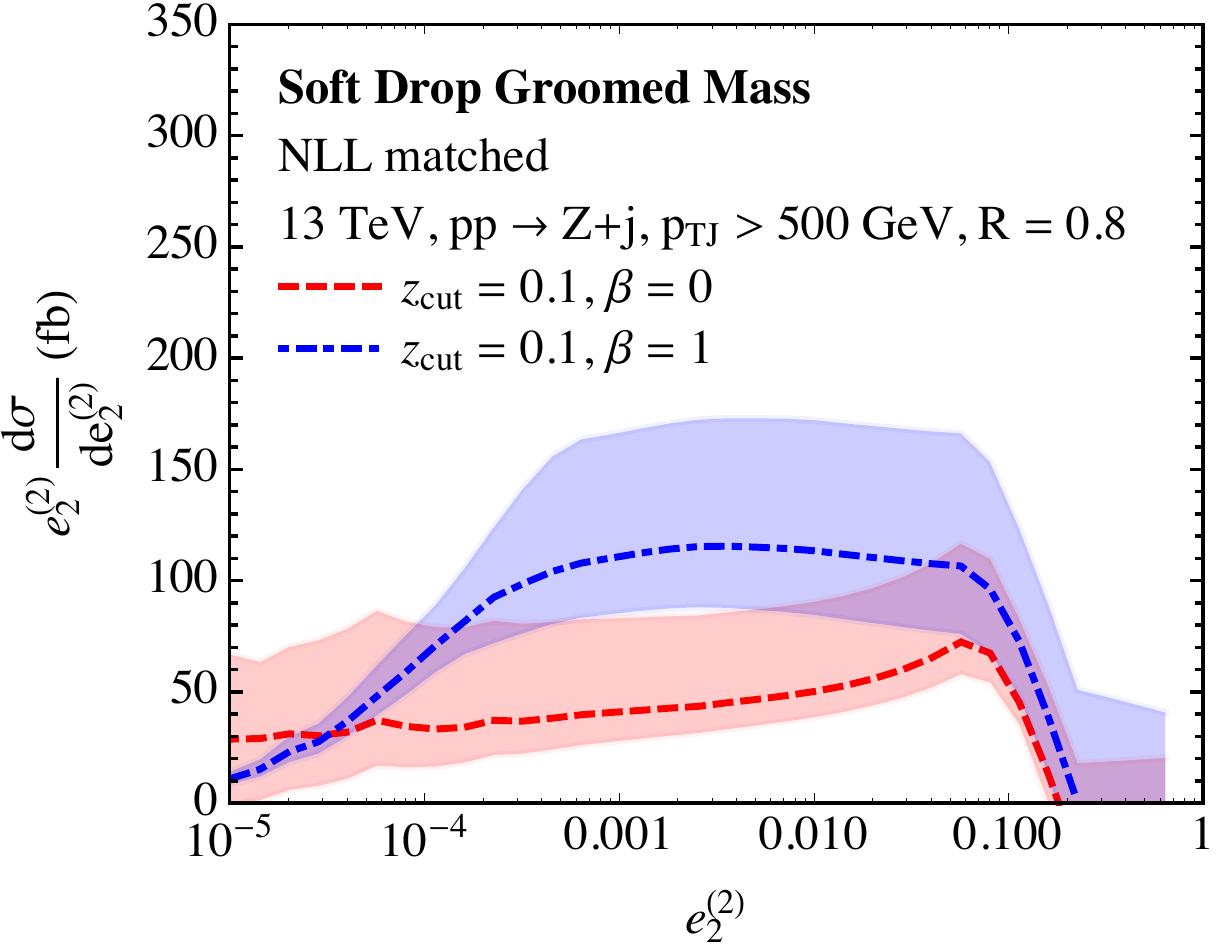}
}\quad
\subfloat[]{\label{fig:ppmatchnnll}
\includegraphics[width=0.45\linewidth]{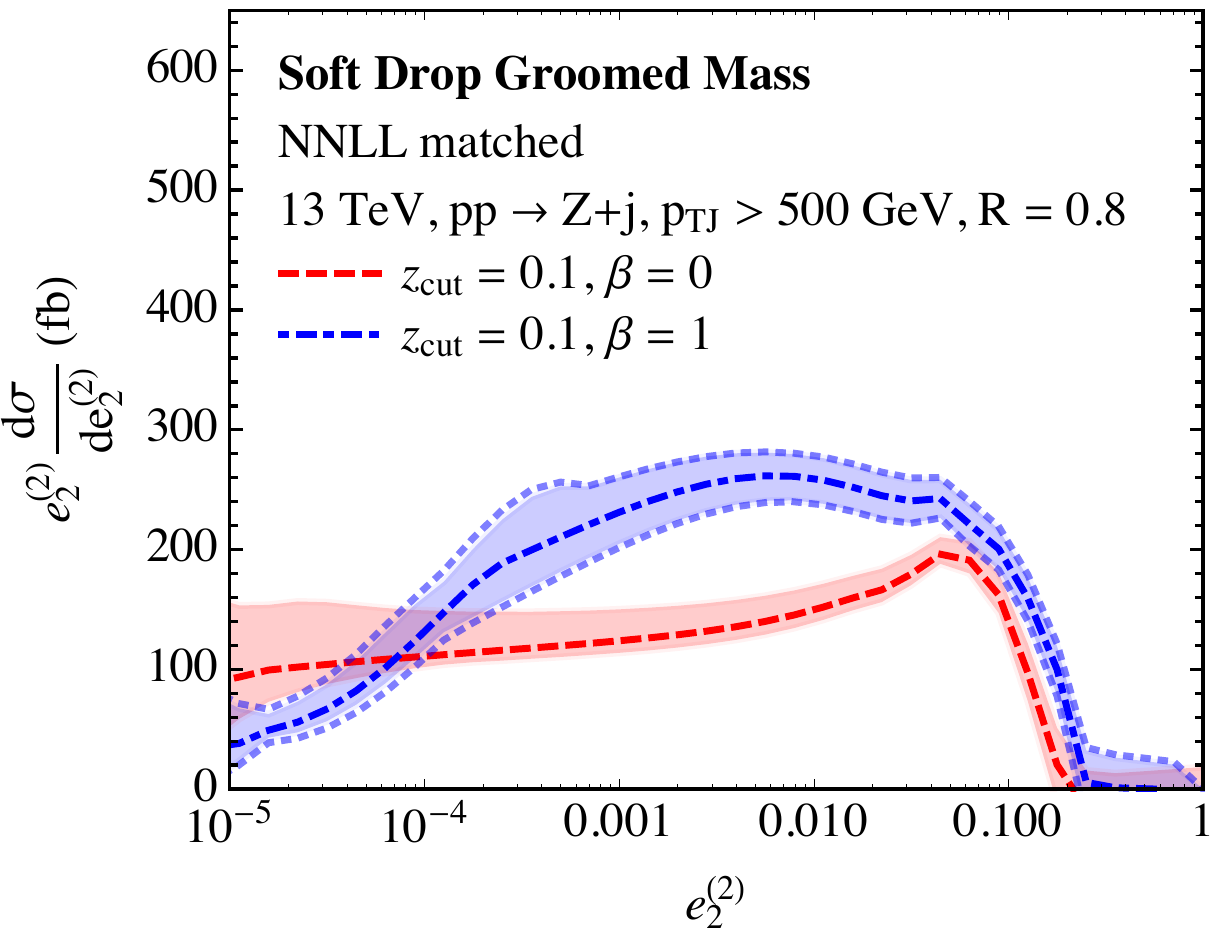}
}
\\
\subfloat[]{\label{fig:ppmatchnll_norm}
\includegraphics[width=0.45\linewidth]{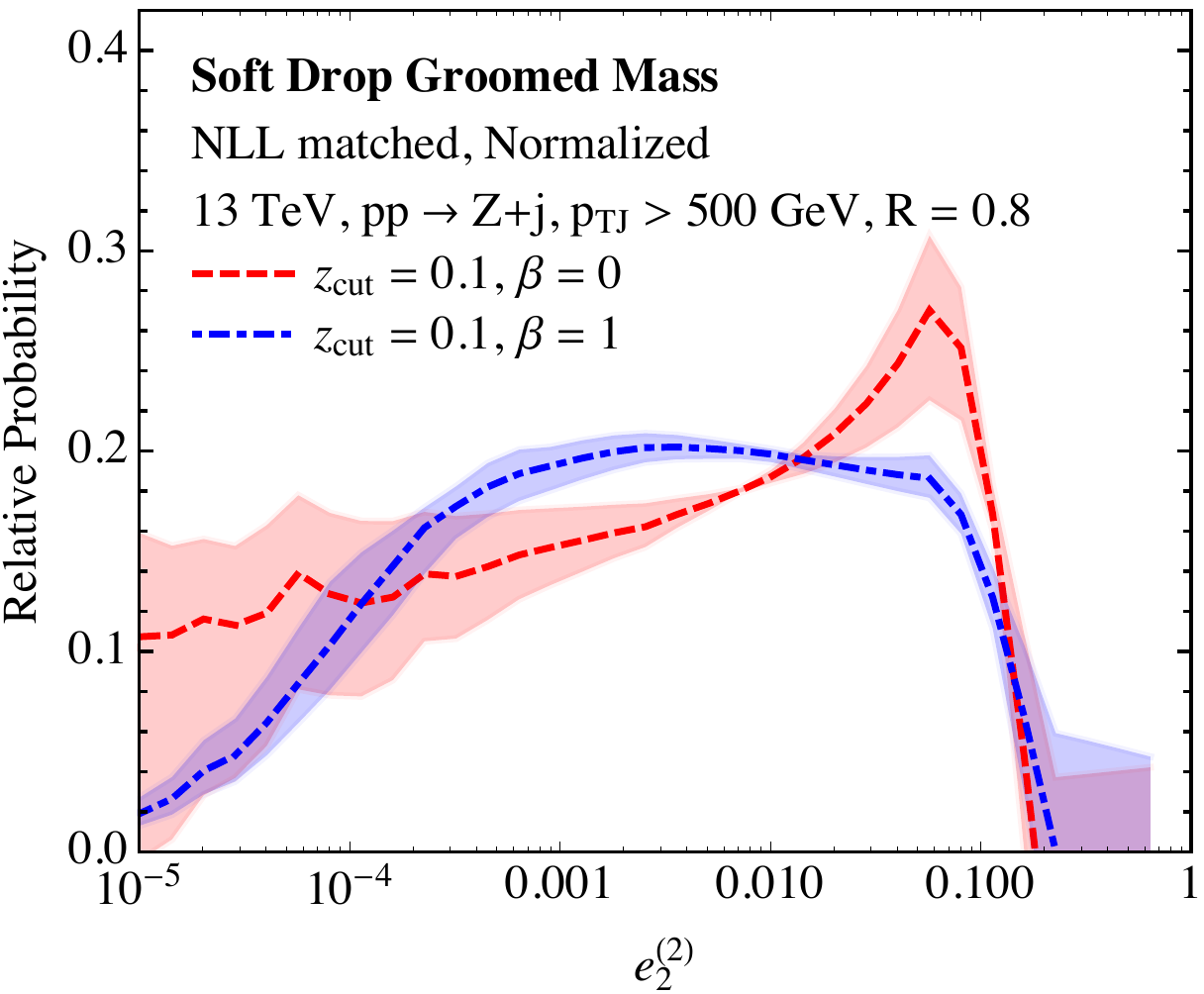}
}\quad
\subfloat[]{\label{fig:ppmatchnnll_norm}
\includegraphics[width=0.45\linewidth]{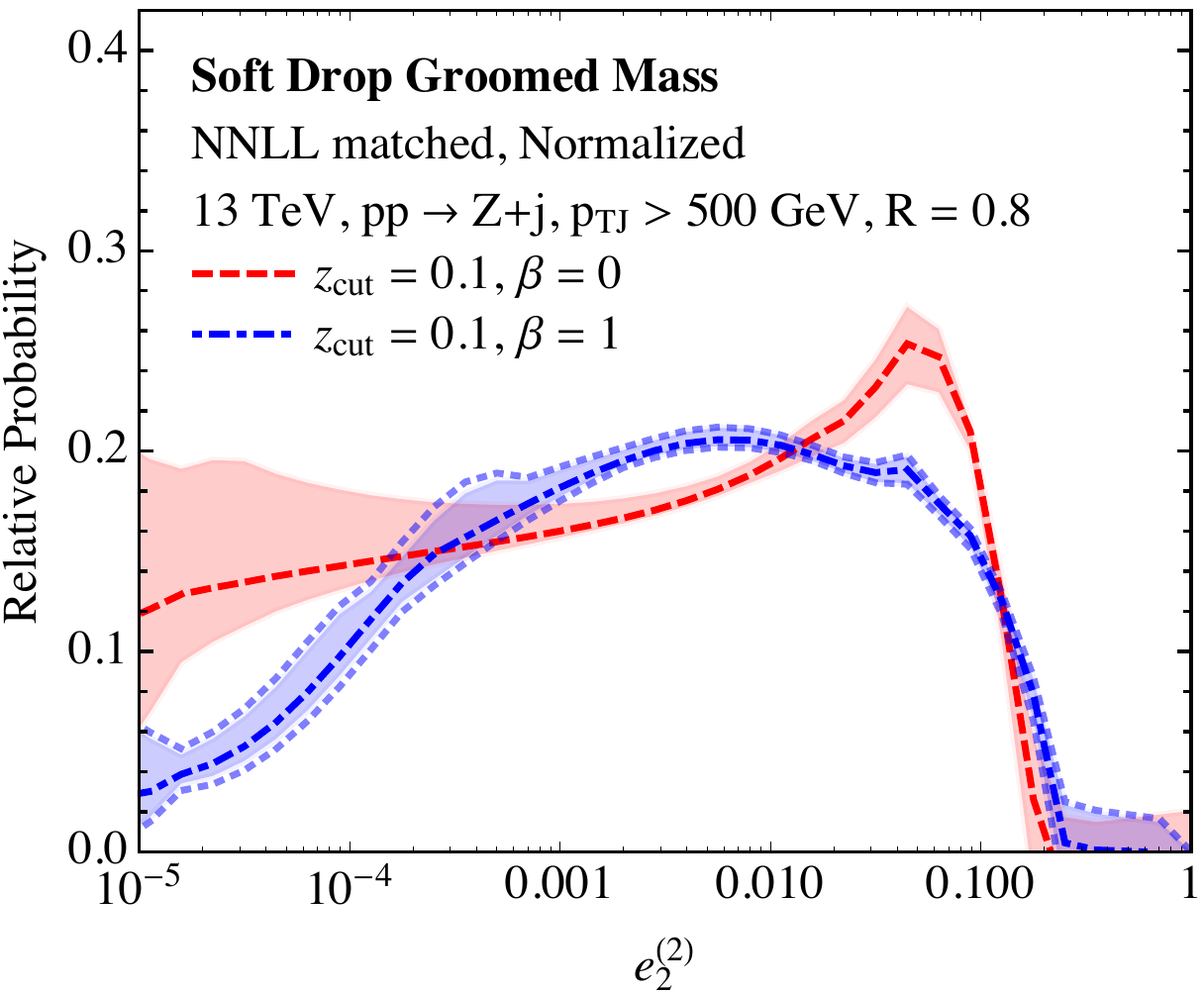}
}
\caption{
NLL matched (left) and NNLL matched (right) distributions for hardest jet $\ecf{2}{2}$ in $pp\to Z+j$ events with soft drop grooming $\zcut = 0.1$ and $\beta = 0$ and $\beta=1$.  Estimates of theoretical uncertainties are represented by the shaded bands.  For soft drop with $\beta = 1$, the dotted lines represent the extent of the theoretical uncertainties when the variation of the two-loop non-cusp anomalous dimension is included.  The distributions in the two upper figures are normalized to the total cross section (in femtobarns), while in the bottom figures, the distributions integrate to the same value over the range $\ecf{2}{2}\in[0.001,0.1]$.
Note the reduction in uncertainties as one moves from NLL to NNLL, and also as one considers the normalized distribution.
}
\label{fig:ppmatch}
\end{figure}

We generate $pp\to Z+j$ events through relative ${\mathcal O}(\alpha_s^2)$ accuracy at the 13 TeV LHC using MSTW 2008 NLO parton distribution functions \cite{Martin:2009iq}.  We require that the $p_T$ of the $Z$ boson is greater than 300 GeV and the absolute value of its pseudorapidity is less than $2.5$.  Jets are clustered with the anti-$k_T$ algorithm with radius $R=0.8$.  We study the hardest jet in these events that satisfies $p_{TJ}>500$ GeV and $|\eta_J|<2.5$.  On these identified jets, we then soft-drop groom and measure $\ecf{2}{\alpha}$ using custom code.  This is an exceptionally computationally demanding procedure at relative ${\mathcal O}(\alpha_s^2)$, due to the complicated phase space of real emissions and the small width of the bins required to calculate the $\ecf{2}{\alpha}$ distribution.  
This precision jet substructure study is only possible because of the development of highly efficient methods for generating fixed-order corrections.

In \Fig{fig:ppmatch} we plot matched distributions for soft-drop $\ecf{2}{2}$ with $\zcut = 0.1$ and both $\beta = 0$ and $\beta = 1$ at NLL and NNLL.  Here, we show both the distributions normalized to the total cross section and normalized over the range $\ecf{2}{2}\in[0.001,0.1]$.  The shaded bands represent estimates of theoretical uncertainties due to residual infrared scale sensitivity.\footnote{The relatively large size of the uncertainty bands for $\ecf{2}{2}\gtrsim 0.1$ is an artifact of our simplistic additive matching.  Additionally, due to the large $K$ factor, the absolute scale of the matched NNLL distribution in \Fig{fig:ppmatchnnll} is roughly twice as large as the matched NLL distribution in \Fig{fig:ppmatchnll}.} We show these bands mainly to allow comparison of the uncertainty remaining at different levels of formal precision.   For the collinear-soft and jet functions in the resummed cross section, we vary the low scales by a factor of two.  To estimate the scale dependence of the 
matching coefficient $\ppmatch_k$ in the resummed cross section is more complicated, and we discuss this in detail in \App{app:ppfact}.  To estimate scale uncertainties in the fixed-order cross section, we vary 
the factorization  and renormalization scales in MCFM by a factor of 2 about $500$ GeV $\simeq p_{TJ}$.  We then take the envelope of all of these scale variations to produce the shaded bands in \Fig{fig:ppmatch}.  For $\beta = 1$ at NNLL, we have also explicitly shown the additional uncertainty due to the two-loop non-cusp anomalous dimension of the collinear-soft function.  In going from NLL to NNLL accuracy, the relative size of the scale uncertainty bands decreases by about a factor of 2 or 3 for both choices of normalization of the distributions.  However, normalizing the distributions over the range $\ecf{2}{2}\in[0.001,0.1]$ dramatically reduces residual scale uncertainties; at NNLL, these normalized distributions have residual scale uncertainties at the 10\% level and smaller.

\subsection{Comparison to Monte Carlo}

We now compare our NNLL resummed and matched calculation of soft-drop groomed $\ecf{2}{2}$ distributions to Monte Carlo simulations.  We generate $pp\to Z+j$ events at the 13 TeV LHC with \herwigpp{2.7.1} and \pythia{8.210}.  To improve statistics somewhat, we have turned off $Z/\gamma$ interference in the Monte Carlos.  The $Z$ boson is forced to decay to electrons, and we require that the invariant mass of the electrons is within 10 GeV of the mass of the $Z$ boson.  We then require that the identified $Z$ boson has $p_{TZ} > $ 300 GeV and $|\eta_Z|<2.5$.  Jets are clustered with \fastjet{3.1.3} using the anti-$k_T$ algorithm with radius $R = 0.8$ and we identify the hardest jet in the event with $p_{TJ}>500$ GeV and $|\eta_J|<2.5$.  We then soft-drop groom this jet and measure $\ecf{2}{2}$.  Both soft drop and the energy correlation functions are implemented using \fastjet{} contrib v.~1.019 \cite{Cacciari:2011ma,fjcontrib}.

\begin{figure}[t]
\centering
\subfloat[]{\label{fig:her_sd0_pp}
\includegraphics[width=0.45\linewidth]{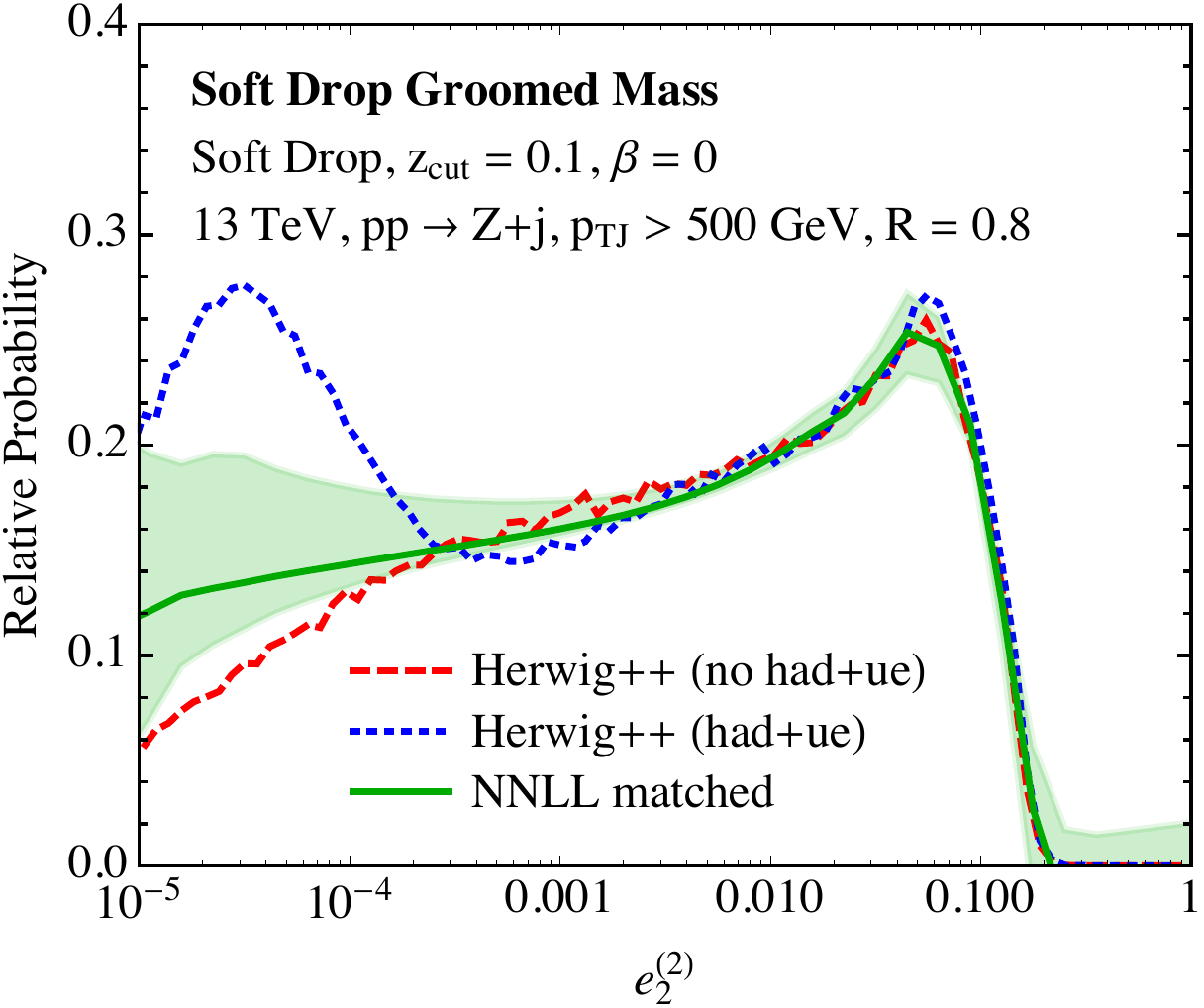}
}\quad
\subfloat[]{\label{fig:py_sd0_pp}
\includegraphics[width=0.45\linewidth]{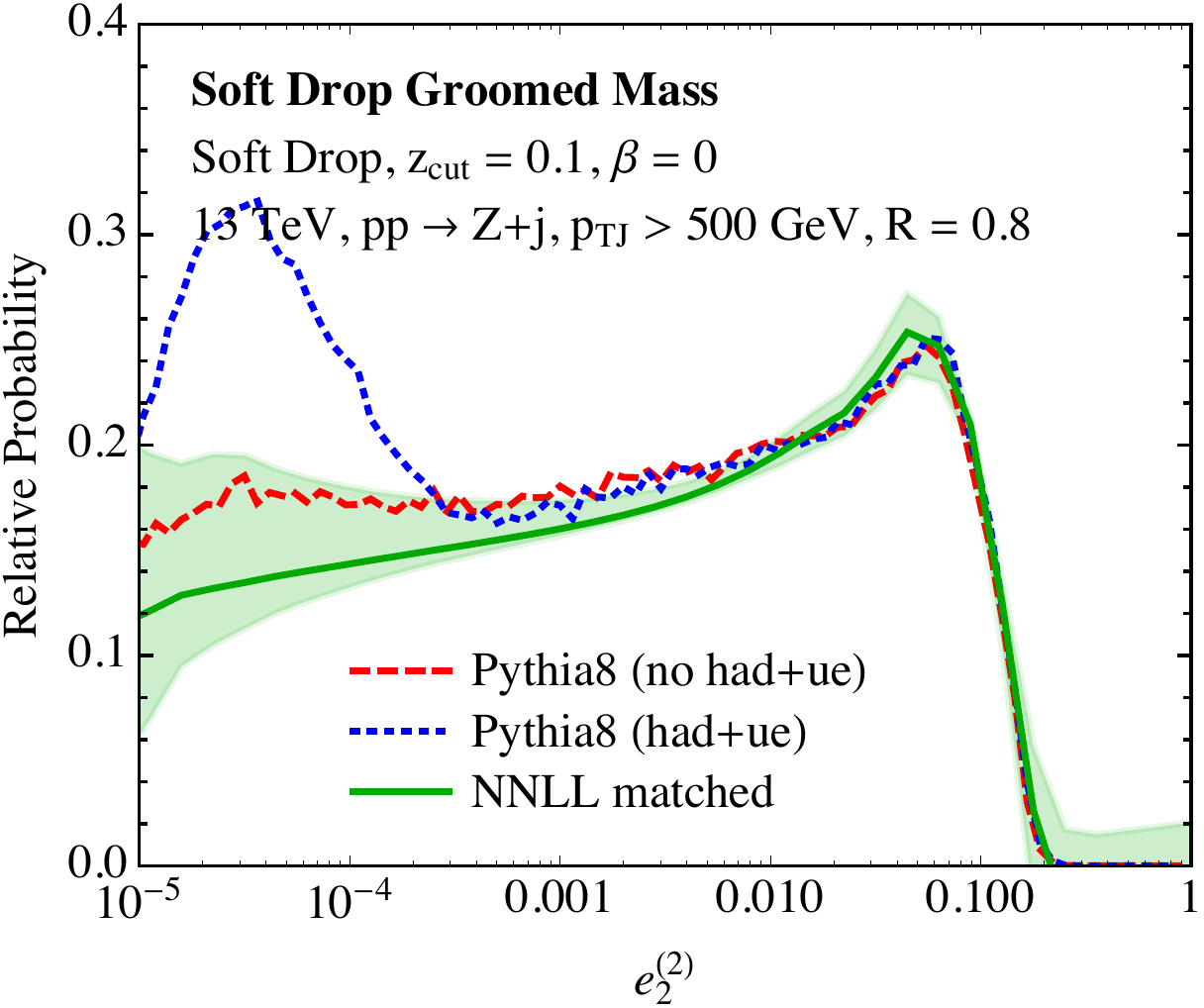}
}\\
\subfloat[]{\label{fig:her_sd1_pp}
\includegraphics[width=0.45\linewidth]{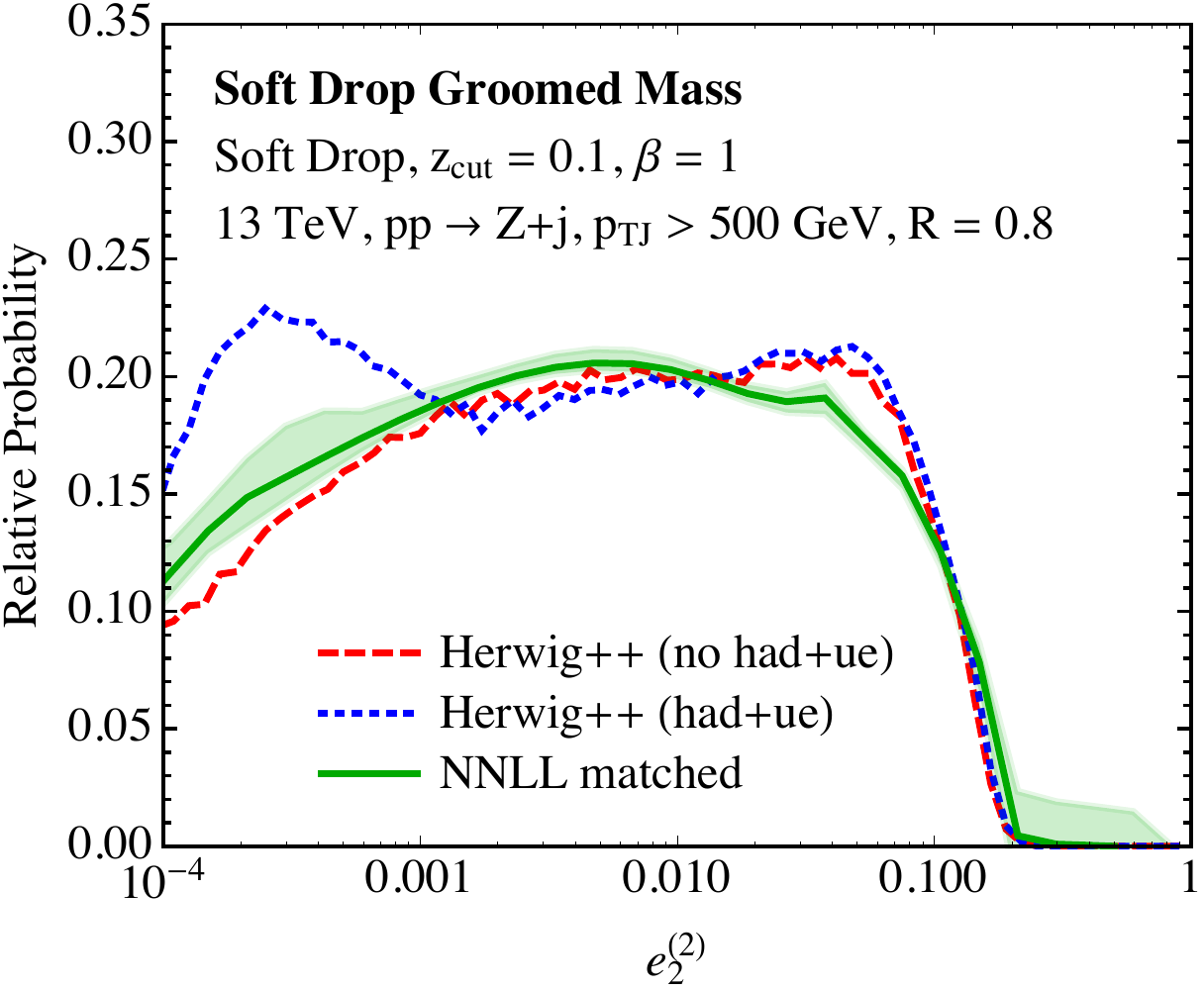}
}\quad
\subfloat[]{\label{fig:py_sd1_pp}
\includegraphics[width=0.45\linewidth]{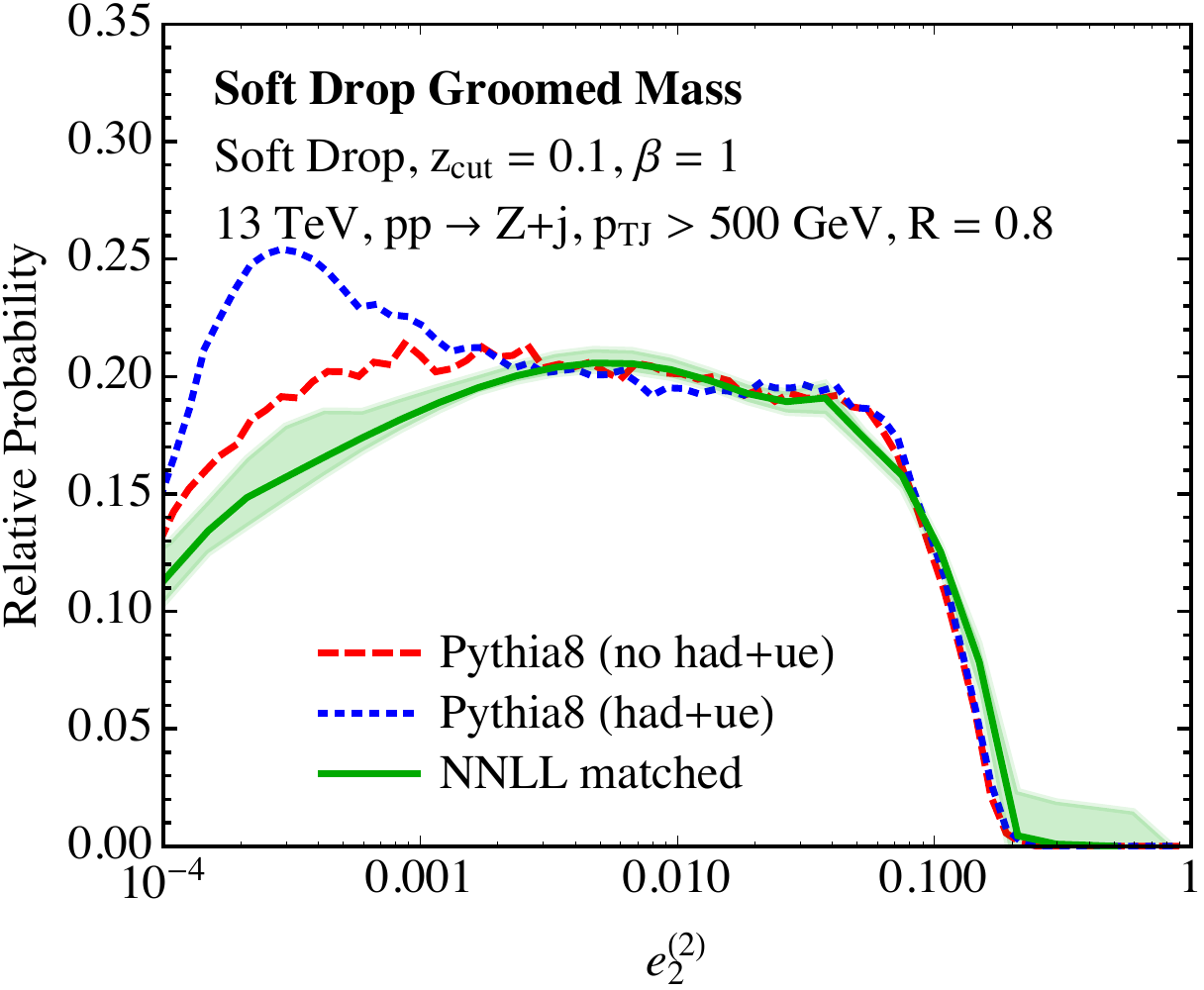}
}
\caption{
Comparison between soft-drop groomed $\ecf{2}{2}$ distributions with $\zcut = 0.1$ and $\beta = 0$ (top) and $\beta = 1$ (bottom) for matched and normalized NNLL, parton-level, and hadron-level Monte Carlo.  All curves integrate to the same value over the range $\ecf{2}{2}\in[0.001,0.1]$.  The uncertainty band for soft drop with $\beta = 1$ at NNLL includes the variation of the two-loop non-cusp anomalous dimension.
}
\label{fig:pp_sd}
\end{figure}

We have generated two samples from both \herwigpp{} and \pythia{} to study the effect of hadronization and underlying event.  One sample is purely parton level: both hadronization and underlying event have been turned off and the other sample is the Monte Carlos run in their default settings, up to the settings of the $Z$ boson mentioned earlier.  The distributions of $\ecf{2}{2}$ measured on soft-drop groomed jets with $\zcut = 0.1$ and both $\beta = 0,1$ are illustrated in \Fig{fig:pp_sd}.  Here, we compare our matched and normalized NNLL calculation to both the parton-level and hadron-level plus underlying event Monte Carlos.  To normalize the Monte Carlo distributions, all curves integrate to the same value on the range $\ecf{2}{2}\in[0.001,0.1]$.

\begin{figure}[t]
\centering
\subfloat[]{\label{fig:pp_comp_sd0}
\includegraphics[width=0.45\linewidth]{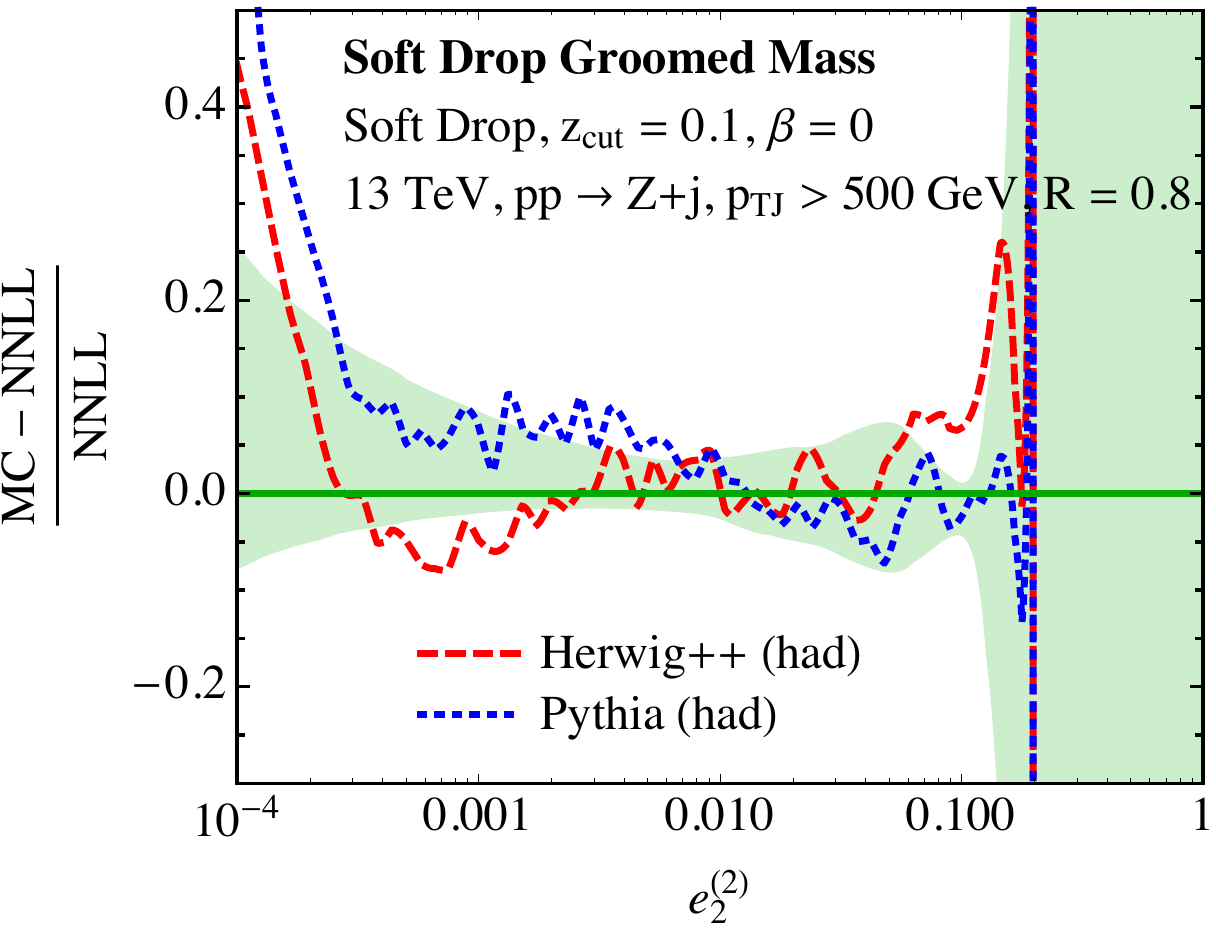}
}\quad
\subfloat[]{\label{fig:pp_comp_sd1}
\includegraphics[width=0.45\linewidth]{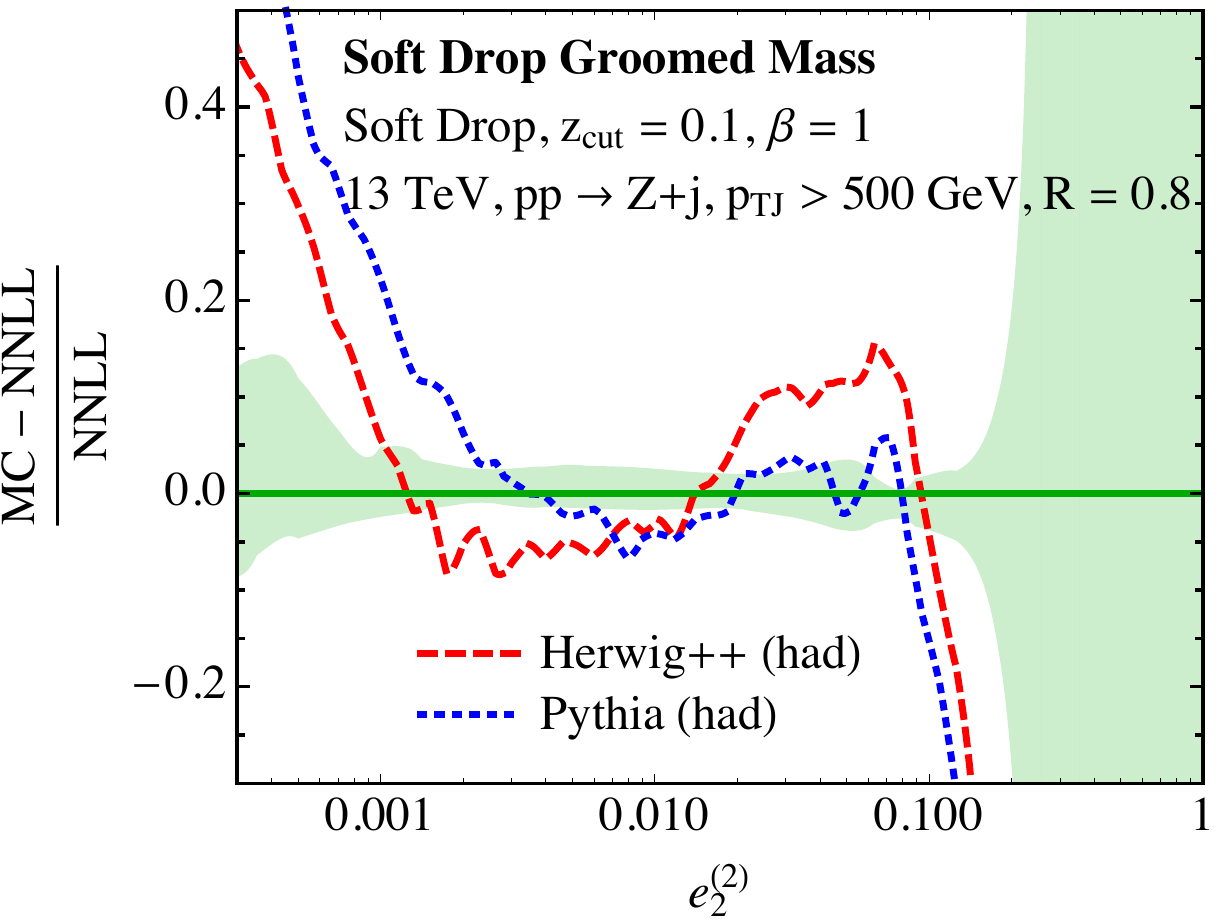}
}
\caption{
Direct comparison of hadron-level output from \herwigpp{} and \pythia{} already shown in \Fig{fig:pp_sd}. 
Soft drop is performed with $\zcut=0.1$ and both $\beta=0$ (left) and $\beta=1$ (right).
Curves are displayed as relative differences between Monte Carlo output and our matched NNLL predictions,
with theoretical uncertainties shown as a shaded band.
}
\label{fig:pp_comp}
\end{figure}

As a more direct comparison of the Monte Carlos, 
\Fig{fig:pp_comp} displays the relative difference between each of the hadron-level Monte Carlos and our matched NNLL predictions,
with our estimates of theoretical uncertainty shown as shaded bands.
Again, soft drop is performed with $\zcut=0.1$, and both $\beta=0$ and $\beta=1$ are shown.
Discrepancies between the Monte Carlo results and our predictions are present but not large.

As observed with jets in $e^+e^-$ collisions, there is good agreement between our precision calculation and the Monte Carlos over a wide dynamic range.  Importantly, this measurement of the soft-drop groomed $\ecf{2}{2}$ is very different from the case in $e^+e^-$.  In $e^+e^-$ collisions we calculated the heavy groomed and ungroomed jet masses. By measuring the heavier of the two jet masses, both masses have to be small, and the observable
is global.
For $pp\to Z+j$ events, we want to make no restrictions on the out-of-jet radiation. Thus although the 
soft drop jet mass is still free of non-global contributions, the ungroomed mass will not be. That is, we do not
have control over all the large logarithms of ungroomed jet mass in $pp\to Z + j$ events, and thus cannot predict
them using our factorized expression, although other approaches are possible.\footnote{Calculations of the ungroomed jet mass in $Z+j$ events
have been done, with varying approaches to handing the non-global contribution~\cite{Chien:2012ur,Jouttenus:2013hs,Dasgupta:2012hg}.}
For this reason, we only show distributions of soft-drop groomed $\ecf{2}{2}$ measurements in $pp\to Z+j$ events.

\Fig{fig:pp_sd} also illustrates that soft drop grooming eliminates sensitivity to both hadronization and underlying event until deep in the infrared.  The parton-level and hadron-level distributions for each Monte Carlo agree almost perfectly until below about $\ecf{2}{2}\lesssim 10^{-3}$.  That hadronization effects are small is expected from our $e^+e^-$ analysis, but this also demonstrates that underlying event effects are negligible.  A similar observation was made in \Ref{Larkoski:2014wba}, though at a much higher jet $p_T$ ($p_T > 3$ TeV).  As in $e^+e^-$ collisions, we expect that the hadronization effects that are observed in the Monte Carlo can be explained by a shape function, though we leave this to future work.

That the shape of the resummed distribution is both completely determined by collinear dynamics and is insensitive to underlying event suggests that by grooming jets with soft drop, we are able to completely isolate factorization-violating effects into an overall normalization.  Therefore, we conjecture that the shape of the leading-power distribution of soft-drop groomed observables as measured in hadron collision events completely factorizes, just like the $p_T$ spectrum in Drell-Yan events \cite{Collins:1984kg}.  We leave a proof of this conjecture to future work.\footnote{Due to the presence of the complicated object $\ppmatch_k(p_{T}^{\min},\eta_{\max},\zcut,R)$, \Eq{eq:factxsecpp} 
is not strictly a factorization theorem. It may not be possible to factorize $\ppmatch_k$ to all orders 
due to the presence of so-called Glauber modes \cite{Collins:1984kg} in the cross section.  While it is beyond the 
scope of this paper, recent work suggests that Glaubers can be included into the cross section directly 
\cite{Rothstein:2016bsq}, and our numerical work indicates that the effect may be absorbable into the normalization.} 
\section{Conclusions}\label{sec:conc}

In this paper, we presented the first calculation for an observable measured exclusively on the constituents of a jet to NNLL accuracy and matched to fixed-order results at ${\mathcal O}(\alpha_s^2)$ relative to the Born process.  The ability to do this calculation required grooming the jet with the soft drop algorithm, which eliminates the complications due to non-global logarithms that afflict ungroomed jet measurements.  The soft drop groomer also significantly reduces nonperturbative effects from hadronization and underlying event, rendering the perturbative calculation of energy correlation functions accurate over several decades.  The insensitivity of soft-drop groomed jet observables to underlying event suggests that the normalized cross section fully factorizes in hadronic scattering events.

To complete the resummed calculation to NNLL accuracy required determining the two-loop non-cusp anomalous dimension for the soft function for which all emissions are removed.  For $\beta = 0$, we were able to use results from the literature to extract the non-cusp anomalous dimension, up to calculable clustering effects.  While not used for results in this paper, the clustering effects when using the anti-$k_T$ algorithm with soft drop are closely related to similar effects found in jet veto calculations.  For soft drop angular exponent $\beta > 0$, we demonstrated a numerical procedure for determining the anomalous dimension using \eventtwo.  This was sufficient to approximate the non-cusp anomalous dimension, but a full calculation of the two-loop soft function for soft drop with $\beta \geq 0$ is desired.

With a complete calculation of the two-loop soft function, including constants, we would be one step closer to resumming to next-to-next-to-next-to-leading logarithmic accuracy (N$^3$LL).  Up to the unknown four-loop cusp anomalous dimension (whose effects have been shown to be small \cite{Becher:2008cf,Abbate:2010xh,Hoang:2014wka}), the only other piece to get to N$^3$LL would be the three-loop non-cusp anomalous dimension of the soft-dropped soft function.  Without an explicit three-loop calculation, this anomalous dimension could in principle be estimated using a technique similar to what we used at two loops, using a fixed-order code like \eerad~\cite{Ridder:2014wza}.  If this is possible, then resummation to this accuracy would potentially reduce residual scale uncertainties to the percent-level, assuming a scaling of uncertainties like observed in going from NLL to NNLL.

For our complete predictions, it was vital to match our resummed calculations to high precision 
fixed-order distributions.  Fixed-order calculations have been traditionally used for observables that are 
inclusive over soft and collinear radiation, like total cross sections or $p_T$ spectra.  The generation of 
fixed-order differential distributions for the plots in this paper required CPU-centuries, which we attained 
only by running on thousands of cores. For calculations of more complicated jet observables, precise fixed order 
computations are likely infeasible with presently available tools.
As jet substructure pushes to higher precision, it will be necessary to have fixed-order calculations that more
efficiently sample the infrared regions of phase space.

The calculations in this paper represent a new frontier of precision QCD.  While jet substructure techniques have been used for some time in experimental analyses at the LHC, they are just now approaching the level of theoretical precision that can be meaningfully compared to data.  By soft-drop grooming jets, we greatly reduce the theoretical challenges, enabling the calculation of a wide range of jet substructure observables to full NNLL accuracy.

\acknowledgments

We thank Simone Marzani, Ian Moult, Ben Nachman, Duff Neill, Iain Stewart, and Hua-Xing Zhu for discussions.  We thank Andrzej Siodmok for help with generation of events in \herwigpp.  A.L.~is supported by the U.S. National Science Foundation, under grant PHY--1419008, the LHC Theory Initiative. The computations in this paper were run on the Odyssey cluster supported by the FAS Division of Science, Research Computing Group at Harvard University.

\appendix

\section{Three-Loop $\beta$-function and Cusp Anomalous Dimension}\label{app:cab}

The $\beta$-function is defined to be
\begin{equation}
\beta(\alpha_s)=\mu\frac{\partial\alpha_s}{\partial\mu} =-2\alpha_s\sum_{n=0}^\infty \beta_n \left(
\frac{\alpha_s}{4\pi}
\right)^{n+1}\,.
\end{equation}
For NNLL resummation, we need the $\beta$-function to three-loop order \cite{Tarasov:1980au,Larin:1993tp}.  The first three coefficients are
\begin{align}
&
\hspace{-0.5cm}
\beta_0 =\frac{11}{3}C_A -\frac{4}{3}T_R n_f \,, \\
&
\hspace{-0.5cm}\beta_1 =\frac{34}{3}C_A^2-4T_R n_f\left(
C_F+\frac{5}{3}C_A
\right) \,,\\
&
\hspace{-0.5cm}\beta_2 = \frac{2857}{54}C_A^3 + T_R n_f \left(
2C_F^2-\frac{205}{9}C_FC_A -\frac{1415}{27}C_A^2
\right) +T_R^2n_f^2\left(
\frac{44}{9}C_F + \frac{158}{27}C_A
\right) \,.
\end{align}

For NNLL resummation, we need the cusp anomalous dimension
\begin{equation}\label{eq:cuspexp}
\Gamma_\text{cusp} = \sum_{n=0}^\infty \Gamma_n \left(
\frac{\alpha_s}{4\pi}
\right)^{n+1}
\end{equation}
to three-loop order.  The first three coefficients of the cusp anomalous dimension are \cite{Korchemsky:1987wg,Vogt:2000ci,Berger:2002sv,Moch:2005tm}:
\begin{align}
\Gamma_0 &= 4 \,,\\
\Gamma_1 &=  4C_A\left(
\frac{67}{9}-\frac{\pi^2}{3}
\right) - \frac{80}{9}T_R n_f\,, \\
\Gamma_2 &= 4C_A^2\left(
\frac{245}{6}-\frac{134\pi^2}{27}+\frac{11\pi^4}{45}+\frac{22}{3}\zeta_3
\right) +32 C_A T_R n_f\left(
-\frac{209}{108}+\frac{5\pi^2}{27}-\frac{7}{3}\zeta_3
\right) \\
&
\hspace{1cm}
+4 C_F T_R n_f\left(
16 \zeta_3 - \frac{55}{3}
\right)-\frac{64}{27}T_R^2 n_f^2
\nonumber \,.
\end{align}

\section{Hard Function}\label{app:hard}

The hard function for dijet production in $e^+e^-$ collisions is defined by the Wilson coefficient for matching the full QCD current onto the SCET dijet operator.  For $e^+e^-\to q\bar q$ events, the Wilson coefficient $C\left(Q^2,\mu\right)$ is
\begin{align}
\langle q\bar q | \bar \psi \Gamma \psi |0\rangle=C\left(Q^2,\mu\right) \langle q\bar q | \bar \chi_n Y^\dagger_n \Gamma Y_{\bar n} \chi_{\bar n} |0\rangle\,.
\end{align}
Here, $\bar \psi \Gamma \psi$ is the QCD current for the production of a $q\bar q$ pair from the vacuum. $\chi_{\bar n}$ is a quark jet operator
collinear quark operator defined in the light-like direction $\bar n$ in SCET.
For calculations at leading power, $\chi_n = W_t^\dagger \psi$, with $W_t$ a Wilson line pointing in some direction $t$ not collinear to $n$ and $\psi$ is an ordinary quark field.
The soft Wilson lines $Y_{n}$ and $Y_{\bar n}$ point in the $n$ and $\bar{n}$ directions respectively.  $\Gamma$ represents a generic Dirac matrix.  We have ignored contraction with the leptonic tensor for simplicity. 
The Wilson lines $Y_{n}$ is defined as
\begin{equation}
Y_{n}(x^\mu)={\bf P} \exp \left( ig \int\limits_0^\infty ds\, n\cdot A(x^\mu+sn^\mu)    \right)\,,
\end{equation}
where  ${\bf P}$ denotes path-ordering. 
 $Y_{\bar n}$ and $W_t$ are defined similarly with $\bar{n}^\mu$ and $t^\mu$ replacing $n^\mu$. In SCET, the gluon fields in the Wilson line are soft gluons for the $Y$'s and collinear
gluons for the $W$'s, but once the sectors are decoupled one can treat any of these gluons simply as a gluon field of full QCD.

The hard function is the square of the Wilson coefficient:
\begin{equation}
H\left(Q^2,\mu \right)=\left |C\left(Q^2,\mu\right) \right|^2\,.
\end{equation}
While we do not present its expression here, the hard function for $e^+e^-\to gg$ events is defined analogously, by matching the Higgs current $F_{\mu\nu}F^{\mu\nu}$ onto SCET.

\subsection{$e^+e^-\to q\bar q$}

The one-loop hard function for the process $e^+e^-\to q\bar q$ is \cite{Bauer:2003di,Manohar:2003vb,Ellis:2010rwa,Bauer:2011uc}
\begin{equation}
    H = 1 + \frac{\alpha_s\,C_F}{2\pi} \left(-L_H^2 - 3\,L_H - 8+\frac{7}{6}\pi^2\right),
\end{equation}
where
\begin{equation}
    L_H = \log \frac{\mu^2}{Q^2}\,.
\end{equation}
The cusp anomalous dimension of the hard function to all orders is
\begin{equation}
\Gamma_H = -2C_F \Gamma_\text{cusp}\,,
\end{equation}
where $\Gamma_\text{cusp}$ is the cusp anomalous dimension defined in \Eq{eq:cuspexp}.  Similar to the cusp anomalous dimension, we define the coefficients of the non-cusp anomalous dimension $\gamma$ via
\begin{equation}\label{eq:gnoncusp}
\gamma = \sum_{n=0}^\infty \gamma^{(n)}\left(
\frac{\alpha_s}{4\pi}
\right)^{n+1}\,.
\end{equation}
Through two-loops, the non-cusp anomalous dimension coefficients of the hard function are \cite{vanNeerven:1985xr,Matsuura:1988sm}
\begin{align}
\gamma_H^{(0)}&=-12 C_F\,,\\
\gamma_H^{(1)}&=\left(
-6+8\pi^2-96\zeta_3
\right)C_F^2+\left(
-\frac{1922}{27}-\frac{22}{3}\pi^2+104 \zeta_3
\right)C_F C_A+\left(
\frac{520}{27}+\frac{8}{3}\pi^2
\right) C_F n_f T_R\,.\nonumber
\end{align}

\subsection{$e^+e^-\to gg$}

In the infinite top quark mass limit or with a finite Yukawa coupling, $e^+e^-$ scattering can produce final state gluon jets.  The hard function for such a process can be extracted from $gg\to H$ calculations.  To all orders, the cusp anomalous dimension of the $e^+e^-\to gg$ hard function is
\begin{equation}
\Gamma_H = -2 C_A \Gamma_\text{cusp}\,,
\end{equation}
where $\Gamma_\text{cusp}$ is the cusp anomalous dimension in \Eq{eq:cuspexp}.  Through two-loops, the coefficients of the non-cusp anomalous dimension are \cite{Harlander:2000mg,Anastasiou:2002wq,Ravindran:2004mb}
\begin{align}\label{eq:gghardanom}
\gamma_H^{(0)} &= -4\beta_0\,,\\
\gamma_H^{(1)} &= \left(
-\frac{236}{9}+8\zeta_3
\right)C_A^2+\left(
-\frac{76}{9}+\frac{2}{3}\pi^2
\right) C_A \beta_0 - 4\beta_1\,. \nonumber
\end{align}

\section{The Global Soft Function}\label{app:softfct}

For arbitrary exponent $\beta$ in the soft-drop groomer, the soft function can be calculated by requiring that soft gluons in measured jets fail the soft drop criterion.  
For hemisphere jets in $e^+e^-\to q\bar q$ events, for example, the soft function is defined by the forward matrix element of soft Wilson lines:
\begin{align}
S_\glob(\zcut)&=\frac{1}{N_C}\text{tr}\langle 0|\text{T}\{Y_{n } Y_{\bar n }\}  \hat\Theta_{SD}\overline{\text{T}}\{Y_{n } Y_{\bar n }\} |0\rangle\,.
\end{align}
Here, $n$ and $\bar n$ are the light-like directions of the $q\bar q$ dipole, T denotes time ordering, and $\hat\Theta_{SD}$ denotes the soft drop groomer operator which requires the final state to fail soft drop.  The action of $\hat\Theta_{SD}$ on soft final states cannot be written in a closed form for an arbitrary final state due to clustering effects, though it can be defined order-by-order.  For example, the matrix element of $\hat\Theta_{SD}$ for $\beta = 0$ on a final state with two soft particles was presented in \Sec{sec:sfanomb0}.

At one-loop for hemisphere jets in $e^+e^-$ collisions, the soft function $S_\glob$ can be calculated from
\begin{align}
\hspace{-0.53cm}
S_\glob&=g^2\mu^{2\epsilon}C_i\int \frac{d^dk}{(2\pi)^d} \frac{n\cdot \bar n}{n \cdot k \, k\cdot \bar n}2\pi \delta(k^2) \Theta(k^0)\Theta(\bar n\cdot k-n\cdot k) \Theta\left(
\zcut\, \frac{Q}{2} \left[
2\frac{n\cdot k}{k^0}
\right]^{\beta/2} - k^0
\right)\nonumber \\
&
\hspace{2cm}
+(n\leftrightarrow \bar n)\,,
\end{align}
where $n,\bar n$ are back-to-back light-like vectors with $n\cdot \bar n=2$.  The requirement $\bar n\cdot k>n\cdot k$ restricts the radiation to lie in one hemisphere, while the requirement 
\begin{equation}
\zcut\,\frac{Q}{2}  \left[
2\frac{n\cdot k}{k^0}
\right]^{\beta/2} > k^0
\end{equation}
 restricts the soft gluon to fail soft drop.  We find
\begin{equation}
    S_\glob = 1 + \frac{\alpha_s\,C_i}{\pi} \left[\frac{1}{2\,(1+\beta)} \, L_S^2 - \frac{\pi^2}{12} \left(\frac{1}{1+\beta} + 2 + \beta\right) \right],
\end{equation}
where $C_i$ is the appropriate color factor ($C_F$ for $e^+e^-\to q\bar q$; $C_A$ for $e^+e^-\to gg$) and
\begin{equation}
    L_S = \log\frac{\mu^2}{Q^2\,(\zcut)^2\, 4^\beta}\,.
\end{equation}
To all orders, the cusp anomalous dimension of the hemisphere wide-angle soft function is
\begin{equation}
\Gamma_S = \frac{2C_i}{1+\beta}\Gamma_\text{cusp}\,,
\end{equation}
where $\Gamma_\text{cusp}$ is the cusp anomalous dimension from \Eq{eq:cuspexp}.  To one-loop order, the non-cusp anomalous dimension is 0: $$\gamma_S^{(0)}=0\,.$$

For NNLL resummation, we need the non-cusp anomalous dimension to two-loop order.  
As discussed in \Sec{sec:sfanomb0}, for soft drop with angular exponent $\beta = 0$, this can be extracted from energy veto calculations, up to clustering effects that we calculated.  
For soft drop with $\beta=0$ and Cambridge/Aachen reclustering, we find the two-loop non-cusp anomalous dimension to be
\begin{equation}\label{eq:sb0anom}
\left.\gamma_S^{(1)}\right|_{\beta = 0}=C_i\left[34.01\,C_F+
\left(
\frac{1616}{27}-56\zeta_3-9.31
\right)C_A-\left(\frac{448}{27}+14.04\right)n_f T_R-\frac{2\pi^2}{3}\beta_0
\right]\,.
\end{equation}

\section{Jet Functions}\label{app:jetfunc}

Here, we present the quark and gluon jet functions on which the energy correlation function $\ecf{2}{\alpha}$ is measured.  The quark jet function, for example, is defined by the forward matrix element:
\begin{align}
J_{q}(\ecf{2}{\alpha})=\frac{(2\pi)^3}{N_C}\text{tr}\langle 0|\frac{\bar n\slash}{2}\chi_{n}(0) \delta(Q-\bar n\cdot{\mathcal P})\delta^{(2)}(\vec{{\mathcal P}}_{\perp})\delta\left(\ecf{2}{\alpha}-\hat e_{2}^{(\alpha)}\right)\bar{\chi}_n(0)|0\rangle \,.
\end{align}
Here, the jet is collinear to the light-like direction $n$, the operator $\delta(Q-\bar n\cdot{\mathcal P})$ restricts the large light-cone component of momentum to be equal to the center-of-mass collision energy $Q$, and $\delta^{(2)}(\vec{{\mathcal P}}_{\perp})$ restricts the jet function to have zero net momentum transverse to the $n$ direction.  The measurement operator is defined by its action on an $n$-particle collinear final state $|X_n\rangle$ as:
\begin{equation}
\hat e_{2}^{(\alpha)}|X_n\rangle =\frac{2^{3\alpha/2}}{Q^2} \sum_{i<j\in X_n} (\bar n\cdot p_i)^{1-\alpha/2}(\bar n\cdot p_j)^{1-\alpha/2}(p_i\cdot p_j)^{\alpha/2}|X_n\rangle\,.
\end{equation}
To write this expression, we have expanded the definition of the energy correlation function from \Sec{sec:e2defs} to leading power with collinear momenta.  The gluon jet function is defined similarly:
\begin{align}
J_{g }(\ecf{2}{\alpha})=
\frac{(2\pi)^3}{N_C}\text{tr}\langle 0|\mathcal{B}_{\perp}^{\mu}(0)\delta(Q-\bar n \cdot{\mathcal P})\delta^{(2)}(\vec{{\mathcal P}}_{\perp})\delta\left(\ecf{2}{\alpha}-\hat e_{2}^{(\alpha)}\right)\mathcal{B}_{\perp\mu}(0)|0\rangle\,,
\end{align}
where $\mathcal{B}_{\perp}^{\mu}$ is the collinear-gauge invariant operator in SCET that creates physical collinear gluons.

The following expressions will be presented in Laplace space, where renormalization is multiplicative and the Laplace space conjugate is $\nu$.  That is,
\begin{equation}
J(\nu) = \int_0^\infty d\ecf{2}{\alpha} \, e^{-\nu \ecf{2}{\alpha}} J(\ecf{2}{\alpha})\,.
\end{equation}
The one-loop quark and gluon jet functions were first calculated in \Ref{Larkoski:2015zka} for jets on which the two-point energy correlation functions with arbitrary angular exponent are measured.

\subsection{Quark Jets}

To one loop, the Laplace-space quark jet function is
\begin{equation}
    J_q(\nu) = 1 + \frac{\alpha_s\,C_F}{2\pi}  \left[\frac{\alpha}{2(\alpha-1)} L_{C}^2 + \frac{3}{2} \, L_{C} + \left(\frac{13}{2} - \frac{12}{2\alpha}\right) - \frac{\pi^2}{12}\left(9-\frac{3}{\alpha-1}-\frac{4}{\alpha}\right)\right],
\end{equation}
where
\begin{equation}
    L_{C} = \log \frac{\mu^2(\nu\,e^{\gamma_E})^{2/\alpha}}{E_J^2}\,.
\end{equation}
To all orders, the cusp anomalous dimension of the quark jet function is
\begin{equation}
\Gamma_C^q = \frac{\alpha}{\alpha-1}C_F\Gamma_\text{cusp}\,,
\end{equation}
where $\Gamma_\text{cusp}$ is the cusp anomalous dimension from \Eq{eq:cuspexp}.  For all $\alpha$, the one-loop non-cusp anomalous dimension is
\begin{equation}
\gamma_C^{q,(0)}=6 C_F\,.
\end{equation}

For NNLL resummation, we also need the two-loop non-cusp anomalous dimension.  
For $\alpha=2$, corresponding to jet mass or thrust, this is known exactly.  
In that case, the non-cusp anomalous dimension is \cite{Neubert:2004dd}
\begin{align}
\left.\gamma_C^{q,(1)}\right|_{\alpha=2}&=C_F\left[
C_F\left(
3-4\pi^2+48\zeta_3
\right)+C_A\left(
\frac{1769}{27}+\frac{22\pi^2}{9}-80\zeta_3
\right)+T_R n_f\left(
-\frac{484}{27}-\frac{8\pi^2}{9}
\right)
\right] \,.
\end{align}

\subsection{Gluon Jets}

To one-loop, the Laplace-space gluon jet function is
\begin{align}
    J_g(\nu) &= \\
    &
    \hspace{-1cm}
    1 + \frac{\alpha_s}{2\pi}  \left[\frac{\alpha\, C_A}{2(\alpha-1)} L_{C}^2 + \frac{\beta_0}{2} \, L_{C} + C_A\left(\frac{67}{9}\frac{\alpha-1}{\alpha} -\frac{\pi^2}{3}\frac{2(\alpha-1)^2-1}{\alpha-1}\right) +n_fT_R\left(
    \frac{26}{9\alpha}-\frac{23}{9}
    \right)\right]\,, \nonumber
\end{align}
where
\begin{equation}
    L_{C} = \log \frac{ \mu^2(\nu\,e^{\gamma_E})^{2/\alpha}}{E_J^2}\,.
\end{equation}
To all orders, the cusp anomalous dimension of the gluon jet function is
\begin{equation}
\Gamma_C^g = \frac{\alpha}{\alpha-1}C_A\Gamma_\text{cusp}\,,
\end{equation}
where $\Gamma_\text{cusp}$ is the cusp anomalous dimension from \Eq{eq:cuspexp}.  For all $\alpha$, the one-loop non-cusp anomalous dimension is
\begin{equation}
\gamma_C^{g,(0)}=2\beta_0\,.
\end{equation}

For NNLL resummation, we also need the two-loop non-cusp anomalous dimension.  
For $\alpha=2$, corresponding to jet mass or thrust, this is known exactly.  
In that case, the non-cusp anomalous dimension is \cite{Becher:2009th}
\begin{align}
\gamma_C^{g,(1)}&=
C_A^2\left(
\frac{2192}{27}-\frac{22\pi^2}{9}-32\zeta_3
\right)+C_A T_R n_f \left(
-\frac{736}{27}+\frac{8\pi^2}{9}
\right)-8C_F T_R n_f \,.
\end{align}

\section{Collinear-Soft Function}\label{app:csoftfct}

The final piece in the factorization theorem is the collinear soft function, defined from soft radiation that is collinear to the jet.  As it describes soft radiation, the collinear-soft function is defined as a forward matrix element of Wilson lines:
\begin{align}
S_{C}(\zcut\ecf{2}{\alpha})&=\frac{1}{N_C}\text{tr}\langle 0|\text{T}\{Y_{n }^\dagger W_{t}\}  \delta\left(
\ecf{2}{\alpha}-\left(
1-\hat\Theta_{SD}
\right)\hat e_2^{(\alpha)}
\right)\overline{\text{T}}\{W_{t}^\dagger Y_{n}\} |0\rangle\,.
\end{align}
The $Y$ and $W$ Wilson lines are the same as the ones in the soft and jet functions respectively, but depend on collinear-soft fields (which, like any of the others, can be treated
as full QCD fields at leading power). 

Now, collinear-soft modes only contribute to $\ecf{2}{\alpha}$ if emissions pass the soft drop groomer: this is denoted by $1-\hat\Theta_{SD}$ in the measurement function.  (Recall that $\hat\Theta_{SD}$ removes emissions from the jet according to soft drop.)  Again, this operator cannot be written in closed form for an arbitrary final state due to clustering effects, but below, we will calculate it explicitly at one-loop.  The $\hat e_2^{(\alpha)}$ measurement operator is defined by its action on an $n$-particle collinear-soft final state $|X_{S,n}\rangle$:
\begin{equation}
\hat e_2^{(\alpha)}|X_{S,n}\rangle =\frac{2^\alpha}{Q}\sum_{i\in X_{S,n}}(\bar n\cdot p_i)^{1-\alpha/2}(n\cdot p_i)^{\alpha/2}
|X_{S,n}\rangle \,.
\end{equation}
This follows from expanding the definition of the energy correlation function from \Sec{sec:e2defs} to leading power with collinear-soft momenta.

This can be calculated at one-loop accuracy from
\begin{align}
\hspace{-0.25cm}
S_{C}&=g^2\mu^{2\epsilon}C_i\int \frac{d^dk}{(2\pi)^d} \frac{n\cdot \bar n}{n \cdot k \, k\cdot \bar n}2\pi \delta(k^2) \Theta(\bar n\cdot k)
\left[ \Theta\left(
\zcut  \left[
4\frac{n\cdot k}{\bar n \cdot k}
\right]^{\beta/2} - \frac{\bar n\cdot k}{Q}
\right)\delta\left(
\ecf{2}{\alpha}
\right)\right.
\\
&
\hspace{3cm}
\left.+\,\Theta\left(
 \frac{\bar n\cdot k}{Q}-\zcut  \left[
4\frac{n\cdot k}{\bar n \cdot k}
\right]^{\beta/2}
\right)\delta\left(
\ecf{2}{\alpha}-\frac{2^\alpha}{Q}(n\cdot k)^{\alpha/2}(\bar n\cdot k)^{1-\alpha/2}
\right)
\right]  \nonumber\\
&=g^2\mu^{2\epsilon}C_i\int \frac{d^dk}{(2\pi)^d} \frac{n\cdot \bar n}{n \cdot k \, k\cdot \bar n}2\pi \delta(k^2) \Theta(\bar n\cdot k)\Theta\left(
 \frac{\bar n\cdot k}{Q}-\zcut  \left[
4\frac{n\cdot k}{\bar n \cdot k}
\right]^{\beta/2}
\right)
\nonumber
\\
&
\hspace{5cm}
\times\left[ \delta\left(
\ecf{2}{\alpha}-\frac{2^\alpha}{Q}(n\cdot k)^{\alpha/2}(\bar n\cdot k)^{1-\alpha/2}
\right)-\delta\left(
\ecf{2}{\alpha}\right)
\right]  \nonumber\,.
\end{align}
where $C_i$ is the color factor of the jet.  In the second equality, we have rearranged the phase space constraints and explicitly removed scaleless integrals.  For this collinear-soft function, at one-loop in Laplace space we find
\begin{equation}
    S_C(\nu) = 1 + \frac{\alpha_s\,C_i}{2\pi} \left[-\frac{\alpha+\beta}{2\,(\alpha-1)(\beta+1)} \, L_{S_C}^2 + \frac{\pi^2}{12} \frac{(\alpha+2+3\,\beta)(\alpha-2-\beta)}{(\alpha+\beta)(\alpha-1)(\beta+1)} \right],
\end{equation}
where
\begin{equation}
    L_{S_C} = \log\frac{\mu^2 \, (\nu\,e^{\gamma_E})^{2\frac{\beta+1}{\alpha+\beta}}}{E_J^2 \, (\zcut)^{2\frac{\alpha-1}{\alpha+\beta}}}\,.
\end{equation}
To all orders, the cusp anomalous dimension of the collinear-soft function is
\begin{equation}
\Gamma_{S_C} = -C_i\frac{\alpha+\beta}{(\alpha-1)(\beta+1)}\Gamma_\text{cusp}\,,
\end{equation}
where $\Gamma_\text{cusp}$ is the cusp anomalous dimension from \Eq{eq:cuspexp}.  To one-loop order, the non-cusp anomalous dimension is 0: $$\gamma_{S_C}^{(0)}=0\,.$$

For NNLL resummation, we need the non-cusp anomalous dimension to two-loop order.  For $\alpha=2$ and $\beta =0$, this can be determined by renormalization group consistency of the cross section directly,
using either the $e^+e^-\to q\bar q$ or the $e^+e^-\to gg$ process.  For soft drop with Cambridge/Aachen reclustering, the two-loop non-cusp anomalous dimension is
\begin{align}
\left.\gamma_{S_C}^{(1)}\right|_{\alpha=2,\beta  =0}=C_i\left[-17.00\, C_F
+\left(
-55.20+\frac{22\pi^2}{9}+56\zeta_3
\right)C_A+\left(
23.61-\frac{8\pi^2}{9}
\right)n_f T_R
\right]\,.
\end{align}

\section{Resummation}\label{app:resum}

Because we work in Laplace space, defined according to
\begin{equation}
    F(\nu) = \int_{0^-}^\infty d\ecf{2}{\alpha} \, e^{-\nu\,\ecf{2}{\alpha}}\,F(\ecf{2}{\alpha})\,,
\end{equation}
the renormalization of all functions in the factorization theorem is multiplicative.  For some function $F$ in the factorization theorem, it generically has the renormalization equation
\begin{equation}
\mu\frac{\partial}{\partial \mu} F(\mu) = \gamma F(\mu) \,,
\end{equation}
where the anomalous dimension of $F$ is $\gamma$.  The anomalous dimension can be written as
\begin{equation}\label{eq:anommaster}
\gamma = \Gamma_F(\alpha_s)\log\frac{\mu^2}{\mu_1^2}+\gamma_F(\alpha_s) \,,
\end{equation}
where $\Gamma_F(\alpha_s)$ is the cusp part of the anomalous dimension, $\mu_1$ is the infrared scale in the logarithm and $\gamma_F(\alpha_s)$ is the non-cusp part of the anomalous dimension.  
The solution\footnote{In the plots of resummed distributions in this paper, we have frozen the strong coupling
at $\mu_\text{NP} = 1$ GeV to keep cross sections finite. In the case of frozen $\alpha_s$, the solution 
to the renormalization group equation for each $F(\mu)$ is quite simple, so we omit the details of the
prescription below $\mu_\text{NP}$ here.}
to the renormalization group equation
can be written more conveniently as an integral with respect to $\alpha_s$, by using the definition of the $\beta$-function as
\begin{equation}
\frac{d\mu}{\mu} = \frac{d\alpha_s}{\beta(\alpha_s)} \,.
\end{equation}
Then, the solution to \Eq{eq:anommaster} can be expressed as
\begin{align}\label{eq:rgsolution}
F(\mu) &= F(\mu_0)\exp\left[
2 \int_{\alpha_s(\mu_0)}^{\alpha_s(\mu)} \frac{d\alpha}{\beta(\alpha)} \Gamma_F(\alpha) \int_{\alpha_s(\mu_0)}^{{\alpha}} \frac{d\alpha'}{\beta(\alpha')} + \int_{\alpha_s(\mu_0)}^{\alpha_s(\mu)} \frac{d\alpha}{\beta(\alpha)} \gamma_F(\alpha)\right.\\
&
\hspace{8cm}
\left.
+\,\log\frac{\mu_0^2}{\mu_1^2}\int_{\alpha_s(\mu_0)}^{\alpha_s(\mu)}\frac{d\alpha}{\beta(\alpha)}\Gamma_F(\alpha)
\right] \,,\nonumber
\end{align}
where $\mu_0$ is a reference scale.

The exponentiated kernels can be explicitly evaluated to any logarithmic accuracy given the anomalous dimensions.  The cusp-part of the anomalous dimension, $\Gamma_F(\alpha_s)$, is proportional to the cusp anomalous dimension, $\Gamma_F(\alpha_s) = d_F \Gamma_\text{cusp}$, where $d_F$ includes an appropriate color factor.  The cusp anomalous dimension has an expansion in $\alpha_s$ given by \Eq{eq:cuspexp}.  The non-cusp anomalous dimension has a similar expansion defined in \Eq{eq:gnoncusp}.  For resummation to NNLL accuracy, we need the $\gamma_0$ and $\gamma_1$ coefficients, corresponding to computing the anomalous dimensions of the functions in the factorization theorem to two-loops.

With these expansions, we are able to explicitly evaluate the exponentiated kernel to NNLL accuracy.  We have:
\begin{align}
K_F(\mu,\mu_0)&\equiv 2\int_{\alpha_s(\mu_0)}^{\alpha_s(\mu)}\frac{d\alpha}{\beta(\alpha)}\Gamma_F(\alpha)\int_{\alpha_s(\mu_0)}^{{\alpha}}\frac{d\alpha'}{\beta(\alpha')}+\int_{\alpha_s(\mu_0)}^{\alpha_s(\mu)}\frac{d\alpha}{\beta(\alpha)}\gamma_F(\alpha) \\
&=C_i\frac{\Gamma_0}{2\beta_0^2}\left\{
\frac{4\pi}{\alpha_s(\mu_0)}\left(
\log \,r+\frac{1}{r}-1
\right)+\left(
\frac{\Gamma_1}{\Gamma_0}-\frac{\beta_1}{\beta_0}
\right)(r-1-\log\, r)-\frac{\beta_1}{2\beta_0}\log^2 r \right.\nonumber\\
&
\hspace{2cm}
\left.
+\frac{\alpha_s(\mu_0)}{4\pi}\left[
\left(
\frac{\Gamma_1\beta_1}{\Gamma_0\beta_0}-\frac{\beta_1^2}{\beta_0^2}
\right)(r-1-r\log\, r)-\left(
\frac{\beta_1^2}{\beta_0^2}-\frac{\beta_2}{\beta_0}
\right)\log\, r\nonumber \right.\right.\\
&
\hspace{2cm}
\left.\left.
+\left(
\frac{\Gamma_2}{\Gamma_0}-\frac{\Gamma_1\beta_1}{\Gamma_0\beta_0}+\frac{\beta_1^2}{\beta_0^2}-\frac{\beta_2}{\beta_0}
\right)\frac{r^2-1}{2}+\left(
\frac{\Gamma_2}{\Gamma_0}-\frac{\Gamma_1\beta_1}{\Gamma_0\beta_0}
\right)(1-r)
\right]
\right\} \nonumber \\
&
\hspace{1cm}
-\frac{\gamma_0}{2\beta_0}\log\, r-\frac{\gamma_0}{2\beta_0}\frac{\alpha_s(\mu_0)}{4\pi}\left(
\frac{\gamma_1}{\gamma_0}-\frac{\beta_1}{\beta_0}
\right)(r-1)\,, \nonumber
\end{align}
where $$r=\frac{\alpha_s(\mu)}{\alpha_s(\mu_0)}\,.$$
The other exponentiated factor is
\begin{align}
\omega_F(\mu,\mu_0)&\equiv \int_{\alpha_s(\mu_0)}^{\alpha_s(\mu)}\frac{d\alpha}{\beta(\alpha)}\Gamma_F(\alpha) \\
&=-C_i\frac{\Gamma_0}{{ 2 }\beta_0}\left\{
\log\,r+\frac{\alpha_s(\mu_0)}{4\pi}\left(
\frac{\Gamma_1}{\Gamma_0}-\frac{\beta_1}{\beta_0}
\right)(r-1) \right.\nonumber\\
&
\hspace{2cm}
\left.
+\frac{1}{2}\frac{\alpha_s^2(\mu_0)}{(4\pi)^2}\left(
\frac{\beta_1^2}{\beta_0^2}-\frac{\beta_2}{\beta_0}+\frac{\Gamma_2}{\Gamma_0}-\frac{\Gamma_1\beta_1}{\Gamma_0\beta_0}
\right)(r^2-1)
\right\} \,.\nonumber
\end{align}
Then, we can write the solution in Laplace space to the renormalization group equation in \Eq{eq:rgsolution} as
\begin{equation}\label{eq:lpsolution}
F(\mu) = e^{K_F(\mu,\mu_0)}F(\mu_0)\left(
\frac{\mu_0^2}{\mu_1^2}
\right)^{\omega_F(\mu,\mu_0)} \,.
\end{equation}

Because the hard function and the wide-angle soft function are independent of the observable $\ecf{2}{\alpha}$, their renormalization group equations are identical in real space and Laplace conjugate space.  For the jet functions and the collinear-soft function, the inverse Laplace transform is non-trivial.

For any of the jet functions appearing in the factorization theorem, the Laplace space solution can be written as
\begin{equation}
J(\nu,\mu) = e^{K_J(\mu,\mu_0)}J(\nu,\mu_0)\left[
\frac{\mu_0^2}{E_J^2}\,(\nu e^{\gamma_E})^{2/\alpha}
\right]^{\omega_J(\mu,\mu_0)}\,.
\end{equation}
Note that the logarithms that appear in the low-scale jet function $J(\nu,\mu_0)$ have the same argument as the factor that is raised to the $\omega_J$ power.  Therefore, using the relationship (noted by \Ref{Becher:2006mr})
\begin{equation}
\frac{\partial^n}{\partial q^n} \nu^q = \nu^q\log^n\nu \,,
\end{equation}
we can re-write the jet function as
\begin{equation}
J(\nu,\mu) = e^{K_J(\mu,\mu_0)}J(L\to \partial_{\omega_J})\left[
\frac{\mu_0^2}{E_J^2}\,(\nu\, e^{\gamma_E})^{2/\alpha}
\right]^{\omega_J(\mu,\mu_0)}\,.
\end{equation}
Here $J(L\to \partial_{\omega_J})$ means that the logarithms in the low-scale jet function $J(\nu,\mu_0)$ are replaced by derivatives with respect to the exponentiated factor $\omega_J(\mu,\mu_0)$.  The exact same replacement can be made for the collinear-soft function.  In that case, we have
\begin{equation}
S_{C}(\nu,z_\text{cut},\mu) = e^{K_{S_C}(\mu,\mu_0)}S_{C}(L\to \partial_{\omega_{S_C}})\left[
\frac{\mu_0^2\,(\nu\, e^{\gamma_E})^{2\frac{\beta+1}{\alpha+\beta}}}
{E_J^2\,(\zcut)^{2\frac{\alpha-1}{\alpha+\beta}}}
\right]^{\omega_{S_C}(\mu,\mu_0)}\,.
\end{equation}

This re-writing of the jet and collinear-soft functions allows for very straightforward inverse Laplace transformation.  In Laplace space, the total differential cross section for left and right hemisphere jets in $e^+e^-$ collisions is
\begin{align}
&\sigma(\nu)\\
&
=\exp\left[
K_{H}(\mu,\mu_{H})+K_{S}(\mu,\mu_{S})+K_{S_C}(\mu,\mu_{S_C}^{(L)})+K_{S_C}(\mu,\mu_{S_C}^{(R)})+K_{J}(\mu,\mu_{J}^{(L)})+K_{J}(\mu,\mu_{J}^{(R)})
\right]\nonumber \\
&
\times
H(Q,\mu_{H})\,S(\zcut,\mu_S)\,S_C(L\to \partial_{\omega_{S_C}}^{(L)})\,S_C(L\to \partial_{\omega_{S_C}}^{(R)})\,J(L\to \partial_{\omega_{J}}^{(L)})\,J(L\to \partial_{\omega_{J}}^{(R)}) \nonumber \\
&
\times
\left(
\frac{\mu_{H}^2}{Q^2}
\right)^{\omega_{H}(\mu,\mu_{H})}\left(
\frac{\mu_{S}^2}{4^\beta z_\text{cut}^2{ Q^2}}
\right)^{\omega_{S}(\mu,\mu_{S})}
\left[
\frac{(\mu_{S_C}^{(L)})^2\,(\nu\, e^{\gamma_E})^{2\frac{\beta+1}{\alpha+\beta}}}
{E_J^2\,(\zcut)^{2\frac{\alpha-1}{\alpha+\beta}}}
\right]^{\omega_{S_C}(\mu,\mu_{S_C}^{(L)})}
\nonumber \\
&
\times
\left[
\frac{(\mu_{S_C}^{(R)})^2\,(\nu\, e^{\gamma_E})^{2\frac{\beta+1}{\alpha+\beta}}}
{E_J^2\,(\zcut)^{2\frac{\alpha-1}{\alpha+\beta}}}
\right]^{\omega_{S_C}(\mu,\mu_{S_C}^{(R)})}\left[
\frac{(\mu_{J}^{(L)})^2}{E_J^2}\,(\nu\, e^{\gamma_E})^{2/\alpha}
\right]^{\omega_{J}(\mu,\mu_{J}^{(L)})}\left[
\frac{(\mu_{J}^{(R)})^2}{E_J^2}\,(\nu\, e^{\gamma_E})^{2/\alpha}
\right]^{\omega_{J}(\mu,\mu_{J}^{(R)})\,.}\nonumber
\end{align}
Note that the inverse Laplace transform commutes with the derivatives, and we have
\begin{equation}
{\cal L}^{-1}[\nu^{q}]=\frac{(\ecf{2}{\alpha})^{-q-1}}{\Gamma(-q)} \,.
\end{equation}
Therefore, the differential cross section in real space can be written as:
\begin{align}
&\ecf{2,L}{\alpha}\ecf{2,R}{\alpha}\frac{d^2\sigma}{d\ecf{2,L}{\alpha}\,d\ecf{2,R}{\alpha}}
\\
&
=\exp\left[
K_{H}(\mu,\mu_{H})+K_{S}(\mu,\mu_{S})+K_{S_C}(\mu,\mu_{S_C}^{(L)})+K_{S_C}(\mu,\mu_{S_C}^{(R)})+K_{J}(\mu,\mu_{J}^{(L)})+K_{J}(\mu,\mu_{J}^{(R)})
\right]\nonumber \\
&
\times
H(Q,\mu_{H})\,S(\zcut,\mu_S)\,S_C(L\to \partial_{\omega_{S_C}}^{(L)})\,S_C(L\to \partial_{\omega_{S_C}}^{(R)})\,J(L\to \partial_{\omega_{J}}^{(L)})\,J(L\to \partial_{\omega_{J}}^{(R)}) \nonumber \\
&
\times
\left(
\frac{\mu_{H}^2}{Q^2}
\right)^{\omega_{H}(\mu,\mu_{H})}\left(
\frac{\mu_{S}^2}{4^\beta z_\text{cut}^2{ Q^2}}
\right)^{\omega_{S}(\mu,\mu_{S})}
\left[
\frac{(\mu_{S_C}^{(L)})^2\,\left(\ecf{2,L}{\alpha}\, e^{-\gamma_E}\right)^{-2\frac{\beta+1}{\alpha+\beta}}}
{E_J^2\,(\zcut)^{2\frac{\alpha-1}{\alpha+\beta}}}
\right]^{\omega_{S_C}(\mu,\mu_{S_C}^{(L)})}
\nonumber \\
&
\times
\left[
\frac{(\mu_{S_C}^{(R)})^2\,\left(\ecf{2,R}{\alpha}\, e^{-\gamma_E}\right)^{-2\frac{\beta+1}{\alpha+\beta}}}
{E_J^2\,(\zcut)^{2\frac{\alpha-1}{\alpha+\beta}}}
\right]^{\omega_{S_C}(\mu,\mu_{S_C}^{(R)})}\!\!\left[
\frac{(\mu_{J}^{(L)})^2}{E_J^2}\,\left( \frac{e^{\gamma_E}}{\ecf{2,L}{\alpha}}\right)^{2/\alpha}
\right]^{\omega_{J}(\mu,\mu_{J}^{(L)})}\!\!\left[
\frac{(\mu_{J}^{(R)})^2}{E_J^2}\,\left( \frac{e^{\gamma_E}}{\ecf{2,R}{\alpha}}\right)^{2/\alpha}
\right]^{\omega_{J}(\mu,\mu_{J}^{(R)})}\nonumber\\
&
\times
\left[\Gamma\left(
-\frac{2(\beta+1)}{\alpha+\beta}\,\omega_{S_C}(\mu,\mu_{S_C}^{(L)})-\frac{2}{\alpha}\,\omega_{J}(\mu,\mu_{J}^{(L)})
\right)\Gamma\left(
-\frac{2(\beta+1)}{\alpha+\beta}\,\omega_{S_C}(\mu,\mu_{S_C}^{(R)})-\frac{2}{\alpha}\,\omega_{J}(\mu,\mu_{J}^{(R)})
\right)\right]^{-1}.\nonumber
\end{align}

\section{Renormalization Group Evolution of $D_k$}\label{app:ppfact}

In this appendix we discuss in detail the renormalization group evolution of the jet flavor coefficient $D_k$ and 
explain the procedure we used to estimate the scale uncertainty introduced by neglecting higher-order terms.

The cross section for soft-drop groomed jets in $pp\to Z+j$ events factorizes in the limit $\ecf{2}{\alpha} \ll \zcut \ll 1$, where
\begin{equation}
\hspace{-0.25cm}\frac{d \sigma_\text{resum}}{ d \ecf{2}{\alpha}}=  \sum_{k = q,\bar q,g} \ppmatch_k(p_{T}^{\min},\eta_{\max},\zcut,R) S_{C,k}(\zcut\ecf{2}{\alpha}) \otimes J_k(\ecf{2}{\alpha})\,.
\end{equation}
The fact that $D_k$ depends on multiple scales prohibits its resummation to all orders. 
Nevertheless, its renormalization scale dependence is completely determined by renormalization group invariance of the cross section. 
We can improve our 
prediction by solving the following renormalization group equation, which holds at leading power in $\zcut$:
\begin{align} \label{renRGE}
\frac{\partial \log D_{k} }{\partial \log \mu} & = - \frac{\partial  \log ( J_k \otimes S_{C,k} )}{ \partial \log \mu} \\ 
&=  \Gamma_{D_{k}} (\alpha_s) \log \left( \frac{\mu^2}{ Q^2   }\right) + \gamma_{D_k} (\alpha_s, \zcut )\,,  \nn 
\end{align}
where $Q= 2 \overline{p}_{TJ}$.   The anomalous dimensions $\Gamma_{D_{k}}$ and $\gamma_{D_k}$ are
\begin{align}
\Gamma_{D_k} &=- \frac{\beta}{1+ \beta} C_k \Gamma_{\text{cusp}}\,,\\
\gamma_{D_k} &= - (\gamma_{J_k} +\gamma_{S_{C,k}})- \frac{C_k}{1+ \beta} \Gamma_{\text{cusp}}
\log  z_\text{cut}^2\,.
\end{align}

Here, $C_k$ is the color Casimir for the jet of flavor $k$.  The anomalous dimension has $\log \zcut$ dependence, which means $Q$ is not a natural scale of $D_k$ where all logarithms are minimized.

Nevertheless, we can still  formally  evolve $D_k$  from a scale $\mu_0\sim Q$ to a renormalization scale $\mu$ common to the jet and collinear-soft function.  Solving the renormalization group evolution \Eq{renRGE}, the improved $D_k$ takes the form
\be\label{eq:rgsolD}
D_k (\mu, \mu_f) \equiv  D_k(\alpha_s, \mu_0, \mu_f)  \left( \frac{\mu_0^2}{Q^2} \right)^{\omega_{D_k}(\mu, \mu_0)} e^{K_{D_k}(\mu, \mu_0)}\,.
\ee 
Here, $\mu_f$ represents the factorization scale; i.e., the scale at which the parton distribution functions in $D_k$ are defined.  The $\omega_{D_k}$ and $K_{D_k}$ functions are defined in \App{app:resum}.  To estimate uncertainties from higher-order corrections due to residual scale dependence in $D_k$, we will vary both $\mu_0$ and $\mu_f$ over the values
\begin{align}
\mu_0 &= \left\{ \frac{Q}{2}, Q, 2 Q  \right\} \,,\\
  \mu_f &= \left\{ \frac{Q}{2}, Q, 2Q \right\}  \,.
\end{align}

For evaluating $D_k$ at fixed-order, we keep the full leading and next-to-leading terms as well as singular terms at the next-to-next-to-leading order in the following expansion of the solution to the renormalization group equation, \Eq{eq:rgsolD}.  Expanding $D_k$ in powers of $\alpha_s$ as
\be
D_k(\alpha_s, \mu, \mu_f) = \sum_{n=0} \left( \frac{\alpha_s(\mu)}{ 4 \pi} \right)^{ n}  D_k^{(n) } (\mu, \mu_f)\,,
\ee
we have the solutions:
\begin{align}
D_k^{ (0)}  &  = c_{D_k}^{(0)}\,,  \\
D_k^{ (1)} &= \Gamma^{(0)}_{D_k}   c_{D_k}^{(0)}   \log^2 \frac{\mu}{ Q} +  
\left( \gamma_{D_k}^{(0)}  c_{D_k}^{(0)} + 2 \beta_0  c_{D_k}^{(0)}\, \right)   \log \frac{\mu}{Q} + 
c_{D_k}^{(1)}\,,   \\
D_k^{ (2)}  &= \frac{1}{2}\left( \Gamma^{(0)}_{D_k} \right)^2 c_{D_k}^{(0)}   \log^4 \frac{\mu}{ \mu_0}  +  \left( \gamma_{D_k}^{(0)}\,  \Gamma_{D_k}^{(0)}\, + \frac{8}{3} \, \beta_0  \Gamma_{D_k}^{(0)}\, \right) c_{D_k}^{(0)}   \log^3 \frac{\mu}{ Q} \\
& \quad +   \left[ \left( \Gamma_{D_k}^{(1)} + \frac{1}{2} \left(\gamma^{(0)}_{D_k}\right)^2  + 3  \beta_0  \gamma^{(0)}_{D_k} \right) c_{D_k}^{(0)}\, + \Gamma^{(0)}_{D_k} \, c_{D_k}^{(1)}   \right] \log^2 \frac{\mu}{ Q}   \\
& \quad + \left[  \left( \gamma^{(1)}_{D_k} + 2 \beta_1  \right) c_{D_k}^{(0)} +  \left( \gamma^{(0)}_{D_k} + 4 \beta_0  \right) c_{D_k}^{(1)}  \right]  \log \frac{\mu}{ Q}  +
c^{(2)}_{D_k} \,.
\end{align}
The non-singular terms $c_{D_k}^{(n)}$ are defined such that at $\mu= Q$,
\be
 D_k^{(n)}(Q, \mu_f) = c_{D_k}^{(n)} (Q,\mu_f)\,.
\ee

Therefore one can  extract the value of $c_{D_k}^{(0)}(Q, \mu_f)$ and $c_{D_k}^{(1)}(Q,\mu_f)$  from MCFM and then extrapolate $D_k$ to arbitrary value of $\mu_0$.
Given the current level of precision of MCFM, this procedure can be done through the next-to-leading order.
At $\cO(\alpha_s^3)$, the $c_{D_k}^{(2)}$ term cannot be determined without the next-to-next-to-leading $pp\to Z+j$ cross section. Note that the size of $c_{D_k}^{(2)}$ is no greater than $\cO(\alpha_s^3 \log^4 \zcut)$.
 Thus we can estimate  that the size of uncertainty  introduced by the unknown higher-loop non-singular term $c^{(2)}_{D_k}$ is roughly a factor of $ 1 \pm \alpha_s^2 \log^4 z_\text{cut} \,e^{\alpha_s^n L^{n+1} + \cdots }$, which is beyond NNLL accuracy.

\bibliography{sd_scet}

\providecommand{\href}[2]{#2}\begingroup\raggedright\begin{thebibliography}{100}

\bibitem{Cacciari:2008gn}
M.~Cacciari, G.~P. Salam, and G.~Soyez, {\it {The Catchment Area of Jets}},
  {\em JHEP} {\bf 04} (2008) 005, [\href{http://arxiv.org/abs/0802.1188}{{\tt
  arXiv:0802.1188}}].

\bibitem{Butterworth:2008iy}
J.~M. Butterworth, A.~R. Davison, M.~Rubin, and G.~P. Salam, {\it {Jet
  substructure as a new Higgs search channel at the LHC}},  {\em
  Phys.Rev.Lett.} {\bf 100} (2008) 242001,
  [\href{http://arxiv.org/abs/0802.2470}{{\tt arXiv:0802.2470}}].

\bibitem{Ellis:2009me}
S.~D. Ellis, C.~K. Vermilion, and J.~R. Walsh, {\it {Recombination Algorithms
  and Jet Substructure: Pruning as a Tool for Heavy Particle Searches}},  {\em
  Phys.Rev.} {\bf D81} (2010) 094023,
  [\href{http://arxiv.org/abs/0912.0033}{{\tt arXiv:0912.0033}}].

\bibitem{Krohn:2009th}
D.~Krohn, J.~Thaler, and L.-T. Wang, {\it {Jet Trimming}},  {\em JHEP} {\bf
  1002} (2010) 084, [\href{http://arxiv.org/abs/0912.1342}{{\tt
  arXiv:0912.1342}}].

\bibitem{Soyez:2012hv}
G.~Soyez, G.~P. Salam, J.~Kim, S.~Dutta, and M.~Cacciari, {\it {Pileup
  subtraction for jet shapes}},  {\em Phys.Rev.Lett.} {\bf 110} (2013), no.~16
  162001, [\href{http://arxiv.org/abs/1211.2811}{{\tt arXiv:1211.2811}}].

\bibitem{Dasgupta:2013ihk}
M.~Dasgupta, A.~Fregoso, S.~Marzani, and G.~P. Salam, {\it {Towards an
  understanding of jet substructure}},  {\em JHEP} {\bf 1309} (2013) 029,
  [\href{http://arxiv.org/abs/1307.0007}{{\tt arXiv:1307.0007}}].

\bibitem{Krohn:2013lba}
D.~Krohn, M.~D. Schwartz, M.~Low, and L.-T. Wang, {\it {Jet Cleansing: Pileup
  Removal at High Luminosity}},  {\em Phys. Rev.} {\bf D90} (2014), no.~6
  065020, [\href{http://arxiv.org/abs/1309.4777}{{\tt arXiv:1309.4777}}].

\bibitem{Larkoski:2014wba}
A.~J. Larkoski, S.~Marzani, G.~Soyez, and J.~Thaler, {\it {Soft Drop}},  {\em
  JHEP} {\bf 1405} (2014) 146, [\href{http://arxiv.org/abs/1402.2657}{{\tt
  arXiv:1402.2657}}].

\bibitem{Berta:2014eza}
P.~Berta, M.~Spousta, D.~W. Miller, and R.~Leitner, {\it {Particle-level pileup
  subtraction for jets and jet shapes}},  {\em JHEP} {\bf 06} (2014) 092,
  [\href{http://arxiv.org/abs/1403.3108}{{\tt arXiv:1403.3108}}].

\bibitem{Cacciari:2014gra}
M.~Cacciari, G.~P. Salam, and G.~Soyez, {\it {SoftKiller, a particle-level
  pileup removal method}},  {\em Eur. Phys. J.} {\bf C75} (2015), no.~2 59,
  [\href{http://arxiv.org/abs/1407.0408}{{\tt arXiv:1407.0408}}].

\bibitem{Bertolini:2014bba}
D.~Bertolini, P.~Harris, M.~Low, and N.~Tran, {\it {Pileup Per Particle
  Identification}},  {\em JHEP} {\bf 10} (2014) 59,
  [\href{http://arxiv.org/abs/1407.6013}{{\tt arXiv:1407.6013}}].

\bibitem{Dasgupta:2013via}
M.~Dasgupta, A.~Fregoso, S.~Marzani, and A.~Powling, {\it {Jet substructure
  with analytical methods}},  {\em Eur.Phys.J.} {\bf C73} (2013), no.~11 2623,
  [\href{http://arxiv.org/abs/1307.0013}{{\tt arXiv:1307.0013}}].

\bibitem{Dasgupta:2015yua}
M.~Dasgupta, A.~Powling, and A.~Siodmok, {\it {On jet substructure methods for
  signal jets}},  {\em JHEP} {\bf 08} (2015) 079,
  [\href{http://arxiv.org/abs/1503.01088}{{\tt arXiv:1503.01088}}].

\bibitem{Dasgupta:2001sh}
M.~Dasgupta and G.~P. Salam, {\it {Resummation of nonglobal QCD observables}},
  {\em Phys. Lett.} {\bf B512} (2001) 323--330,
  [\href{http://arxiv.org/abs/hep-ph/0104277}{{\tt hep-ph/0104277}}].

\bibitem{Bauer:2000yr}
C.~W. Bauer, S.~Fleming, D.~Pirjol, and I.~W. Stewart, {\it {An Effective field
  theory for collinear and soft gluons: Heavy to light decays}},  {\em
  Phys.Rev.} {\bf D63} (2001) 114020,
  [\href{http://arxiv.org/abs/hep-ph/0011336}{{\tt hep-ph/0011336}}].

\bibitem{Bauer:2001ct}
C.~W. Bauer and I.~W. Stewart, {\it {Invariant operators in collinear effective
  theory}},  {\em Phys.Lett.} {\bf B516} (2001) 134--142,
  [\href{http://arxiv.org/abs/hep-ph/0107001}{{\tt hep-ph/0107001}}].

\bibitem{Bauer:2001yt}
C.~W. Bauer, D.~Pirjol, and I.~W. Stewart, {\it {Soft collinear factorization
  in effective field theory}},  {\em Phys.Rev.} {\bf D65} (2002) 054022,
  [\href{http://arxiv.org/abs/hep-ph/0109045}{{\tt hep-ph/0109045}}].

\bibitem{Bauer:2002nz}
C.~W. Bauer, S.~Fleming, D.~Pirjol, I.~Z. Rothstein, and I.~W. Stewart, {\it
  {Hard scattering factorization from effective field theory}},  {\em
  Phys.Rev.} {\bf D66} (2002) 014017,
  [\href{http://arxiv.org/abs/hep-ph/0202088}{{\tt hep-ph/0202088}}].

\bibitem{Frye:2016okc}
C.~Frye, A.~J. Larkoski, M.~D. Schwartz, and K.~Yan, {\it {Precision physics
  with pile-up insensitive observables}},
  \href{http://arxiv.org/abs/1603.06375}{{\tt arXiv:1603.06375}}.

\bibitem{Banfi:2004yd}
A.~Banfi, G.~P. Salam, and G.~Zanderighi, {\it {Principles of general
  final-state resummation and automated implementation}},  {\em JHEP} {\bf
  0503} (2005) 073, [\href{http://arxiv.org/abs/hep-ph/0407286}{{\tt
  hep-ph/0407286}}].

\bibitem{Jankowiak:2011qa}
M.~Jankowiak and A.~J. Larkoski, {\it {Jet Substructure Without Trees}},  {\em
  JHEP} {\bf 1106} (2011) 057, [\href{http://arxiv.org/abs/1104.1646}{{\tt
  arXiv:1104.1646}}].

\bibitem{Larkoski:2013eya}
A.~J. Larkoski, G.~P. Salam, and J.~Thaler, {\it {Energy Correlation Functions
  for Jet Substructure}},  {\em JHEP} {\bf 1306} (2013) 108,
  [\href{http://arxiv.org/abs/1305.0007}{{\tt arXiv:1305.0007}}].

\bibitem{Bauer:2011uc}
C.~W. Bauer, F.~J. Tackmann, J.~R. Walsh, and S.~Zuberi, {\it {Factorization
  and Resummation for Dijet Invariant Mass Spectra}},  {\em Phys.Rev.} {\bf
  D85} (2012) 074006, [\href{http://arxiv.org/abs/1106.6047}{{\tt
  arXiv:1106.6047}}].

\bibitem{Procura:2014cba}
M.~Procura, W.~J. Waalewijn, and L.~Zeune, {\it {Resummation of
  Double-Differential Cross Sections and Fully-Unintegrated Parton Distribution
  Functions}},  {\em JHEP} {\bf 02} (2015) 117,
  [\href{http://arxiv.org/abs/1410.6483}{{\tt arXiv:1410.6483}}].

\bibitem{Larkoski:2015zka}
A.~J. Larkoski, I.~Moult, and D.~Neill, {\it {Non-Global Logarithms,
  Factorization, and the Soft Substructure of Jets}},  {\em JHEP} {\bf 09}
  (2015) 143, [\href{http://arxiv.org/abs/1501.04596}{{\tt arXiv:1501.04596}}].

\bibitem{Larkoski:2015kga}
A.~J. Larkoski, I.~Moult, and D.~Neill, {\it {Analytic Boosted Boson
  Discrimination}},  \href{http://arxiv.org/abs/1507.03018}{{\tt
  arXiv:1507.03018}}.

\bibitem{Becher:2015hka}
T.~Becher, M.~Neubert, L.~Rothen, and D.~Y. Shao, {\it {An Effective Field
  Theory for Jet Processes}},  \href{http://arxiv.org/abs/1508.06645}{{\tt
  arXiv:1508.06645}}.

\bibitem{Chien:2015cka}
Y.-T. Chien, A.~Hornig, and C.~Lee, {\it {Soft-collinear mode for jet cross
  sections in soft collinear effective theory}},  {\em Phys. Rev.} {\bf D93}
  (2016), no.~1 014033, [\href{http://arxiv.org/abs/1509.04287}{{\tt
  arXiv:1509.04287}}].

\bibitem{vonManteuffel:2013vja}
A.~von Manteuffel, R.~M. Schabinger, and H.~X. Zhu, {\it {The Complete Two-Loop
  Integrated Jet Thrust Distribution In Soft-Collinear Effective Theory}},
  {\em JHEP} {\bf 03} (2014) 139, [\href{http://arxiv.org/abs/1309.3560}{{\tt
  arXiv:1309.3560}}].

\bibitem{Banfi:2012yh}
A.~Banfi, G.~P. Salam, and G.~Zanderighi, {\it {NLL+NNLO predictions for
  jet-veto efficiencies in Higgs-boson and Drell-Yan production}},  {\em JHEP}
  {\bf 06} (2012) 159, [\href{http://arxiv.org/abs/1203.5773}{{\tt
  arXiv:1203.5773}}].

\bibitem{Becher:2012qa}
T.~Becher and M.~Neubert, {\it {Factorization and NNLL Resummation for Higgs
  Production with a Jet Veto}},  {\em JHEP} {\bf 07} (2012) 108,
  [\href{http://arxiv.org/abs/1205.3806}{{\tt arXiv:1205.3806}}].

\bibitem{Banfi:2012jm}
A.~Banfi, P.~F. Monni, G.~P. Salam, and G.~Zanderighi, {\it {Higgs and Z-boson
  production with a jet veto}},  {\em Phys. Rev. Lett.} {\bf 109} (2012)
  202001, [\href{http://arxiv.org/abs/1206.4998}{{\tt arXiv:1206.4998}}].

\bibitem{Becher:2013xia}
T.~Becher, M.~Neubert, and L.~Rothen, {\it {Factorization and
  $N^{3}LL_{p}$+NNLO predictions for the Higgs cross section with a jet veto}},
   {\em JHEP} {\bf 10} (2013) 125, [\href{http://arxiv.org/abs/1307.0025}{{\tt
  arXiv:1307.0025}}].

\bibitem{Stewart:2013faa}
I.~W. Stewart, F.~J. Tackmann, J.~R. Walsh, and S.~Zuberi, {\it {Jet $p_T$
  resummation in Higgs production at $NNLL'+NNLO$}},  {\em Phys. Rev.} {\bf
  D89} (2014), no.~5 054001, [\href{http://arxiv.org/abs/1307.1808}{{\tt
  arXiv:1307.1808}}].

\bibitem{Catani:1996vz}
S.~Catani and M.~H. Seymour, {\it {A General algorithm for calculating jet
  cross-sections in NLO QCD}},  {\em Nucl. Phys.} {\bf B485} (1997) 291--419,
  [\href{http://arxiv.org/abs/hep-ph/9605323}{{\tt hep-ph/9605323}}]. [Erratum:
  Nucl. Phys.B510,503(1998)].

\bibitem{Bell:2015lsf}
G.~Bell, R.~Rahn, and J.~Talbert, {\it {Automated Calculation of Dijet Soft
  Functions in Soft-Collinear Effective Theory}},  in {\em {Proceedings, 12th
  International Symposium on Radiative Corrections (Radcor 2015) and LoopFest
  XIV (Radiative Corrections for the LHC and Future Colliders)}}, 2015.
\newblock \href{http://arxiv.org/abs/1512.06100}{{\tt arXiv:1512.06100}}.

\bibitem{Berger:2003iw}
C.~F. Berger, T.~Kucs, and G.~F. Sterman, {\it {Event shape / energy flow
  correlations}},  {\em Phys. Rev.} {\bf D68} (2003) 014012,
  [\href{http://arxiv.org/abs/hep-ph/0303051}{{\tt hep-ph/0303051}}].

\bibitem{Almeida:2008yp}
L.~G. Almeida, S.~J. Lee, G.~Perez, G.~F. Sterman, I.~Sung, and J.~Virzi, {\it
  {Substructure of high-$p_T$ Jets at the LHC}},  {\em Phys. Rev.} {\bf D79}
  (2009) 074017, [\href{http://arxiv.org/abs/0807.0234}{{\tt
  arXiv:0807.0234}}].

\bibitem{Ellis:2010rwa}
S.~D. Ellis, C.~K. Vermilion, J.~R. Walsh, A.~Hornig, and C.~Lee, {\it {Jet
  Shapes and Jet Algorithms in SCET}},  {\em JHEP} {\bf 1011} (2010) 101,
  [\href{http://arxiv.org/abs/1001.0014}{{\tt arXiv:1001.0014}}].

\bibitem{Larkoski:2014uqa}
A.~J. Larkoski, D.~Neill, and J.~Thaler, {\it {Jet Shapes with the Broadening
  Axis}},  {\em JHEP} {\bf 1404} (2014) 017,
  [\href{http://arxiv.org/abs/1401.2158}{{\tt arXiv:1401.2158}}].

\bibitem{Chien:2012ur}
Y.-T. Chien, R.~Kelley, M.~D. Schwartz, and H.~X. Zhu, {\it {Resummation of Jet
  Mass at Hadron Colliders}},  {\em Phys. Rev.} {\bf D87} (2013), no.~1 014010,
  [\href{http://arxiv.org/abs/1208.0010}{{\tt arXiv:1208.0010}}].

\bibitem{Jouttenus:2013hs}
T.~T. Jouttenus, I.~W. Stewart, F.~J. Tackmann, and W.~J. Waalewijn, {\it {Jet
  mass spectra in Higgs boson plus one jet at next-to-next-to-leading
  logarithmic order}},  {\em Phys.Rev.} {\bf D88} (2013), no.~5 054031,
  [\href{http://arxiv.org/abs/1302.0846}{{\tt arXiv:1302.0846}}].

\bibitem{Dasgupta:2012hg}
M.~Dasgupta, K.~Khelifa-Kerfa, S.~Marzani, and M.~Spannowsky, {\it {On jet mass
  distributions in Z+jet and dijet processes at the LHC}},  {\em JHEP} {\bf
  1210} (2012) 126, [\href{http://arxiv.org/abs/1207.1640}{{\tt
  arXiv:1207.1640}}].

\bibitem{Feige:2012vc}
I.~Feige, M.~D. Schwartz, I.~W. Stewart, and J.~Thaler, {\it {Precision Jet
  Substructure from Boosted Event Shapes}},  {\em Phys.Rev.Lett.} {\bf 109}
  (2012) 092001, [\href{http://arxiv.org/abs/1204.3898}{{\tt
  arXiv:1204.3898}}].

\bibitem{Catani:1991hj}
S.~Catani, Y.~L. Dokshitzer, M.~Olsson, G.~Turnock, and B.~R. Webber, {\it {New
  clustering algorithm for multi - jet cross-sections in e+ e- annihilation}},
  {\em Phys. Lett.} {\bf B269} (1991) 432--438.

\bibitem{Catani:1993hr}
S.~Catani, Y.~L. Dokshitzer, M.~H. Seymour, and B.~R. Webber, {\it
  {Longitudinally invariant $K_t$ clustering algorithms for hadron hadron
  collisions}},  {\em Nucl. Phys.} {\bf B406} (1993) 187--224.

\bibitem{Ellis:1993tq}
S.~D. Ellis and D.~E. Soper, {\it {Successive combination jet algorithm for
  hadron collisions}},  {\em Phys. Rev.} {\bf D48} (1993) 3160--3166,
  [\href{http://arxiv.org/abs/hep-ph/9305266}{{\tt hep-ph/9305266}}].

\bibitem{Dokshitzer:1997in}
Y.~L. Dokshitzer, G.~D. Leder, S.~Moretti, and B.~R. Webber, {\it {Better jet
  clustering algorithms}},  {\em JHEP} {\bf 08} (1997) 001,
  [\href{http://arxiv.org/abs/hep-ph/9707323}{{\tt hep-ph/9707323}}].

\bibitem{Wobisch:1998wt}
M.~Wobisch and T.~Wengler, {\it {Hadronization corrections to jet
  cross-sections in deep inelastic scattering}},  in {\em {Monte Carlo
  generators for HERA physics. Proceedings, Workshop, Hamburg, Germany,
  1998-1999}}, 1998.
\newblock \href{http://arxiv.org/abs/hep-ph/9907280}{{\tt hep-ph/9907280}}.

\bibitem{Wobisch:2000dk}
M.~Wobisch, {\em {Measurement and QCD analysis of jet cross-sections in deep
  inelastic positron proton collisions at s**(1/2) = 300-GeV}}.
\newblock PhD thesis, Aachen, Tech. Hochsch., 2000.

\bibitem{Cacciari:2008gp}
M.~Cacciari, G.~P. Salam, and G.~Soyez, {\it {The Anti-k(t) jet clustering
  algorithm}},  {\em JHEP} {\bf 0804} (2008) 063,
  [\href{http://arxiv.org/abs/0802.1189}{{\tt arXiv:0802.1189}}].

\bibitem{Fleming:2007qr}
S.~Fleming, A.~H. Hoang, S.~Mantry, and I.~W. Stewart, {\it {Jets from massive
  unstable particles: Top-mass determination}},  {\em Phys. Rev.} {\bf D77}
  (2008) 074010, [\href{http://arxiv.org/abs/hep-ph/0703207}{{\tt
  hep-ph/0703207}}].

\bibitem{Schwartz:2007ib}
M.~D. Schwartz, {\it {Resummation and NLO matching of event shapes with
  effective field theory}},  {\em Phys. Rev.} {\bf D77} (2008) 014026,
  [\href{http://arxiv.org/abs/0709.2709}{{\tt arXiv:0709.2709}}].

\bibitem{Beneke:1997zp}
M.~Beneke and V.~A. Smirnov, {\it {Asymptotic expansion of Feynman integrals
  near threshold}},  {\em Nucl.Phys.} {\bf B522} (1998) 321--344,
  [\href{http://arxiv.org/abs/hep-ph/9711391}{{\tt hep-ph/9711391}}].

\bibitem{Becher:2014oda}
T.~Becher, A.~Broggio, and A.~Ferroglia, {\it {Introduction to Soft-Collinear
  Effective Theory}},  \href{http://arxiv.org/abs/1410.1892}{{\tt
  arXiv:1410.1892}}.

\bibitem{Feige:2013zla}
I.~Feige and M.~D. Schwartz, {\it {An on-shell approach to factorization}},
  {\em Phys. Rev.} {\bf D88} (2013), no.~6 065021,
  [\href{http://arxiv.org/abs/1306.6341}{{\tt arXiv:1306.6341}}].

\bibitem{Feige:2014wja}
I.~Feige and M.~D. Schwartz, {\it {Hard-Soft-Collinear Factorization to All
  Orders}},  {\em Phys. Rev.} {\bf D90} (2014), no.~10 105020,
  [\href{http://arxiv.org/abs/1403.6472}{{\tt arXiv:1403.6472}}].

\bibitem{Feige:2015rea}
I.~Feige, M.~D. Schwartz, and K.~Yan, {\it {Removing phase-space restrictions
  in factorized cross sections}},  {\em Phys. Rev.} {\bf D91} (2015) 094027,
  [\href{http://arxiv.org/abs/1502.05411}{{\tt arXiv:1502.05411}}].

\bibitem{Hornig:2011iu}
A.~Hornig, C.~Lee, I.~W. Stewart, J.~R. Walsh, and S.~Zuberi, {\it {Non-global
  Structure of the $O({\alpha}_s^2)$ Dijet Soft Function}},  {\em JHEP} {\bf
  1108} (2011) 054, [\href{http://arxiv.org/abs/1105.4628}{{\tt
  arXiv:1105.4628}}].

\bibitem{Hornig:2011tg}
A.~Hornig, C.~Lee, J.~R. Walsh, and S.~Zuberi, {\it {Double Non-Global
  Logarithms In-N-Out of Jets}},  {\em JHEP} {\bf 1201} (2012) 149,
  [\href{http://arxiv.org/abs/1110.0004}{{\tt arXiv:1110.0004}}].

\bibitem{Kelley:2011ng}
R.~Kelley, M.~D. Schwartz, R.~M. Schabinger, and H.~X. Zhu, {\it {The two-loop
  hemisphere soft function}},  {\em Phys. Rev.} {\bf D84} (2011) 045022,
  [\href{http://arxiv.org/abs/1105.3676}{{\tt arXiv:1105.3676}}].

\bibitem{Schwartz:2014wha}
M.~D. Schwartz and H.~X. Zhu, {\it {Non-global Logarithms at 3 Loops, 4 Loops,
  5 Loops and Beyond}},  {\em Phys.Rev.} {\bf D90} (2014) 065004,
  [\href{http://arxiv.org/abs/1403.4949}{{\tt arXiv:1403.4949}}].

\bibitem{Khelifa-Kerfa:2015mma}
K.~Khelifa-Kerfa and Y.~Delenda, {\it {Non-global logarithms at finite N$_{c}$
  beyond leading order}},  {\em JHEP} {\bf 03} (2015) 094,
  [\href{http://arxiv.org/abs/1501.00475}{{\tt arXiv:1501.00475}}].

\bibitem{Caron-Huot:2015bja}
S.~Caron-Huot, {\it {Resummation of non-global logarithms and the BFKL
  equation}},  \href{http://arxiv.org/abs/1501.03754}{{\tt arXiv:1501.03754}}.

\bibitem{Neill:2015nya}
D.~Neill, {\it {The Edge of Jets and Subleading Non-Global Logs}},
  \href{http://arxiv.org/abs/1508.07568}{{\tt arXiv:1508.07568}}.

\bibitem{Almeida:2014uva}
L.~G. Almeida, S.~D. Ellis, C.~Lee, G.~Sterman, I.~Sung, and J.~R. Walsh, {\it
  {Comparing and counting logs in direct and effective methods of QCD
  resummation}},  {\em JHEP} {\bf 04} (2014) 174,
  [\href{http://arxiv.org/abs/1401.4460}{{\tt arXiv:1401.4460}}].

\bibitem{Korchemsky:1987wg}
G.~Korchemsky and A.~Radyushkin, {\it {Renormalization of the Wilson Loops
  Beyond the Leading Order}},  {\em Nucl.Phys.} {\bf B283} (1987) 342--364.

\bibitem{Vogt:2000ci}
A.~Vogt, {\it {Next-to-next-to-leading logarithmic threshold resummation for
  deep inelastic scattering and the Drell-Yan process}},  {\em Phys.Lett.} {\bf
  B497} (2001) 228--234, [\href{http://arxiv.org/abs/hep-ph/0010146}{{\tt
  hep-ph/0010146}}].

\bibitem{Berger:2002sv}
C.~F. Berger, {\it {Higher orders in A(alpha(s))/[1-x]+ of nonsinglet partonic
  splitting functions}},  {\em Phys.Rev.} {\bf D66} (2002) 116002,
  [\href{http://arxiv.org/abs/hep-ph/0209107}{{\tt hep-ph/0209107}}].

\bibitem{Moch:2005tm}
S.~Moch, J.~Vermaseren, and A.~Vogt, {\it {Three-loop results for quark and
  gluon form-factors}},  {\em Phys.Lett.} {\bf B625} (2005) 245--252,
  [\href{http://arxiv.org/abs/hep-ph/0508055}{{\tt hep-ph/0508055}}].

\bibitem{Tarasov:1980au}
O.~Tarasov, A.~Vladimirov, and A.~Y. Zharkov, {\it {The Gell-Mann-Low Function
  of QCD in the Three Loop Approximation}},  {\em Phys.Lett.} {\bf B93} (1980)
  429--432.

\bibitem{Larin:1993tp}
S.~Larin and J.~Vermaseren, {\it {The Three loop QCD Beta function and
  anomalous dimensions}},  {\em Phys.Lett.} {\bf B303} (1993) 334--336,
  [\href{http://arxiv.org/abs/hep-ph/9302208}{{\tt hep-ph/9302208}}].

\bibitem{vanNeerven:1985xr}
W.~L. van Neerven, {\it {Dimensional Regularization of Mass and Infrared
  Singularities in Two Loop On-shell Vertex Functions}},  {\em Nucl. Phys.}
  {\bf B268} (1986) 453.

\bibitem{Matsuura:1988sm}
T.~Matsuura, S.~C. van~der Marck, and W.~L. van Neerven, {\it {The Calculation
  of the Second Order Soft and Virtual Contributions to the Drell-Yan
  Cross-Section}},  {\em Nucl. Phys.} {\bf B319} (1989) 570.

\bibitem{Bauer:2003pi}
C.~W. Bauer and A.~V. Manohar, {\it {Shape function effects in B ---> X(s)
  gamma and B ---> X(u) l anti-nu decays}},  {\em Phys. Rev.} {\bf D70} (2004)
  034024, [\href{http://arxiv.org/abs/hep-ph/0312109}{{\tt hep-ph/0312109}}].

\bibitem{Bosch:2004th}
S.~W. Bosch, B.~O. Lange, M.~Neubert, and G.~Paz, {\it {Factorization and shape
  function effects in inclusive B meson decays}},  {\em Nucl. Phys.} {\bf B699}
  (2004) 335--386, [\href{http://arxiv.org/abs/hep-ph/0402094}{{\tt
  hep-ph/0402094}}].

\bibitem{Becher:2010pd}
T.~Becher and G.~Bell, {\it {The gluon jet function at two-loop order}},  {\em
  Phys. Lett.} {\bf B695} (2011) 252--258,
  [\href{http://arxiv.org/abs/1008.1936}{{\tt arXiv:1008.1936}}].

\bibitem{Becher:2006qw}
T.~Becher and M.~Neubert, {\it {Toward a NNLO calculation of the anti-B --->
  X(s) gamma decay rate with a cut on photon energy. II. Two-loop result for
  the jet function}},  {\em Phys. Lett.} {\bf B637} (2006) 251--259,
  [\href{http://arxiv.org/abs/hep-ph/0603140}{{\tt hep-ph/0603140}}].

\bibitem{Neubert:2004dd}
M.~Neubert, {\it {Renormalization-group improved calculation of the B ---> X(s)
  gamma branching ratio}},  {\em Eur. Phys. J.} {\bf C40} (2005) 165--186,
  [\href{http://arxiv.org/abs/hep-ph/0408179}{{\tt hep-ph/0408179}}].

\bibitem{Becher:2006mr}
T.~Becher, M.~Neubert, and B.~D. Pecjak, {\it {Factorization and Momentum-Space
  Resummation in Deep-Inelastic Scattering}},  {\em JHEP} {\bf 01} (2007) 076,
  [\href{http://arxiv.org/abs/hep-ph/0607228}{{\tt hep-ph/0607228}}].

\bibitem{Catani:1999ss}
S.~Catani and M.~Grazzini, {\it {Infrared factorization of tree level QCD
  amplitudes at the next-to-next-to-leading order and beyond}},  {\em
  Nucl.Phys.} {\bf B570} (2000) 287--325,
  [\href{http://arxiv.org/abs/hep-ph/9908523}{{\tt hep-ph/9908523}}].

\bibitem{Chien:2010kc}
Y.-T. Chien and M.~D. Schwartz, {\it {Resummation of heavy jet mass and
  comparison to LEP data}},  {\em JHEP} {\bf 1008} (2010) 058,
  [\href{http://arxiv.org/abs/1005.1644}{{\tt arXiv:1005.1644}}].

\bibitem{Bahr:2008pv}
M.~Bahr, S.~Gieseke, M.~Gigg, D.~Grellscheid, K.~Hamilton, et~al., {\it
  {Herwig++ Physics and Manual}},  {\em Eur.Phys.J.} {\bf C58} (2008) 639--707,
  [\href{http://arxiv.org/abs/0803.0883}{{\tt arXiv:0803.0883}}].

\bibitem{Bellm:2013hwb}
J.~Bellm et~al., {\it {Herwig++ 2.7 Release Note}},
  \href{http://arxiv.org/abs/1310.6877}{{\tt arXiv:1310.6877}}.

\bibitem{Sjostrand:2006za}
T.~Sjostrand, S.~Mrenna, and P.~Z. Skands, {\it {PYTHIA 6.4 Physics and
  Manual}},  {\em JHEP} {\bf 0605} (2006) 026,
  [\href{http://arxiv.org/abs/hep-ph/0603175}{{\tt hep-ph/0603175}}].

\bibitem{Sjostrand:2014zea}
{\it {An Introduction to PYTHIA 8.2}},  {\em Comput. Phys. Commun.} {\bf 191}
  (2015) 159--177, [\href{http://arxiv.org/abs/1410.3012}{{\tt
  arXiv:1410.3012}}].

\bibitem{Giele:2007di}
W.~T. Giele, D.~A. Kosower, and P.~Z. Skands, {\it {A simple shower and
  matching algorithm}},  {\em Phys. Rev.} {\bf D78} (2008) 014026,
  [\href{http://arxiv.org/abs/0707.3652}{{\tt arXiv:0707.3652}}].

\bibitem{Giele:2011cb}
W.~T. Giele, D.~A. Kosower, and P.~Z. Skands, {\it {Higher-Order Corrections to
  Timelike Jets}},  {\em Phys. Rev.} {\bf D84} (2011) 054003,
  [\href{http://arxiv.org/abs/1102.2126}{{\tt arXiv:1102.2126}}].

\bibitem{Hartgring:2013jma}
L.~Hartgring, E.~Laenen, and P.~Skands, {\it {Antenna Showers with One-Loop
  Matrix Elements}},  {\em JHEP} {\bf 10} (2013) 127,
  [\href{http://arxiv.org/abs/1303.4974}{{\tt arXiv:1303.4974}}].

\bibitem{Larkoski:2013yi}
A.~J. Larkoski, J.~J. Lopez-Villarejo, and P.~Skands, {\it {Helicity-Dependent
  Showers and Matching with VINCIA}},  {\em Phys. Rev.} {\bf D87} (2013), no.~5
  054033, [\href{http://arxiv.org/abs/1301.0933}{{\tt arXiv:1301.0933}}].

\bibitem{Catani:1990rr}
S.~Catani, B.~R. Webber, and G.~Marchesini, {\it {QCD coherent branching and
  semiinclusive processes at large x}},  {\em Nucl. Phys.} {\bf B349} (1991)
  635--654.

\bibitem{Dokshitzer:1995ev}
Y.~L. Dokshitzer, V.~A. Khoze, and S.~I. Troian, {\it {Specific features of
  heavy quark production. LPHD approach to heavy particle spectra}},  {\em
  Phys. Rev.} {\bf D53} (1996) 89--119,
  [\href{http://arxiv.org/abs/hep-ph/9506425}{{\tt hep-ph/9506425}}].

\bibitem{Cacciari:2011ma}
M.~Cacciari, G.~P. Salam, and G.~Soyez, {\it {FastJet User Manual}},  {\em
  Eur.Phys.J.} {\bf C72} (2012) 1896,
  [\href{http://arxiv.org/abs/1111.6097}{{\tt arXiv:1111.6097}}].

\bibitem{Korchemsky:1999kt}
G.~P. Korchemsky and G.~F. Sterman, {\it {Power corrections to event shapes and
  factorization}},  {\em Nucl. Phys.} {\bf B555} (1999) 335--351,
  [\href{http://arxiv.org/abs/hep-ph/9902341}{{\tt hep-ph/9902341}}].

\bibitem{Korchemsky:2000kp}
G.~P. Korchemsky and S.~Tafat, {\it {On power corrections to the event shape
  distributions in QCD}},  {\em JHEP} {\bf 10} (2000) 010,
  [\href{http://arxiv.org/abs/hep-ph/0007005}{{\tt hep-ph/0007005}}].

\bibitem{Stewart:2014nna}
I.~W. Stewart, F.~J. Tackmann, and W.~J. Waalewijn, {\it {Dissecting Soft
  Radiation with Factorization}},  {\em Phys.Rev.Lett.} {\bf 114} (2015), no.~9
  092001, [\href{http://arxiv.org/abs/1405.6722}{{\tt arXiv:1405.6722}}].

\bibitem{Becher:2014aya}
T.~Becher, R.~Frederix, M.~Neubert, and L.~Rothen, {\it {Automated NNLL $+$ NLO
  resummation for jet-veto cross sections}},  {\em Eur. Phys. J.} {\bf C75}
  (2015), no.~4 154, [\href{http://arxiv.org/abs/1412.8408}{{\tt
  arXiv:1412.8408}}].

\bibitem{Farhi:2015jca}
D.~Farhi, I.~Feige, M.~Freytsis, and M.~D. Schwartz, {\it {Streamlining
  resummed QCD calculations using Monte Carlo integration}},
  \href{http://arxiv.org/abs/1507.06315}{{\tt arXiv:1507.06315}}.

\bibitem{Banfi:2006hf}
A.~Banfi, G.~P. Salam, and G.~Zanderighi, {\it {Infrared safe definition of jet
  flavor}},  {\em Eur. Phys. J.} {\bf C47} (2006) 113--124,
  [\href{http://arxiv.org/abs/hep-ph/0601139}{{\tt hep-ph/0601139}}].

\bibitem{Campbell:2002tg}
J.~M. Campbell and R.~K. Ellis, {\it {Next-to-leading order corrections to
  $W^+$ 2 jet and $Z^+$ 2 jet production at hadron colliders}},  {\em Phys.
  Rev.} {\bf D65} (2002) 113007,
  [\href{http://arxiv.org/abs/hep-ph/0202176}{{\tt hep-ph/0202176}}].

\bibitem{Campbell:2003hd}
J.~M. Campbell, R.~K. Ellis, and D.~L. Rainwater, {\it {Next-to-leading order
  QCD predictions for $W$ + 2 jet and $Z$ + 2 jet production at the CERN LHC}},
   {\em Phys. Rev.} {\bf D68} (2003) 094021,
  [\href{http://arxiv.org/abs/hep-ph/0308195}{{\tt hep-ph/0308195}}].

\bibitem{Martin:2009iq}
A.~D. Martin, W.~J. Stirling, R.~S. Thorne, and G.~Watt, {\it {Parton
  distributions for the LHC}},  {\em Eur. Phys. J.} {\bf C63} (2009) 189--285,
  [\href{http://arxiv.org/abs/0901.0002}{{\tt arXiv:0901.0002}}].

\bibitem{fjcontrib}
``Fastjet contrib.'' \url{http://fastjet.hepforge.org/contrib/}.

\bibitem{Collins:1984kg}
J.~C. Collins, D.~E. Soper, and G.~F. Sterman, {\it {Transverse Momentum
  Distribution in Drell-Yan Pair and W and Z Boson Production}},  {\em Nucl.
  Phys.} {\bf B250} (1985) 199.

\bibitem{Rothstein:2016bsq}
I.~Z. Rothstein and I.~W. Stewart, {\it {An Effective Field Theory for Forward
  Scattering and Factorization Violation}},
  \href{http://arxiv.org/abs/1601.04695}{{\tt arXiv:1601.04695}}.

\bibitem{Becher:2008cf}
T.~Becher and M.~D. Schwartz, {\it {A precise determination of $\alpha_s$ from
  LEP thrust data using effective field theory}},  {\em JHEP} {\bf 07} (2008)
  034, [\href{http://arxiv.org/abs/0803.0342}{{\tt arXiv:0803.0342}}].

\bibitem{Abbate:2010xh}
R.~Abbate, M.~Fickinger, A.~H. Hoang, V.~Mateu, and I.~W. Stewart, {\it {Thrust
  at N$^3$LL with Power Corrections and a Precision Global Fit for
  alphas(mZ)}},  {\em Phys. Rev.} {\bf D83} (2011) 074021,
  [\href{http://arxiv.org/abs/1006.3080}{{\tt arXiv:1006.3080}}].

\bibitem{Hoang:2014wka}
A.~H. Hoang, D.~W. Kolodrubetz, V.~Mateu, and I.~W. Stewart, {\it
  {$C$-parameter distribution at N$^3$LL? including power corrections}},  {\em
  Phys. Rev.} {\bf D91} (2015), no.~9 094017,
  [\href{http://arxiv.org/abs/1411.6633}{{\tt arXiv:1411.6633}}].

\bibitem{Ridder:2014wza}
A.~Gehrmann-De~Ridder, T.~Gehrmann, E.~Glover, and G.~Heinrich, {\it {EERAD3:
  Event shapes and jet rates in electron-positron annihilation at order
  $\alpha_s^3$}},  {\em Comput.Phys.Commun.} {\bf 185} (2014) 3331,
  [\href{http://arxiv.org/abs/1402.4140}{{\tt arXiv:1402.4140}}].

\bibitem{Bauer:2003di}
C.~W. Bauer, C.~Lee, A.~V. Manohar, and M.~B. Wise, {\it {Enhanced
  nonperturbative effects in Z decays to hadrons}},  {\em Phys. Rev.} {\bf D70}
  (2004) 034014, [\href{http://arxiv.org/abs/hep-ph/0309278}{{\tt
  hep-ph/0309278}}].

\bibitem{Manohar:2003vb}
A.~V. Manohar, {\it {Deep inelastic scattering as x ---> 1 using soft collinear
  effective theory}},  {\em Phys. Rev.} {\bf D68} (2003) 114019,
  [\href{http://arxiv.org/abs/hep-ph/0309176}{{\tt hep-ph/0309176}}].

\bibitem{Harlander:2000mg}
R.~V. Harlander, {\it {Virtual corrections to g g ---> H to two loops in the
  heavy top limit}},  {\em Phys. Lett.} {\bf B492} (2000) 74--80,
  [\href{http://arxiv.org/abs/hep-ph/0007289}{{\tt hep-ph/0007289}}].

\bibitem{Anastasiou:2002wq}
C.~Anastasiou and K.~Melnikov, {\it {Pseudoscalar Higgs boson production at
  hadron colliders in NNLO QCD}},  {\em Phys. Rev.} {\bf D67} (2003) 037501,
  [\href{http://arxiv.org/abs/hep-ph/0208115}{{\tt hep-ph/0208115}}].

\bibitem{Ravindran:2004mb}
V.~Ravindran, J.~Smith, and W.~L. van Neerven, {\it {Two-loop corrections to
  Higgs boson production}},  {\em Nucl. Phys.} {\bf B704} (2005) 332--348,
  [\href{http://arxiv.org/abs/hep-ph/0408315}{{\tt hep-ph/0408315}}].

\bibitem{Becher:2009th}
T.~Becher and M.~D. Schwartz, {\it {Direct photon production with effective
  field theory}},  {\em JHEP} {\bf 02} (2010) 040,
  [\href{http://arxiv.org/abs/0911.0681}{{\tt arXiv:0911.0681}}].

\end{thebibliography}\endgroup

\end{document}